\documentclass[prb,aps,twocolumn,showpacs,amsmath,amssymb,floatfix,eqsecnum,superscriptaddress]{revtex4-1}
\usepackage{graphicx}
\usepackage{times}
\usepackage{color}
\usepackage{amsmath}
\usepackage{amstext}
\usepackage{amssymb}
\usepackage[colorlinks,citecolor=blue]{hyperref}
\usepackage{graphicx}
\usepackage{amsmath}
\usepackage{amstext}
\usepackage{amssymb}
\usepackage{amsfonts}
\usepackage{longtable,booktabs}
\usepackage{hyperref}\usepackage{url}%Turn on when output to pdf file
\usepackage{subfigure}%
\usepackage{dsfont}

\usepackage{amsbsy}
\usepackage{dcolumn}
\usepackage{amsthm}
\usepackage{bm}
\usepackage{esint}
\usepackage{multirow}
\usepackage{hyperref}
\hypersetup{
    colorlinks=true,
    linkcolor=blue,
    filecolor=magenta,
    urlcolor=cyan,
}
\usepackage{cleveref}

\usepackage{mathrsfs}
\usepackage{amsfonts}
\usepackage{amsbsy}
\usepackage{dcolumn}
\usepackage{bm}
\usepackage{multirow}
\usepackage{color}
\usepackage{slashed}

\begin{document}

\def \beq {\begin{equation}}
\def \eeq {\end{equation}}
\def \beqa {\begin{eqnarray}}

\def \eeqa {\end{eqnarray}}
\newcommand \dg {\dagger}
\newcommand \up {\uparrow}
\newcommand \down {\downarrow}
\newcommand \al {\alpha}
\newcommand \be {\beta}
\newcommand \sg {\sigma}
\newcommand \ran {\rangle}
\newcommand \lan {\langle}
\newcommand \un {\underline}
\newcommand \ep {\epsilon}
\newcommand \lam {\lambda}
\newcommand \pd {\partial}
\newcommand \pdb {\bar{\partial}}
\newcommand \mb {\mathbf}
\newcommand \mbs {\boldsymbol}
\newcommand \yhat {\mathbf{\hat{y}}}
\newcommand \tb {\bar{t}}
\newcommand \nnb {\nonumber}
\newcommand \D {\Delta}
\newcommand \T {\hat{T}}
\newcommand \K {\hat{K}}
\newcommand \II {\mathbb{I}}
\newcommand \gamt {\tilde{\gamma}}
\newcommand \gam {\gamma}
\newcommand \dt {\tilde{d}}
\newcommand \Om {\Omega}
\newcommand \J {\mathcal{J}}
\newcommand \vphi {\varphi}
\newcommand \XX {\mathbb{X}}
\def\S{\textrm{S}}
\def\ga{\gamma}
\def\Ga{\Gamma}
\def\ka{\kappa}
\def\sgn{{\rm sgn}}
\def\tcrc{\textcolor{red}{citation}}
\def\HG{\mathcal{H}_G}
\def\distAB{\text{dist}(A,B)}
\def\Tr{\text{Tr}}
\def\Vol{\text{Vol}}
\def\ShannonTO{4}

\title{Topological Electromagnetic Responses of Bosonic Quantum Hall, Topological Insulator, and Chiral Semi-Metal phases in All Dimensions}

\author{Matthew F. Lapa}
\affiliation{Department of Physics and Institute for Condensed Matter Theory, University of Illinois at Urbana-Champaign, 61801-3080}
\author{Chao-Ming Jian}
\affiliation{Kavli Institute for Theoretical Physics, University of California, Santa Barbara, California, 93106, USA}
\affiliation{Station Q, Microsoft Research, Santa Barbara, California 93106-6105, USA}
\author{Peng Ye}
\affiliation{Department of Physics and Institute for Condensed Matter Theory, University of Illinois at Urbana-Champaign, 61801-3080}
\author{Taylor L. Hughes}
\affiliation{Department of Physics and Institute for Condensed Matter Theory, University of Illinois at Urbana-Champaign, 61801-3080}

\date{\today}

\begin{abstract}

We calculate the topological part of the electromagnetic response of Bosonic Integer Quantum Hall (BIQH) phases in odd (spacetime) dimensions, and Bosonic Topological Insulator (BTI) and Bosonic chiral semi-metal (BCSM) phases in even dimensions. To do this we use the Nonlinear Sigma Model (NLSM) description of bosonic symmetry-protected topological (SPT) phases, and the method of gauged Wess-Zumino (WZ) actions. We find the surprising result that for BIQH states in dimension $2m-1$ ($m=1,2,\dots$), the bulk response to an electromagnetic field $A_{\mu}$ is characterized by a Chern-Simons term for $A_{\mu}$ with a level quantized in integer multiples of $m!$ (factorial). We also show that BTI states (which have an extra $\mathbb{Z}_2$ symmetry) can exhibit a $\mathbb{Z}_2$ breaking Quantum Hall effect on their boundaries, with this boundary Quantum Hall effect described by a Chern-Simons term at level $\frac{m!}{2}$. We show that the factor of $m!$ can be understood by requiring gauge invariance of the exponential of the Chern-Simons term on a general Euclidean manifold, and we also use this argument to characterize the electromagnetic \emph{and} gravitational responses of fermionic SPT phases with $U(1)$ symmetry in all odd dimensions. We then use our gauged boundary actions for the BIQH and BTI states to (i) construct a bosonic analogue of a chiral semi-metal (BCSM) in even dimensions, (ii) show that the boundary of the BTI state exhibits a bosonic analogue of the parity anomaly of Dirac fermions in odd dimensions, and (iii) study anomaly inflow at domain walls on the boundary of BTI states. In a series of Appendices we derive important formulas and additional results. In particular, in Appendix~\ref{app:EC} we use the connection between equivariant cohomology and gauged WZ actions to give a mathematical interpretation of the actions for the BIQH and BTI boundaries constructed in this paper.

\end{abstract}

\maketitle 

\tableofcontents

\section{Introduction}
\label{sec:intro}

In the years since the theoretical prediction and experimental discovery of the electron topological insulators\cite{kanehasan,qi2011topological}, 
the study of symmetry-protected topological (SPT) phases of matter 
\cite{senthil2015review,gu2009tensor,pollmann2010entanglement,chen2013symmetry,chen2012symmetry} 
has emerged as an extremely rich subfield of condensed matter physics,
with interesting and surprising connections to high-energy physics and mathematics. 
Although there has been tremendous progress in the understanding of these states of matter, some basic issues about these
phases are still the subject of intense investigation. As illustrative examples we point to the 
question of which theories can describe a surface termination of the time-reversal invariant electron topological insulator in
three spatial
dimensions\cite{son2015composite,wang2015dual,metlitski2016particle,mross2015,seiberg2016gapped,seiberg2016duality,karch2016particle}, 
as well as the analogous question for the surface of the bosonic topological insulator in three spatial 
dimensions\cite{xu2015,lapa2016bosonic}. 

A very useful definition of an SPT phase is as follows\cite{levin2012braiding}. 
Consider a quantum many-body system with Hamiltonian $H$, where $H$
has the symmetries of a group $G$ and a gapped spectrum. Then the ground state $|\Psi\ran$ of $H$ represents an SPT phase 
if it satisfies several properties. First, $|\Psi\ran$ should be unique independently of the topology of the (closed) 
spatial manifold that $H$ is
defined on. This ensures that the ground state of $H$ does not represent a phase with topological order (no excitations with 
fractional charge or statistics, etc.). 
Second, $|\Psi\ran$ should be invariant under the action of $G$, i.e., $U(g)|\Psi\ran = |\Psi\ran$ for any $g\in G$, where
$U(g)$ is a representation of $G$ on the Hilbert space of the system. 
This means that the ground state of $H$ does not spontaneously 
break the symmetry of the group $G$. Finally, $|\Psi\ran$ cannot be continuously tuned to a \emph{trivial}
product state (e.g., by adding terms to the Hamiltonian) without (i) breaking the symmetry of $G$, or (ii) closing the gap in the 
spectrum of $H$. Despite the lack of anyon excitations in the bulk, interesting degrees of freedom will in general be present at the boundary of
an SPT phase.

In this paper we focus our primary attention on bosonic SPT phases
and, in particular, on those bosonic SPT phases which are analogues of more familiar topological phases of fermions. We
are especially interested in the Bosonic Integer Quantum Hall (BIQH) 
effect\cite{SenthilLevin,VL2012,grover2013,regnault2013,he2015,moller2009,moller2015,ye2013,liu2014,furukawa2013integer,wu2013quantum}, 
a bosonic analog of the ordinary $\nu=1$ 
Integer Quantum Hall effect of fermions in three (spacetime) dimensions, and in the time-reversal invariant Bosonic Topological Insulator 
(BTI)\cite{VS2013,MKF2013,SenthilWang2013,ye2015vortex}, 
a bosonic analogue of the time-reversal invariant electron topological insulator in four dimensions. 
In fact, the main goal
of our paper is to consider generalizations of the BIQH and BTI states to \emph{all} odd and even spacetime dimensions, 
respectively, and then to study the physical properties of these higher-dimensional states. 
The reader should note that in the remainder of this paper the word 
``dimension'' always refers to the \emph{spacetime} dimension. We always write ``spatial dimension'' when we want to discuss the
dimension of space only.

BIQH phases are protected only by $U(1)$ charge-conservation symmetry, while the BTI phase
is protected by the symmetry group $U(1)\rtimes \mathbb{Z}_2$, where, as we discuss later, the $\mathbb{Z}_2$ symmetry is unitary 
charge-conjugation symmetry $\mathbb{Z}^C_2$ in dimensions $2,6,10$, etc., and anti-unitary time-reversal symmetry 
$\mathbb{Z}^T_2$ in dimensions $4,8,12,$ etc. The symbol ``$\rtimes$" means that the $U(1)$ and $\mathbb{Z}_2$ symmetry 
operations do not commute with each other. Since both of these phases have $U(1)$ charge-conservation symmetry, they
can both be coupled to an external electromagnetic field $A_{\mu}$. One can then study the electromagnetic response of these
states.

One of our main results 
 in this paper is an explicit derivation of the (topological part of) the
electromagnetic response of BIQH phases in all odd dimensions and BTI phases in all even dimensions. From a physical standpoint the
magnitude of the electromagnetic response is extremely interesting, as it is known already in three dimensions that the requirement 
that a BIQH state have no topological order places a constraint on the allowed values of the Hall conductance of any putative 
BIQH state\cite{SenthilLevin}. In particular, the Hall conductance must be a multiple of $2$ (in units of $\frac{e^2}{h}$), i.e.,
a BIQH state has twice the Hall conductance that a free fermion Integer Quantum Hall state can have.
In higher dimensions we also find that the electromagnetic response of the BIQH state is some integer multiple of the minimum
value which can be realized by free fermions, and we find analogous results in even dimensions for BTI states. 

To calculate the electromagnetic response of these states, we need a concrete model to work with. For reasons to be discussed in the
next section, we choose to use the Nonlinear Sigma Model (NLSM) description of bosonic SPT 
phases\cite{xu2013wave,CenkeClass1,CenkeClass2,CenkeBraiding,CenkeLine,CenkeSU2N}. This allows us to use the 
theory of gauged Wess-Zumino (WZ) actions\cite{witten1983global,manes1985,wu1985cohomology,HullSpence1,WittenHolo,HullSpence2} 
to study the 
boundary of these states, and from our study of the boundary we are able to deduce the bulk response. As a byproduct, our 
explicit construction of gauged WZ actions for the boundaries of these states allows us to study several physical properties of
these states in more detail. We show that the boundary theory for the BTI displays a bosonic analogue of the parity
anomaly for Dirac fermions in odd
dimensions\cite{Redlich,niemi1983axial,alvarez1985anomalies,witten2015fermion,witten2016parity}, and we also use the
boundary theory of the BIQH state to construct effective theories for bosonic analogues of 
Weyl (or chiral) semi-metals in all even dimensions. 

For the case of the BIQH state, we also provide an
alternative derivation of the response by requiring the gauge invariance of (the exponential of) the Chern-Simons 
functional describing the electromagnetic response of the state. We also use this gauge invariance argument to derive and discuss
the electromagnetic and gravitational responses of Fermionic Integer Quantum Hall (FIQH) phases in different dimensions. This gauge invariance argument provides us with a general understanding of the difference in the quantization of response coefficients of BIQH and FIQH phases.

Before moving on, we take this opportunity to provide some justification for our study of bosonic SPT phases in dimensions higher
than the physically relevant dimensions of two, three, and four. Studying a state of matter in generic dimensions can often
reveal underlying organizational principles or mathematical structures which cannot be seen by studying low-dimensional examples on
their own. An obvious example of this is the periodic table of topological insulators and 
superconductors\cite{kitaev2009periodic,ryu2010topological}, which exhibits an eightfold periodicity in the dimension of space 
(i.e., the pattern does not completely develop if one considers only low dimensions). 
In the case of bosonic SPT phases, low-dimensional examples suggest that the response of the bosonic analogue of a given
fermionic state (Integer Quantum Hall or electron topological insulator) is twice that of its fermionic counterpart. 
However, our results in this paper clearly show that this is \emph{not} the case in higher dimensions. Finally, it is also worth 
mentioning that many new insights on four-dimensional physics can be gained by imagining that our four-dimensional spacetime
is the boundary of a five-dimensional SPT phase\cite{kravec2013gauge,wen2013lattice,you2015interacting}. 

This paper is organized as follows. First, in Sec.~\ref{sec:summary} we outline our basic approach
and summarize our main results.
In Sec.~\ref{sec:background} we review the relevant background information on 
BIQH and BTI phases, the NLSM description of SPT phases, and the method of gauged WZ actions. In Sec.~\ref{sec:BIQH}
we construct the gauged WZ action for the boundary of the BIQH phase, and we use the anomaly of the gauged boundary
action to deduce the bulk response of the BIQH phase. We also give an alternative derivation of the BIQH response which
relies on only the bulk physics of the NLSM.  In Sec.~\ref{sec:gauge-invariance} we use a general gauge
invariance argument to understand the electromagnetic response of BIQH states, and also the electromagnetic and 
gravitational responses of FIQH states in odd dimensions.  In particular, we illuminate the important differences between the quantization of response coefficients in BIQH and FIQH phases. In Sec.~\ref{sec:BTI} we construct the gauged WZ action for the
boundary of the BTI phase, and we use the gauged boundary action to study the symmetry-breaking BIQH response  
of the BTI boundary. In Sec.~\ref{sec:applications} we use the results from Sec.~\ref{sec:BIQH} and Sec.~\ref{sec:BTI} 
to (i) construct effective theories for bosonic analogues of Weyl, or chiral, semi-metals in all even dimensions, 
(ii) show that the boundary of a BTI
state displays an analogue of the parity anomaly for Dirac fermions in odd dimensions, and (iii) study the physics of symmetry-breaking
domain walls on the boundary of BTI states. Sec.~\ref{sec:conclusion} presents our conclusions. Finally, in a series of 
Appendices we examine the results of the paper from a more mathematical point of view, and also derive several important 
formulas which are used throughout the paper.

\section{Basic approach and Summary of Results}
\label{sec:summary}

In this section we outline our basic approach to calculating the electromagnetic response of higher-dimensional bosonic SPT phases,
and then we present our results. In this paper we work in units where $\hbar=e=1$, where $e$ is the charge of the basic
particles (bosons or fermions) which make up the state we are interested in. To restore $e$ in any formula one can simply replace
$A_{\mu}$ (the external electromagnetic field) with $e A_{\mu}$.

Let us first discuss the general form that the topological part of the electromagnetic response is
expected to take for BIQH and BTI states.
In odd dimensions, the response of a higher-dimensional analogue of a Quantum Hall state
to an external field $A=A_{\mu}dx^{\mu}$ (we use differential form notation) is characterized by a Chern-Simons (CS) term 
$S_{CS}[A]$ in the effective action for the external field. In $2m-1$ dimensions this term takes the form 
\beq
	S_{CS}[A]= \frac{N_{2m-1}}{(2\pi)^{m-1} m!}\int_{\mathcal{M}} A\wedge F^{m-1}\ , \label{eq:CS-term}
\eeq
where $N_{2m-1}$ is called the \emph{level} of the CS term, $F=dA$, $F^{m-1}$ is shorthand for the 
wedge product of $F$ with itself $m-1$ times, and $\mathcal{M}$ represents the spacetime manifold. Let
us also note here that all actions in the paper are written down in Minkowski signature (real time) except in Sec.~\ref{sec:gauge-invariance} and
Appendix~\ref{app:CP}, where we consider CS and other terms in Euclidean spacetimes.
On the other hand, the response of an analogue of a topological insulator in $2m$ dimensions is characterized by a ``Chern character" term (we avoid using ``theta-term" here since that name is also used for a type of topological term in the NLSM action),
\beq
	S_{CC}[A]= \frac{\Theta_{2m}}{(2\pi)^m m!}\int_{\mathcal{M}}F^m \ . \label{eq:chern-character}
\eeq
Here the coefficient $\Theta_{2m}$ should be interpreted as an angular variable, although its period is not necessarily $2\pi$. We
call this term a ``Chern character" term as the quantity $\frac{1}{m!}\left(\frac{F}{2\pi}\right)^m$ appears as the 
$m^{th}$ term in the expansion of the total Chern character $\text{ch}[F]= e^{\frac{F}{2\pi}}$ of a $U(1)$ principal bundle
with curvature $F.$~\cite{EGH} Since locally we can write $F^m = d( A\wedge F^{m-1})$, we see that for a BTI
state with a boundary, the term $S_{CC}[A]$ can be interpreted as a CS term at level $\frac{\Theta_{2m}}{2\pi}$ on the 
$(2m-1)$-dimensional boundary of the BTI state (more precisely, this is only true when the bulk field configuration $F$ has vanishing topological 
contributions).

For the analogues of Integer Quantum Hall states of fermions (FIQH states) in odd dimensions, the level $N_{2m-1}$ of the CS term can
be any integer\cite{bernevig2002effective,karabali2006quantum,hasebe2014higher,karabali2016geometry}, 
while for free fermion topological insulators, and their generalizations to higher dimensions, the angle $\Theta_{2m}$ is $2\pi$-periodic
and the value which represents a non-trivial topological insulator state is $\Theta_{2m}=\pi$\cite{QHZ2008} (the result for
fermionic topological insulators in any even dimension is easily established using the axial anomaly for a Dirac fermion in
$2m$ dimensions). For bosonic SPT phases in low dimensions we know that $N_3= 2k$, $k\in\mathbb{Z}$ for BIQH states in 
three dimensions\cite{SenthilLevin,VL2012}, that $\Theta_4$ has $4\pi$-periodicity, and $\Theta_4= 2\pi$ 
for the non-trivial BTI state in four dimensions\cite{VS2013,MKF2013,ye2014constructing,ye2013symmetry}. 

One of the main purposes of this paper is to calculate the values of the
response coefficients $N_{2m-1}$ and $\Theta_{2m}$ for BIQH and BTI states in all dimensions. 
There are (at least) two ways that one could go about doing this. The first way would be to formulate
a general physical argument based, for example, on the consistency of the value of $N_{2m-1}$ or $\Theta_{2m}$ 
and the fact that a bosonic SPT state should have no fractionalized excitations, and in this way determine a constraint on the possible 
values of $N_{2m-1}$ or $\Theta_{2m}$. In fact,
such an argument has already been given for the BIQH state in the case $m=2$ (three spacetime dimensions). 
In Ref.~\onlinecite{SenthilLevin} the authors showed that if the response coefficient $N_3$ (which is just the 
Hall conductance in units of $\frac{e^2}{h}$) is odd, then 
the underlying theory must contain an excitation of charge one (in units of the charge $e$ of the underlying bosons) with fermionic
exchange statistics. An excitation with fermionic statistics is not allowed in a state of bosons which has no fractionalized 
excitations, and so the authors of Ref.~\onlinecite{SenthilLevin} concluded that $N_3$ must be an even integer for BIQH states
in three dimensions. Generalizing this argument to higher dimensions clearly represents a significant conceptual 
difficulty, as in higher dimensions one is probably forced to consider generalized braiding processes for extended objects such
as string or membrane excitations\cite{wang2014braiding,jian2014layer,kong2014braided}. For this reason we do not pursue this approach in 
this work, and instead use a second method.

The second method for answering this question, and the method that we choose to use, is to (i) start with a concrete field-theoretic model which 
is believed to accurately describe the low-energy physics of a BIQH or BTI state in the relevant dimension, (ii) couple this model 
to the 
external field $A$, and (iii) directly calculate the electromagnetic response for this particular model. 
In the literature there are two main kinds of  field-theoretic models that can describe SPT phases: topological quantum field 
theory (TQFT) in terms of gauge field variables (e.g., Chern-Simons theory in three 
dimensions\cite{VL2012,ye2013,cheng2014topological,gu2016multikink}
and twisted gauge theory\cite{kapustin2014anomalies,ning2016symmetry,ye2016twisted} in four dimensions\cite{ye2016topological,ye2015vortex} )
and the Nonlinear Sigma Model (NLSM) description in terms of  constrained bosonic fields 
\cite{xu2013wave,CenkeClass1,CenkeClass2,CenkeBraiding,CenkeLine,CenkeSU2N}. In both approaches the bulk topological 
order is trivial but global symmetry is imposed nontrivially on the field variables. In this paper we choose to use
the NLSM description since this description can be easily generalized to any spacetime dimension.

In the NLSM description, a bulk bosonic SPT phase in $d+1$
spacetime dimensions is described by an $O(d+2)$ NLSM with topological theta term having coefficient $\theta= 2\pi k$ where 
$k \in \mathbb{Z}$. In this description the boundary of the SPT phase is then
described by an $O(d+2)$ NLSM with Wess-Zumino (WZ) term, 
where the coefficient of the WZ term, known as the \emph{level} of the WZ term, is equal to $k$. 
Conventionally, writing down the
WZ term in the boundary theory requires defining an extension of the NLSM field into an auxiliary direction of spacetime. 
In a series of works\cite{xu2013wave,CenkeClass1,CenkeClass2,CenkeBraiding,CenkeLine,CenkeSU2N}, 
the NLSM description has been shown to accurately describe the structure of the ground
state wave function of SPT phases\cite{levin2012braiding}, the point and loop braiding statistics of excitations in gauged SPT 
phases\cite{levin2012braiding,wang2014braiding,jian2014layer,wang2015non,wang2016quantum}, the decorated
domain wall construction of SPT phases\cite{chen2014symmetry}, as well as several other properties of these phases. 
In addition, a mathematical classification of 
bosonic SPT phases based on the NLSM description has been shown to be completely identical to the group cohomology 
classification\cite{chen2013symmetry} in situations where both classification schemes can be applied. 
In fact, there is even a concrete procedure for 
calculating the cocycle which classifies an SPT phase in the group cohomology approach by starting with the NLSM description of
that SPT phase\cite{ElseNayak}. Additional applications of NLSMs to the study of SPT phases with translation symmetry and to exotic 
quantum phase transitions in Weyl semi-metals were considered recently in Refs.~\onlinecite{you2016stripe,you2016decorated}.
However, despite the many successes of the NLSM description, deriving the electromagnetic response of a bosonic SPT phase \emph{directly} from its 
NLSM description remains a difficult problem. 
In the few instances in which the response
of an SPT phase has been determined from its NLSM description it has been by an indirect method such as an appeal to 
gauge invariance of the final effective action\cite{liu2013symmetry}, a dual vortex description of the theory\cite{VS2013}, 
or a description of the NLSM involving
auxiliary fermions which also carry charge of the external field $A$\cite{CenkeDual,lapa2016bosonic}. The descriptions in terms of 
auxiliary fermions are in turn 
based on a set of formulas due to Abanov and Wiegmann\cite{AbanovWiegmann} which allow one to generate an $O(d+2)$ NLSM 
with theta term by coupling the NLSM field to a set of auxiliary fermions and then integrating out those fermions.

In this paper we overcome this difficulty and give a direct computation of the response of higher-dimensional generalizations
of BIQH and BTI states in all dimensions from their NLSM description. To do this we use a two-pronged approach. First, instead
of focusing on the bulk of the SPT phase, we study the boundary, and in particular, the behavior of
the gauged boundary theory. In the case of the BIQH state we find that the boundary has a perturbative $U(1)$ anomaly, which
we explicitly calculate. Since the CS action changes by a boundary term under a gauge transformation, requiring the entire system 
(bulk plus boundary) to be gauge-invariant allows us to determine the bulk response coefficient $N_{2m-1}$ from the boundary 
anomaly. In the BTI case we show that the boundary exhibits a Quantum Hall response when the
associated discrete symmetry (e.g., time-reversal in four dimensions) of the BTI state is broken. Again, from this boundary response we can
directly read off the coefficient $\Theta_{2m}$ using the fact that for a system
with boundary, the action $S_{CC}[A]$ is equivalent to a CS action with level $\frac{\Theta_{2m}}{2\pi}$ on the boundary of the BTI.

To study the boundary theory coupled to the external field electromagnetic $A$ we use the method of gauged WZ actions
\cite{witten1983global,manes1985,HullSpence1,WittenHolo,HullSpence2} (see also
Refs.~\onlinecite{moon2015competing,StoneLopes2016} for some recent applications of gauged WZ actions in condensed
matter physics). This machinery can be applied to this problem since, in the NLSM description, the boundary of an SPT phase in $d+1$ 
dimensions is described by an $O(d+2)$ NLSM with WZ term. Therefore we require knowledge of the proper way to gauge a 
WZ action in order to gauge the boundary theory of the SPT phase. 
For readers who are familiar with gauged WZ actions it is also worth remarking that all terms
in the gauged actions we 
write down (with the sole exception of the original un-gauged WZ term) are expressed as integrals of 
fields only over the physical boundary spacetime. That is, we do not assume an extension of the external field $A$ into the
auxiliary direction of spacetime which is used to write down the WZ term. This is to be contrasted with the general approach 
of Ref.~\onlinecite{HullSpence2}, in which all terms in the gauged action are written as integrals over the extended spacetime,
and an analogue of the method used to obtain the Chern-Simons form from the Chern character must then be used to reduce the 
terms in the action to integrals only over the physical spacetime. 
This difficulty usually prevents one from writing down an explicit local (i.e., not 
involving integrals over the extended spacetime) form for the gauged WZ action in any dimension. We emphasize that 
here we do not encounter this difficulty. For the BIQH and BTI systems that we study, we give explicit local expressions for the 
gauged boundary action in all dimensions.

In Sec.~\ref{sec:BIQH} we use this method to derive the unusual result that for BIQH states in $2m-1$ dimensions the level of the 
CS term in the effective action for $A$ is quantized as
\beq
	N_{2m-1}= (m!) k\ ,\ k\in\mathbb{Z}\ ,
\eeq
where $m!$ denotes the factorial of $m$. 
This general formula agrees with existing results for the cases of three\cite{VL2012,SenthilLevin,ye2013} and five\cite{CenkeDual} dimensions 
($m=2$ and $m=3$, respectively), and gives a prediction for all higher odd dimensions. In this case we also provide an alternative
derivation of the value of $N_{2m-1}$ using only the NLSM description of the \emph{bulk} of the BIQH state, 
which confirms our result derived using the anomaly of the boundary theory.

 Next, in Sec.~\ref{sec:gauge-invariance} we show that the BIQH response computed in Sec.~\ref{sec:BIQH} can 
be understood by requiring the exponential of the CS response action for the BIQH state to be invariant under large $U(1)$ 
gauge transformations when the response theory is formulated on general closed, compact Euclidean manifolds. Furthermore, we apply these gauge invariance arguments to study the electromagnetic 
\emph{and} gravitational responses of fermionic SPT phases with $U(1)$ symmetry in odd dimensions, and point out the distinctive features between the bosonic and fermionic cases.

Moving on to the BTI case, we show in Sec.~\ref{sec:BTI}, using the NLSM description of the BTI phase,  
that the non-trivial BTI state in $2m$ dimensions is characterized by a coefficient
\beq
	\Theta_{2m}= 2\pi\left(\frac{m!}{2}\right)\ .
\eeq 
Again, this general formula agrees with the known answer in four dimensions\cite{VS2013,MKF2013,ye2014constructing,ye2013symmetry} 
($m=2$) and gives a prediction 
for all higher even dimensions. It also suggests that the period of the parameter $\Theta_{2m}$ is $2\pi (m!)$ for BTI states
in $2m$ dimensions.

In Sec.~\ref{sec:applications} we use the gauged boundary actions derived in Sec.~\ref{sec:BIQH} and Sec.~\ref{sec:BTI}
to derive several other interesting results. 
First, we construct an effective theory for a bosonic analogue of a two-node
Weyl (or chiral) semi-metal in all even dimensions $d$ using two copies of the boundary action for the BIQH state. 
We refer to this state as a bosonic chiral semi-metal (BCSM).
The theory that we construct has an electromagnetic response of the form ($\mathbb{R}^{d-1,1}$ is $d$-dimensional
Minkowski spacetime)
\beq
	S^{(b)}_{eff}[A,B] = -2\left(\frac{d}{2}+1\right)\frac{1}{(2\pi)^{\frac{d}{2}}} \int_{\mathbb{R}^{d-1,1}} B\wedge A\wedge (dA)^{\frac{d}{2}-1}
\eeq
where $B= B_{\mu}dx^{\mu}$ is a one-form
 whose components $B_{\mu}$ represent the separation in energy and momentum of
the two copies of the BIQH boundary theory (in the fermionic case the components of $B_{\mu}$ specify the separation in 
energy and momentum of the two Weyl cones). This response is larger than the response of the 
fermionic chiral semi-metal in the same dimension by a factor of $\left(\frac{d}{2}+1\right)!$. This factor
turns out to be identical to the factor of $m!$ discussed earlier for the BIQH state, since our semi-metal theory in
$d$ dimensions is constructed from two copies of the boundary theory for the BIQH state in $d+1=2m-1$ dimensions.
Next, we show that the boundary theory of the BTI exhibits a bosonic analogue of the parity anomaly of a single 
Dirac
fermion in odd dimensions\cite{Redlich,niemi1983axial,alvarez1985anomalies,witten2015fermion,witten2016parity}. 
This parity anomaly is essentially the statement that although the boundary theory of the BTI is gauge-invariant and 
possesses the $\mathbb{Z}_2$ symmetry of the BTI state, the $\mathbb{Z}_2$ symmetry can be
\emph{spontaneously} broken at the boundary of the BTI, resulting in a half-quantized BIQH response on the boundary.
 This anomaly then
provides strong evidence that the boundary theory of the BTI (with the symmetries of the BTI phase)
cannot be realized intrinsically in $2m-1$ dimensions.  Finally, we analyze the physics of symmetry-breaking domain walls on the
boundary of the BTI state, and we show that the physics of such domain walls provides a nice example of the phenomenon of
\emph{anomaly inflow}\cite{callan1985anomalies} in bosonic SPT phases.

The Appendices of the paper contain several additional results, most of a more mathematical nature. 
In Appendix~\ref{app:EC} we use the well-known connection between gauged WZ actions and equivariant cohomology to 
understand the mathematical structure of the gauged WZ actions that we construct for the boundaries of BIQH and BTI states.
In particular, we show that the construction of these actions is related to the mathematical problem of constructing an
\emph{equivariant extension} of the volume form for the sphere $S^{2m-1}$ (in the BIQH case) or $S^{2m}$ in the BTI case, 
and we study this mathematical problem in detail.
In Appendix~\ref{app:CP} we show an example of the computation of the Chern character for the field strength 
$F$ on the complex projective space $C\mathbb{P}^m$.  This example serves to illustrate the necessity of the peculiar 
quantization of the CS level required for gauge invariance of the CS term on general manifolds as derived in
Sec.~\ref{sec:gauge-invariance}. In Appendix~\ref{app:dim-red} we discuss a dimensional reduction procedure which allows
one to obtain the response action for the BTI phase from the response action for the BIQH phase in one higher dimension.
In Appendix~\ref{app:dim-red-NLSM} 
we derive a general dimensional reduction formula for topological theta terms in NLSMs. Finally, in Appendix~\ref{app:O2} 
we compute the electromagnetic response of the $O(2)$ NLSM in one dimension.

\section{Background}
\label{sec:background}

In this section we introduce the relevant background material necessary for understanding the later sections of the paper.
We start with a brief review of the physics of the BIQH and BTI states, and also present definitions of higher-dimensional 
generalizations of these states. We then review the NLSM description of the bulk and boundary of bosonic SPT phases,
and discuss the specifics of the NLSM descriptions of the BIQH and BTI states that we study in this paper. Finally, we give a
general discussion of the tool of gauged WZ actions, and we describe in concrete terms the procedure that we use in this 
paper to construct gauged WZ actions for the boundaries of BIQH and BTI states. 

\subsection{BIQH and BTI phases}

In its original formulation\cite{VL2012,SenthilLevin}, 
the BIQH phase was conceived of as a gapped quantum phase of bosons in three spacetime dimensions 
which exhibits a non-zero Hall conductance, but does not have any bulk topological order. As an SPT phase it is protected by
only charge-conservation symmetry, i.e., we have $G= U(1)$ where $G$ is the symmetry group of the SPT phase.
Physically, the BIQH state is characterized by a CS
term in the effective action for the external field $A$,
\beq
	S_{eff}[A]= \frac{N_3}{4\pi}\int_{\mathcal{M}}A\wedge dA\ , \label{eq:BIQH-action}
\eeq
in which the coefficient $N_3$ (which is just the Hall conductance in units of $\frac{e^2}{h}$) 
is quantized in integer multiples of $2$. The authors of Ref.~\onlinecite{SenthilLevin} gave a
very appealing physical argument for why the value of $N_3=1$ is not allowed if the BIQH state is required to have no 
fractionalized excitations, and we now briefly review their argument. 
Consider a hypothetical BIQH state on flat space, and a configuration of $A$ in which a thin tube of $2\pi$ flux pierces
the spatial surface. According to the action $S_{eff}[A]$, the point in space where the flux tube pierces the plane will 
bind a charge equal to $N_3$. Now one invokes a standard argument\footnote{In fact, this statement is only true on a lattice when we can couple to a compact $U(1)$ gauge field, or in the continuum when the level of the CS term
is an integer. To see what can go wrong, consider $N_3= \frac{1}{q}$ for $q\in\mathbb{Z}$. Then the object created by threading a thin $2\pi$ flux tube has charge $\frac{1}{q}$ and so only $q$ such fluxes are a physical excitation of the system, in the sense that all states in the
physical Hilbert space of the quantum mechanical system should have integer charge.} that $2\pi$ flux is gauge-equivalent to zero flux,
and so the point-like excitation created by 
threading the flux is in fact an excitation of the BIQH fluid and not an external defect.
One can therefore ask about the phase obtained by the wavefunction of the system after a process in which
two such excitations are exchanged. 
By the Aharanov-Bohm effect, taking one excitation completely around another results in a
statistical phase of $2\pi N_3$. The phase for an exchange process is therefore half of that, $\vartheta_{ex}= \pi N_3$. 
From this result the authors of Ref.~\onlinecite{SenthilLevin} concluded that the state described by the effective action of
Eq.~\eqref{eq:BIQH-action} contains a fermionic excitation if $N_3$ is odd, and so $N_3$ must be an even integer in order
for the action of Eq.~\eqref{eq:BIQH-action} to represent the electromagnetic response of a BIQH phase.

In this paper we consider generalizations of the BIQH state to all odd spacetime dimensions. One definition of a BIQH state
in $2m-1$ dimensions which is sufficient for our purposes is that a BIQH state is an SPT phase of bosons which is 
protected by the symmetry group $G= U(1)$, where $U(1)$ is charge conservation symmetry, and which exhibits a CS 
response to an applied electromagnetic field $A$ of the form of Eq.~\eqref{eq:CS-term}.
We should also mention here that in odd dimensions there is a countable infinity of different BIQH states, i.e.,
these states have a $\mathbb{Z}$ classification\cite{chen2013symmetry,CenkeClass1}. This means that the
coefficient $N_{2m-1}$ takes on a countable infinity of values which all have the form of some particular number times an integer.

On the other hand, the BTI phase\cite{chen2013symmetry,VS2013,ye2015vortex} 
is a bosonic analogue of the time-reversal invariant electron topological insulator in four spacetime dimensions. 
As an SPT phase it is protected by the symmetry group $G= U(1)\rtimes \mathbb{Z}^T_2$, where $U(1)$ represents charge
conservation and $\mathbb{Z}^T_2$ is time-reversal symmetry. If we write $\mathbb{Z}^T_2= (1,\mathcal{T})$ where
$\mathcal{T}$ is the time-reversal operator, then we have $\mathcal{T}^2= 1$ for the BTI. This should be contrasted
with the relation $\mathcal{T}^2= (-1)^F$ which holds for the electron topological insulator, where $F$ is the fermion number.
The semi-direct product ``$\rtimes$" indicates that the $U(1)$ and $\mathbb{Z}^T_2$ symmetries do not commute with
each other. In the next subsection we will see an explicit representation of the action of the group $U(1)\rtimes \mathbb{Z}^T_2$
on the fields in the NLSM description of the BTI. 

The bulk of the BTI phase is characterized by an effective action for $A$ of the Chern character form
\beq
	S_{eff}[A]= \frac{\Theta_4}{8 \pi^2}\int_{\mathcal{M}}F\wedge F\ ,
\eeq
where $F=dA,$ and  $\Theta_4=2\pi$ for the BTI 
(compare with $\Theta_4=\pi$ for the electron topological insulator\cite{QHZ2008}). The parameter $\Theta_4$ has
$2\pi$-periodicity in the case of the electron topological insulator\cite{QHZ2008} but $4\pi$-periodicity in the BTI 
case\cite{VS2013,MKF2013}. One way
to understand this effective action is to consider what happens when the spacetime $\mathcal{M}$ has a boundary 
$\pd\mathcal{M}$. In this case, if the bulk field configuration $F$ is topologically trivial, then we can write 
$F\wedge F= d(A\wedge dA)$ to find
\beq
	S_{eff}[A]= \frac{\Theta_4}{2\pi}\frac{1}{4 \pi}\int_{\pd\mathcal{M}}A\wedge dA\ ,
\eeq
which is equivalent to a Quantum Hall state with Hall conductance $\sigma_H= \frac{\Theta_4}{2\pi}$
on the boundary of $\mathcal{M}$. In particular, for the BTI we have $\Theta_4=2\pi$ so that the surface of the BTI exhibits a 
\emph{half-quantized} BIQH effect (i.e., $\sigma_H=1$ on the surface). Such a surface Quantum Hall response breaks 
the time-reversal symmetry of the BTI. 

Now we turn to the question of how to generalize the BTI state to all even dimensions. 
The main issue with generalizing the BTI state to all even dimensions is that the
discrete part of the symmetry group $G$, which was anti-unitary 
time-reversal symmetry $\mathbb{Z}^T_2$ in four dimensions, should
be chosen differently when the spacetime dimension is equal to zero or two modulo four. Whenever the spacetime dimension
is equal to zero modulo four we choose the discrete part of $G$ to be anti-unitary time-reversal symmetry 
$\mathbb{Z}^T_2$. On the other hand, whenever the spacetime dimension is equal to two modulo four we choose the discrete
part of $G$ to be \emph{unitary} charge-conjugation (or particle-hole) symmetry $\mathbb{Z}^C_2$. This choice is
consistent with the results of the group cohomology\cite{chen2013symmetry} and NLSM\cite{CenkeClass1} 
classifications of SPT phases in these dimensions, and with the 
symmetries which protect the fermion topological insulators in two and four spacetime dimensions, respectively\cite{QHZ2008}. 

We therefore choose to use the following definition of a BTI phase in all even dimensions. A BTI phase in spacetime dimension
$2m$ is an SPT phase of bosons with symmetry group
\beq
	G= \begin{cases}
	U(1)\rtimes \mathbb{Z}^T_2 &,\ m=\text{even} \\
	U(1)\rtimes \mathbb{Z}^C_2 &,\ m=\text{odd}
\end{cases},
\eeq 
and which exhibits a bulk response to an external field $A$ of the form of Eq.~\eqref{eq:chern-character}. As we noted earlier, 
when the spacetime $\mathcal{M}$ has a boundary $\pd\mathcal{M}$, and when the field configuration $F$ is topologically trivial, this 
bulk response is equivalent to a boundary Quantum Hall response of the form of Eq.~\eqref{eq:CS-term} with coefficient 
$N_{2m-1}= \frac{\Theta_{2m}}{2\pi}$.
In addition, this boundary Quantum Hall response breaks the $\mathbb{Z}^T_2$ symmetry (for $m$ even) or $\mathbb{Z}^C_2$ 
symmetry (for $m$ odd) of the BTI phase. When we discuss the BTI phase in a general dimension $2m$, and when we do not
have a particular $m$ in mind, we just write $\mathbb{Z}_2$ for the discrete part of $G$. However, the reader should always
keep in mind that the $\mathbb{Z}_2$ symmetry is different for the cases of $m$ even and $m$ odd as discussed in this 
subsection. 

Finally, we also mention that based on the group cohomology\cite{chen2013symmetry} and NLSM\cite{CenkeClass1}  
classification schemes, only the smallest value of $\Theta_{2m}$ is expected to represent a non-trivial BTI phase in $2m$
dimensions. This can be understood as follows. For SPT phases with $U(1)\rtimes \mathbb{Z}^T_2$
symmetry in four dimensions the group cohomology and NLSM classifications predict a $(\mathbb{Z}_2)^2$ classification. 
One of these $\mathbb{Z}_2$ factors corresponds to the BTI state, while the other corresponds to a state in which the
$U(1)$ symmetry plays no role\cite{MKF2013} (so this second state cannot be interpreted as an insulator). 
This means that there is only a single non-trivial BTI state in four dimensions. In 
addition, in two dimensions the classification for SPTs with $U(1)\rtimes \mathbb{Z}^C_2$ symmetry is $\mathbb{Z}_2,$ and
the $U(1)$ symmetry does play a role in the non-trivial phase, so we identify that phase with the BTI phase in two dimensions. 
Based on this evidence we expect the existence of a single non-trivial BTI phase to generalize to all even dimensions.
In the context of the NLSM classification this can be understood as coming from the fact 
that in $2m$ dimensions the $O(2m+1)$ NLSM theory with $\theta= 2\pi k$ can be smoothly connected to the theory with 
$\theta=2\pi (k\pm 2)$ (see, e.g., the discussion in Ref.~\onlinecite{CenkeClass1}). 

\subsection{NLSM description of the bulk and boundary of SPT phases}

We now give a brief review of the NLSM description of SPT states, which was presented in its fully developed form in 
Ref.~\onlinecite{CenkeClass1}. Let us consider bosonic 
SPT phases in $d+1$ spacetime dimensions. The spacetime coordinates are $x^{\mu}$, $\mu=0,\dots,d$ ($x^0=t$ is the time 
coordinate), and for now we focus on the case of
flat Minkowski spacetime $\mathbb{R}^{d,1}$
with the mostly minus metric $\eta= \text{diag}(1,-1,\dots,-1)$. Following the prescription of Ref.~\onlinecite{CenkeClass1}, a 
bosonic SPT phase in this dimension is described by an $O(d+2)$ NLSM with topological theta term  where
the coefficient of the theta term is given by $\theta=2\pi k$ with $k\in\mathbb{Z}$. The $O(d+2)$ NLSM is a theory of a
$(d+2)$-component unit vector field $\mb{n}$ (i.e., $\mb{n}\cdot\mb{n}=1$) with components $n_a$, $a=1,\dots,d+2$. 
Because of the constraint, the configuration space (or target space) of 
the NLSM field is the $d+1$-dimensional sphere $S^{d+1}$. Latin indices $a,b,c,\dots$, which label components of 
$n_a$, can be raised and lowered with the Euclidean metrics $\delta^{ab}$, $\delta_{ab}$, and so $n^a$ and $n_a$ are
numerically equal to each other. In what follows we use the summation convention for any indices (Latin or Greek) 
which appear once in an upper position and once in a lower position in any expression. 

The NLSM action describing the SPT phase is
\beq
	S_{bulk}[\mb{n}]= \int d^{d+1}x \ \frac{1}{2g}(\pd^{\mu}n^a)(\pd_{\mu}n_a) + S_{\theta}[\mb{n}]\ , \label{eq:NLSM-action}
\eeq
where $g>0$ is the coupling constant of the NLSM (with units of (length)$^{d-1}$),
and $S_{\theta}[\mb{n}]$ is the 
theta term. To write the theta term in a compact
way we first introduce some notation. Let $\omega_{d+1}$ be the volume form on $S^{d+1}$. Explicitly, we have
\beq
	\omega_{d+1}= \sum_{a=1}^{d+2} (-1)^{a-1} n_a dn_1 \wedge \cdots \wedge \overline{dn_a}\wedge \cdots\wedge dn_{d+2}\ , \label{eq:volume-form}
\eeq
where the overline means to omit that term from the wedge product. 
We also use the notation $\mathcal{A}_{d+1} \equiv \text{Area}[S^{d+1}]=\frac{2\pi^{\frac{d+2}{2}}}{\Gamma(\frac{d+2}{2})}$
for the area of the sphere $S^{d+1}$.
In terms of these quantities, the theta term can be written
compactly in differential form notation as
\beq
	 S_{\theta}[\mb{n}]= \frac{\theta}{\mathcal{A}_{d+1}}\int_{\mathbb{R}^{d,1}} \mb{n}^*\omega_{d+1}\ ,
\eeq
where $\mb{n}^*\omega_{d+1}$ denotes the pull-back to spacetime of the form $\omega_{d+1}$ via the map 
$\mb{n}: \mathbb{R}^{d,1}\to S^{d+1}$. In coordinates this becomes
\begin{widetext}
\beq
	S_{\theta}[\mb{n}]= \frac{\theta}{\mathcal{A}_{d+1}}\int d^{d+1}x\ \ep^{a_1\cdots a_{d+2}} n_{a_1}\pd_{x^0}n_{a_2}\pd_{x^1}n_{a_3}\cdots\pd_{x^d}n_{a_{d+2}}\ . \label{eq:theta-term-coords}
\eeq
\end{widetext}

 For the description of SPT phases we have $\theta=2\pi k$ for integer 
$k$. The reason for choosing $\theta=2\pi k$ is that at these values of $\theta$ the NLSM is expected to flow
to a disordered ($g\to \infty)$ fixed point under the Renormalization Group\cite{CenkeClass1}.
In addition we note that the full action of Eq.~\eqref{eq:NLSM-action} (including theta term) has an $SO(d+2)$ global
symmetry, where the action of the group on the NLSM field is given by $n_a \to {R_a}^b n_b$ for any matrix $R \in SO(d+2)$.
When the coefficient $\theta$ is set to zero this symmetry is promoted to an $O(d+2)$ global symmetry (under a general 
transformation $R\in O(d+2)$ the theta term transforms only by acquiring the sign $\text{det}[R]=\pm 1$). 
The fixed point theory (with
$g\to\infty$ at $\theta=2\pi k$) is gapped and has a unique ground state which does not break the $SO(d+2)$ symmetry of the NLSM with 
theta term\cite{xu2013wave}. This property of the disordered ground state of the NLSM at $\theta=2\pi k$ is one of the main reasons why
these field theories are useful for describing SPT phases.

SPT phases are classified according to their symmetry group $G$. In the NLSM description of Ref.~\onlinecite{CenkeClass1}
this symmetry is encoded in a homomorphism $\sigma: G \to O(d+2)$, which maps $g\in G$ to some $(d+2)\times(d+2)$ matrix 
$\sigma(g) \in O(d+2)$. We refer to such a 
$\sigma$ as a \emph{symmetry assignment}. According to the NLSM classification of SPT phases, if $g\in G$ represents an
\emph{internal} unitary symmetry operation (i.e., $g$ does not have any action on the spacetime coordinates) then $\sigma$ should 
be chosen so that $\text{det}[\sigma(g)]= 1$. In this case it is then clear that the action of $g$ leaves the theta term
invariant. On the other hand, if $g\in G$ represents the time-reversal operation, then $\sigma$ should be chosen so that 
$\text{det}[\sigma(g)]= -1$. Since the time-reversal operation also sends $t\to-t$ (in addition to its action on the components of
the NLSM field), the minus sign in the theta term from $\text{det}[\sigma(g)]$ will be canceled by the minus sign from sending 
$\pd_t \to -\pd_t$. Thus, choosing $\text{det}[\sigma(g)]=-1$
in this case ensures that the theta term is invariant under the time-reversal transformation.

Not all NLSMs with a symmetry assignment will describe a non-trivial SPT phase.
For example an NLSM with a symmetry assignment $\sigma$ will describe a trivial
phase if there exists a vector $\mb{v}$ such that $\sigma(g) \mb{v}= \mb{v}$ $\forall g\in G$. This is because in this case
we are allowed to add a term $\mb{n}\cdot\mb{v}$ to the NLSM action without breaking the symmetry of the group $G$. Such
a term will then drive the system into a trivial phase in which $\mb{n}$ is parallel or anti-parallel to $\mb{v}$ at all points in space.
If a vector $\mb{v}$ with this property does not exist, then the NLSM with symmetry assignment $\sigma$ can describe a non-trivial 
SPT phase.

When an SPT phase has a bulk description in terms of an $O(d+2)$ NLSM with theta term and theta angle $\theta=2\pi k$, its
$d$-dimensional boundary is described by an $O(d+2)$ NLSM with Wess-Zumino (WZ) term at level $k$. Let us for simplicity study
the boundary perpendicular to the $x^d$ direction, so on the boundary we have coordinates $x^{\mu}$, $\mu=0,\dots,d-1$, 
and the boundary spacetime is $\mathbb{R}^{d-1,1}$.
To write down the WZ term we need to extend the field configuration $n_a$ 
into a fictitious extra dimension of the boundary spacetime. 
We take $s\in[0,1]$ to be the coordinate for this extra direction,
and define $\mathcal{B}= [0,1]\times \mathbb{R}^{d-1,1}$ to be the extended boundary spacetime. 
Let $\tilde{n}_a(x^{\mu},s)$ be an extension of the field $n_a$ into the $s$ direction. It is typical to choose boundary 
conditions in the extra direction so that $\tilde{n}_a(x^{\mu},1)= \delta_{a,1}$ (i.e., a trivial configuration) and
$\tilde{n}_a(x^{\mu},0)= n_a(x^{\mu})$ so that the physical boundary spacetime is located at $s=0$. 
Then the action for the boundary theory takes the form
\beq
	S_{bdy}[\mb{n}]= \int d^{d}x \ \frac{1}{2g_{bdy}}(\pd^{\mu}n^a)(\pd_{\mu}n_a) + S_{WZ}[\mb{n}]\ , \label{eq:NLSM-action-bdy}\ 
\eeq
where the WZ term is
\beq
	S_{WZ}[\mb{n}]= \frac{2\pi k}{\mathcal{A}_{d+1}}\int_{\mathcal{B}} \tilde{\mb{n}}^*\omega_{d+1}\ . \label{eq:WZ-term1}
\eeq
Here $g_{bdy}$ is the coupling constant for the boundary theory, and the WZ term now involves the pull-back of 
$\omega_{d+1}$ to $\mathcal{B}$ (the extended boundary spacetime) via the map $\tilde{\mb{n}}: \mathcal{B}\to S^{d+1}$. 
Again, in coordinates this takes the form
\begin{widetext}
\beq
	S_{WZ}[\mb{n}]= \frac{2\pi k}{\mathcal{A}_{d+1}}\int_0^1 ds \int d^{d}x\ \ep^{a_1\cdots a_{d+2}} \tilde{n}_{a_1}\pd_{s}\tilde{n}_{a_2}\pd_{x^0}\tilde{n}_{a_3}\cdots\pd_{x^{d-1}}\tilde{n}_{a_{d+2}}\ .
\eeq
\end{widetext}

We now discuss the specific symmetry assignments $\sigma: G\to O(d+2)$ which will be used to construct NLSM descriptions
of BIQH states in odd spacetime dimensions and BTI states in even spacetime dimensions. We start with the case of BIQH states
in $2m-1$ spacetime dimensions. In this case the integer $m$ is related to $d$ by the relation $2m=d+2$, and the
BIQH state is described by an $O(2m)$ NLSM with theta term. In the BIQH case the symmetry group is just $G=U(1)$
and the particular $U(1)$ symmetry that we are interested in
is embedded in the full $O(2m)$ group as follows. We first combine pairs of the $2m$ components $n_a$ of the NLSM field to create the $m$ 
boson fields
\beq
	b_{\ell}= n_{2\ell-1} + i n_{2\ell}\ ,\ \ell = 1,\dots,m\ .
\eeq
Then the $U(1)$ symmetry we consider acts on the NLSM field as
\beq
	U(1): b_{\ell} \to e^{i\xi}b_{\ell}, \forall \ell\ , \label{eq:U1}
\eeq
where $\xi$ is a constant parameter. We can consider the fields $b_{\ell}$ to be $m$ complex scalar 
fields of charge $1$, but subject to the constraint
$\sum_{\ell=1}^m |b_{\ell}|^2= 1$, which is equivalent to the constraint $\mb{n}\cdot\mb{n}=1$ for the NLSM field $n_a$. 
This choice of $U(1)$ transformation, and the corresponding pairing of the components of $\mb{n}$ into 
the bosons $b_{\ell}$, is convenient, but it is not unique. 
Since the NLSM action with theta term (or WZ term) is still invariant 
under the group $SO(2m)$, we can do any change of basis $n^a \to {M_a}^b n_b$ with $M \in SO(2m)$ to obtain a theory with a 
different action of the $U(1)$ symmetry, but with the same physical properties. As discussed above, the most important property 
of the symmetry assignment is that there should not be any vector $\mb{v}$ that remains fixed under the $U(1)$ action. 
Indeed, if such a $\mb{v}$ exists then the NLSM with this symmetry assignment describes a trivial phase. 
The choice above satisfies this requirement.

For the case of BTI states in even dimensions $2m,$ the integer $m$ is instead related to $d$ by the formula $2m+1=d+2$, so 
that these states are described by $O(2m+1)$ NLSMs with theta term. As we discussed in the previous 
subsection the symmetry group in this case is $G= U(1)\rtimes \mathbb{Z}^T_2$ for $m$ even and 
$G= U(1)\rtimes \mathbb{Z}^C_2$ for $m$ odd. To define the symmetry assignment $\sigma$ in this case we again take pairs of the first
$2m$ components of the NLSM field and combine them into bosons $b_{\ell}$, $\ell=1,\dots,m$ as done for the BIQH case. The $U(1)$ symmetry
we consider again acts as in Eq.~\eqref{eq:U1} on these bosons, but leaves the final component $n_{2m+1}$ of the NLSM field
fixed. Finally, in the BTI case the additional discrete $\mathbb{Z}_2$ symmetry (which is either
$\mathbb{Z}^T_2$ or $\mathbb{Z}^C_2$ depending on the parity of $m$) is taken to act on the NLSM field as
\begin{subequations}
\label{eq:TC-trans}
\beqa
	\mathbb{Z}_2: n_a &\to& n_a\ ,\ a= 1,3,\dots,2m-1,\  \\ 
	n_a &\to& -n_a\ ,\ a= 2,4,\dots,2m,2m+1\ .
\eeqa
\end{subequations}
In the case where the $\mathbb{Z}_2$ symmetry is time-reversal $\mathbb{Z}^T_2$, we also need to send 
$t\to -t$ in the argument of $n_a$ and in the action. 
Under the transformation in Eq.~\eqref{eq:TC-trans}, the theta term of the NLSM picks up the 
sign $(-1)^{m+1}$. So we see that for $m$ odd the theta term in the NLSM automatically has this symmetry, while in the case of 
$m$ even it must be supplemented with the replacement $t\to -t$, which gives an extra minus sign in the theta term. So the
NLSM has the internal, unitary $\mathbb{Z}^C_2$ particle-hole symmetry in the case of $m$ odd, 
while in the case of $m$ even it has the anti-unitary time-reversal symmetry $\mathbb{Z}^T_2$.

Now that we know how the fields in the NLSM description transform under the $U(1)$ symmetry of the BIQH and BTI
phases, we can considering coupling the NLSM theory, and in particular the boundary theory which involves a WZ term, to
the external electromagnetic field $A= A_{\mu}dx^{\mu}$. In order to do this, we are going to need the tool of gauged WZ 
actions. 

\subsection{Gauged Wess-Zumino actions}

We now give a discussion of the theory of gauged WZ actions, mostly focusing on the general philosophy behind the construction
of a gauged WZ action. The details of this construction will be worked out explicitly for the boundary theories of the BIQH and BTI 
phases in all dimensions in later sections of this paper. In addition, in Appendix~\ref{app:EC} we review the relation
between gauged WZ actions and equivariant cohomology, and we re-examine the gauged WZ
actions constructed in this paper from this more mathematical point of view.

Before we start, let us note that the kinetic term for the NLSM is easily gauged using 
ordinary minimal coupling (also known as a ``Peierls substitution" in a condensed matter context). 
In fact, the gauged kinetic term is most simply written in terms of the $b_{\ell}$ as
\beq
	S_{kin,gauged}[\mb{n},A]= \int d^d x\ \frac{1}{2g_{bdy}}\sum_{\ell=1}^{m} (D^{\mu} b_{\ell})^*(D_{\mu}b_{\ell})\ ,
\eeq
for the boundary of the BIQH state ($d+2 = 2m$), or
\beqa
	S_{kin,gauged}[\mb{n},A] &=& \int d^d x\ \frac{1}{2g_{bdy}}\Bigg[\sum_{\ell=1}^{m} (D^{\mu} b_{\ell})^*(D_{\mu}b_{\ell})   \nnb \\
&+& (\pd^{\mu} n_{2m+1})(\pd_{\mu} n_{2m+1}) \Bigg]\ ,
\eeqa
for the boundary of a BTI state ($d+2 = 2m+1$), where $D_{\mu}= \pd_{\mu}- i A_{\mu}$ is the usual covariant derivative. 
Note here that since we are only interested in enforcing a $U(1)$ subroup of the full $SO(d+2)$ symmetry group of the
NLSM, we could allow a different boundary coupling constant $g_{bdy,\ell}$ for each species $b_{\ell}$ of boson. This type of anisotropy
in the coupling constant will not affect the results in the rest of the paper, since those results only depend on the form of the WZ term.

Gauging the WZ term is more subtle. The main problem we face in attempting to gauge this term is the fact that the WZ term is
written as an integral of an expression involving the field $\tilde{n}_a$ over the $(d+1)$-dimensional extended spacetime 
$\mathcal{B}$. One method\cite{HullSpence2} 
for gauging a WZ term involves defining an extension $\tilde{A}$ of the gauge field $A$ into the 
extra $s$-direction, and then applying the usual minimal coupling procedure (but using the extended field $\tilde{A}$) inside the 
WZ term. This has the effect of replacing the integrand $\tilde{\mb{n}}^*\omega_{d+1}$ of the WZ term 
in Eq.~\eqref{eq:WZ-term1} with $\tilde{\mb{n}}^*\omega^{\tilde{A}}_{d+1}$, where 
$\omega^{\tilde{A}}_{d+1}$ represents the volume form on $S^{d+1}$ but with the ordinary exterior derivative $d$ 
replaced with a gauge-covariant exterior derivative $D$ (the precise form of $D$ is not important for the general discussion here).
However, minimal coupling alone is not sufficient, as varying the minimally-coupled WZ action
does not lead to $d$-dimensional equations of motion, i.e., the resulting equations of motion depend on the extensions 
$\tilde{n}_a$ and $\tilde{A}$. To remedy this the authors of Ref.~\onlinecite{HullSpence2} used the following prescription. 
They suggested that one should add a second term $U(\tilde{n}_a,\tilde{A})$ to the integrand of the WZ term such that the 
combination $\omega^{\tilde{A}}_{d+1}+U$ is a closed form on the extended spacetime. Since a closed form is locally exact 
(i.e.,
a closed form $\omega$ can be written as $\omega= d\gamma_i$ for some $\gamma_i$ on each coordinate patch $\mathcal{U}_i$
of the manifold),
variation of this new WZ term leads to $d$-dimensional equations of motion on each coordinate patch of
the original spacetime manifold. 
There is, however, one conceptual issue with
this method, which the authors of Ref.~\onlinecite{HullSpence2} point out (see their discussion in the paragraph after equation
4.7). The problem is that in the usual setup of the WZ term, the form $\omega^{\tilde{A}}_{d+1}$ (and also $\omega_{d+1}$)
is a $(d+1)$-form
on the $(d+1)$-dimensional extended spacetime $\mathcal{B}$, and so it is trivially closed. Therefore in order to apply the
method of Ref.~\onlinecite{HullSpence2} one has to imagine that the extended spacetime $\mathcal{B}$ is 
embedded in a spacetime $\mathcal{X}$ of even higher dimension so that $d\omega^{\tilde{A}}_{d+1}$ is not trivially equal
to zero.

From this discussion it is clear that gauging a WZ is in general a difficult procedure. However,
for the problems encountered in this paper, in which we only deal with a $U(1)$ subgroup of the full $O(d+2)$ symmetry of the 
NLSM theories, we do not need the complicated machinery developed in Ref.~\onlinecite{HullSpence2}. 
Instead, we use the following concrete procedure (which is similar in spirit to the methods used
in Refs.~\onlinecite{HullSpence1,WittenHolo}) to gauge the $U(1)$ symmetry of our theories. 
First we consider how the WZ term changes under the 
transformation $b_{\ell} \to e^{i\xi}b_{\ell}$ (with a spacetime-dependent $\xi$).
We will see that it changes by a term which is a total derivative, which means that the change of the WZ term can be
written as an integral only over the physical boundary spacetime $\mathbb{R}^{d-1,1}$ instead of over the extended
spacetime $\mathcal{B}$. Next we attempt to cancel this change in
the action by adding an integral over spacetime of the NLSM field coupled to $A$. 
We will see that this procedure usually needs to be iterated several times because the counterterms that we add to the 
action may not transform nicely under a gauge transformation, where ``nicely" is defined below by 
Eq.~\eqref{eq:gauge-trans}. We use the following criterion, inspired by the discussion in Ref.~\onlinecite{WittenHolo}, 
for determining when the action has been properly gauged. 

\textit{Gauging principle:} 
The correctly gauged action $S_{gauged}[\mb{n}, A]$, if it is not completely gauge-invariant, must
transform under a gauge transformation $b_{\ell} \to e^{i\xi}b_{\ell}$, $A\to A+d\xi$, as
\beq
	S_{gauged}[\mb{n}, A] \to S_{gauged}[\mb{n}, A] + \delta_{\xi} S_{gauged}[A,\xi]\ , \label{eq:gauge-trans}
\eeq
where we have used the notation $\delta_{\xi} S_{gauged}$ to indicate the change in 
$S_{gauged}$ under a gauge transformation. The key point here is that
the change in the action under a gauge transformation depends only on $A$ and $\xi$, but not on the matter field
$\mb{n}$. 

Let us also note here that in this paper we use the word ``anomaly" to refer to the change in the action
(or action plus path integral measure) under a $U(1)$ gauge transformation. There is no anomaly if the action (plus path
integral measure) is gauge-invariant. The gauging principle stated above then simply asserts that the anomaly 
$\delta_{\xi} S_{gauged}[A,\xi]$ of the gauged action $S_{gauged}[\mb{n}, A]$ should only depend on $A$ and $\xi$.

We will see in the following sections 
that we may need to add several counterterms to the WZ action to get Eq.~\eqref{eq:gauge-trans} to 
hold. In the BIQH case the correctly gauged action still transforms under a gauge transformation, 
and so the $U(1)$ symmetry of the boundary theory of the BIQH phase is anomalous. This fact is what allows us to deduce
the bulk CS response of the BIQH state. On the other hand, for the surface of the BTI it is possible to construct a completely
gauge-invariant action. However, from the form of the gauge-invariant action we will be able to see that if the NLSM field
condenses in a way that preserves the $U(1)$ symmetry, but breaks the $\mathbb{Z}_2$ symmetry of the BTI phase, then
the surface of the BTI will exhibit a $\mathbb{Z}_2$ symmetry-breaking Quantum Hall response.

\section{Electromagnetic response of BIQH states in all odd dimensions}
\label{sec:BIQH}

In this section we construct the gauged WZ action for the boundary of BIQH states in all odd dimensions. The action 
we construct satisfies the gauging principle of
Eq.~\eqref{eq:gauge-trans}, but is still not completely gauge-invariant, as evidenced in the $U(1)$ anomaly of the boundary theory of the BIQH state. We then use the $U(1)$ anomaly of the gauged boundary action to calculate the bulk CS response of the 
BIQH state in all odd dimensions. As we discussed in the introduction, we find that for the BIQH state in $2m-1$ dimensions the
level $N_{2m-1}$ of the CS term appearing in the effective action is quantized in units of $m!$. We then give a more intuitive
derivation of the BIQH response using only the dimensional reduction properties of CS terms and of theta terms in NLSMs. 
This second derivation relies on results which we derive in Appendices~\ref{app:dim-red-NLSM} and \ref{app:O2}.
This intuitive picture confirms our more technical derivation using gauged WZ actions.

 The result in this section is related to the results of several other sections of this paper. 
In the next section, Sec.~\ref{sec:gauge-invariance}, we show that the factor of $m!$ 
for the CS response of the BIQH state computed in this section can be understood by requiring that  partition functions containing the CS response action be invariant
under large $U(1)$ gauge transformations on general Euclidean manifolds.
Later, in Appendix~\ref{app:EC}, we re-examine the gauged WZ action constructed in this section in light of the well-known 
connection 
between gauged WZ actions and equivariant cohomology of the target space of the NLSM. The construction of a gauged WZ action
for the boundary of the BIQH state is equivalent to the problem of constructing an \emph{equivariant extension} (with respect to the 
$U(1)$ symmetry) of the volume form $\omega_{2m-1}$ for $S^{2m-1}$. In Appendix~\ref{app:EC} we attempt to construct
such an extension, and then show that the construction fails at the last step. The fact that such an extension does not exist
is mathematically equivalent to our finding that the gauged action for the boundary 
of the BIQH state still has a $U(1)$ anomaly. In Appendix~\ref{app:EC} we also show that the 
differential forms $\Omega^{(r)}$, which appear later in this section in the counterterms of Eq.~\eqref{eq:BIQH-cts},
are the same forms which appear in the construction of the equivariant extension of $\omega_{2m-1}$ (although the extension fails 
at the last step in this case as mentioned above).

%We begin this section by giving a walk-through of the construction of the gauged boundary
%action for the simple example of the edge theory of the ordinary BIQH state in three spacetime dimensions. After this warm-up
%we present the construction of the gauged boundary action for a BIQH state in any odd dimension $2m-1$. We then show how the 
%anomaly of the gauged boundary action allows one to deduce the
%bulk CS response of the BIQH state.

Let us make a few remarks on the notation used in this section and in later sections of the paper. 
In what follows we omit the pull-back symbol $\mb{n}^*$ so as 
not to clutter the notation, but one should always remember that
the integrand of any integral should be pulled back to spacetime (or the extended spacetime, in which case one would write
$\tilde{\mb{n}}^*$). 
In addition we will express many quantities in terms of the
integer $m$ instead of $d$. Recall that these are related by $2m=d+2$ in the BIQH case.
So for example we write the WZ term as
\beq
	S_{WZ}[\mb{n}]= \frac{2\pi k}{\mathcal{A}_{2m-1}}\int_{\mathcal{B}} \omega_{2m-1}\ .
\eeq
For later use we also define several differential forms which are constructed from the components of the NLSM field. 
We define the one form $\mathcal{J}_{\ell}$ and two form $\mathcal{K}_{\ell}$ 
by
\begin{subequations}
\label{eq:J-K-def}
\beqa
	\mathcal{J}_{\ell} &=& n_{2\ell-1}dn_{2\ell}- n_{2\ell}dn_{2\ell-1}  \\
	\mathcal{K}_{\ell} &=& dn_{2\ell-1}\wedge dn_{2\ell}\ .
\eeqa
\end{subequations}
Under a gauge transformation $b_{\ell}\to e^{i\xi}b_{\ell}$ these forms transform as
\begin{subequations}
\label{eq:J-K-var}
\beqa
	\mathcal{J}_{\ell} &\to& \mathcal{J}_{\ell}  + (n_{2\ell-1}^2 + n_{2\ell}^2)d\xi  \\    
	\mathcal{K}_{\ell} &\to& \mathcal{K}_{\ell}+ (n_{2\ell-1}dn_{2\ell-1} + n_{2\ell}dn_{2\ell})\wedge d\xi\ .  
\eeqa
\end{subequations}
We also note here that
\beq
	\mathcal{K}_{\ell}= \frac{1}{2}d\mathcal{J}_{\ell}\ ,
\eeq
and so
\beq
	d\mathcal{K}_{\ell}= 0\ ,
\eeq
i.e., $\mathcal{K}_{\ell}$ is an \emph{exact} differential form.

\subsection{$O(4)$ NLSM with WZ term in two spacetime dimensions}

Before presenting the gauged action for any integer $m$, we warm up with an explicit calculation for the simplest possible case, 
which is the $O(4)$ NLSM with WZ term which appears at the two-dimensional boundary of the 
BIQH state in three dimensions. 
We also mention here that an $O(4)$ NLSM with WZ term in two dimensions is equivalent to 
a model of an $SU(2)$ matrix field $U= n_4\mathbb{I} + \sum_{a=1}^3 n_a \sigma^a$ (where $\sigma^a$ are the
three Pauli matrices) with WZ term for $U$, so the analysis in this subsection is actually a special case of the analysis done in 
Refs.~\onlinecite{HullSpence1,WittenHolo}. Although we focus on the case of a continuous symmetry (namely the $U(1)$ charge
conservation symmetry), we also note here that anomalies in the two-dimensional boundary theories of SPT phases protected by the symmetry
of a \emph{finite} abelian group were considered previously in Ref.~\onlinecite{wang2015bosonic}.

In the $O(4)$ case the volume form can be written as
\beq
	\omega_3= \mathcal{J}_1\wedge\mathcal{K}_2 + \mathcal{J}_2\wedge\mathcal{K}_1 \ .
\eeq
Under the transformation $b_{\ell} \to e^{i\xi}b_{\ell}$ we have
\beqa
	\delta_{\xi}\omega_3 &=&  \mathcal{K}_1\wedge d\xi + \mathcal{K}_2\wedge d\xi \nnb \\
	&=& \frac{1}{2}d\mathcal{J}_1\wedge d\xi + \frac{1}{2}d\mathcal{J}_2\wedge d\xi  \nnb \\
	&=& \frac{1}{2}d\left[ \mathcal{J}_1\wedge d\xi + \mathcal{J}_2\wedge d\xi \right] \ ,
\eeqa
which is a total derivative.
So we find (neglecting any terms coming from the boundary of the physical spacetime $\mathbb{R}^{1,1}$)
\beq
	\delta_{\xi} S_{WZ}[\mb{n}]= \frac{2\pi k}{\mathcal{A}_{3}}\frac{1}{2}\int_{\mathbb{R}^{1,1}} \left( \mathcal{J}_1+ \mathcal{J}_2\right)\wedge d\xi \ .
\eeq
We attempt to cancel this variation by adding the counterterm
\beq 
	S^{(1)}_{ct}[\mb{n},A] = -\frac{2\pi k}{\mathcal{A}_{3}}\frac{1}{2}\int_{\mathbb{R}^{1,1}} \left( \mathcal{J}_1+ \mathcal{J}_2\right)\wedge A\ .
\eeq
It is clear that when we send $A\to A+d\xi$ in $S^{(1)}_{ct}$ it will cancel the gauge variation of the WZ term. 

At this point our candidate for the gauged WZ term is then
\beq
	S_{WZ,gauged}[\mb{n}, A] = S_{WZ}[\mb{n}] + S^{(1)}_{ct}[\mb{n},A]\ .
\eeq
However, this action is not completely gauge-invariant, and under a gauge transformation we find
\beqa
	\delta_{\xi} S_{WZ,gauged}[\mb{n},A] &=& -\frac{2\pi k}{\mathcal{A}_{3}}\frac{1}{2}\int_{\mathbb{R}^{1,1}} \left( \delta_{\xi}\mathcal{J}_1+ \delta_{\xi}\mathcal{J}_2\right)\wedge A \nnb \\
	&=& -\frac{2\pi k}{\mathcal{A}_{3}}\frac{1}{2}\int_{\mathbb{R}^{1,1}} d\xi \wedge A \nnb \\
	&=& -\frac{ k}{2\pi}\int_{\mathbb{R}^{1,1}} d\xi \wedge A \nnb \\
	&=& k\int_{\mathbb{R}^{1,1}} \xi \left(\frac{F}{2\pi}\right) \ , \label{eq:O4-anomaly}
\eeqa
where we used the formula for $\delta_{\xi} \mathcal{J}_{\ell}$ from Eq.~\eqref{eq:J-K-var}, 
the fact that $\mb{n}$ is a unit vector field, 
$\mathcal{A}_{3}= 2\pi^2$, and also performed an integration by parts in the last line ($F=dA$).
We conclude that the $U(1)$ symmetry here is anomalous and, since the kinetic term has been made completely gauge-invariant,
the total anomaly of the boundary theory is given by Eq.~\eqref{eq:O4-anomaly}. We also note
that the anomaly in  Eq.~\eqref{eq:O4-anomaly} is exactly what is needed to cancel the gauge variation of the bulk CS
action of Eq.~\eqref{eq:BIQH-action} with $N_3= -2k$.

\subsection{The $O(2m)$ NLSM with WZ term in $2m-2$ spacetime dimensions}

Now we move on to the general case of an $O(2m)$ NLSM with WZ term on the $2m-2$ dimensional boundary of a BIQH
state in $2m-1$ dimensions (recall that $m$ is related to the integer $d$ in the BIQH case by $d= 2m-2$, so that
$d$ is also the dimension of the boundary spacetime). In this case we find
that a total of $m-1$ counterterms are needed in order for the gauged WZ action
to transform as in Eq.~\eqref{eq:gauge-trans} under a gauge transformation. To start we note that the volume form
$\omega_{2m-1}$ can be re-written using the forms $\mathcal{J}_{\ell}$ and $\mathcal{K}_{\ell}$ as
\beq
	\omega_{2m-1}= \frac{1}{(m-1)!}\sum_{\ell_1,\dots,\ell_m=1}^m \mathcal{J}_{\ell_1}\wedge \mathcal{K}_{\ell_2}\wedge \dots \wedge \mathcal{K}_{\ell_m}\ .
\eeq
To see it, simply note that if any of $\ell_2,\dots,\ell_m$ are equal to each other or to $\ell_1$ then the wedge product
vanishes. So each index $\ell_s$ can be summed over the full range of $1$ to $m$. However, this means that we are actually
over-counting in the sum over all $\ell_s$. This is not a problem though as $\mathcal{K}_{\ell_s}$ can be commuted past each 
other in the wedge products (they are all two-forms), so all we need to do to remedy this is to divide by the factor of $(m-1)!$, 
where $m-1$ is the number of factors of $\mathcal{K}_{\ell}$ appearing in the expression.

Now for any integer $r$ in the range $0,\dots,m-1,$ we introduce the form
\beq
	\Omega^{(r)}= \sum_{\ell_1,\dots,\ell_{m-r}=1}^m \mathcal{J}_{\ell_1}\wedge \mathcal{K}_{\ell_2}\wedge \dots \wedge \mathcal{K}_{\ell_{m-r}}\ . \label{eq:omega-forms}
\eeq
In particular, we have $\omega_{2m-1}= \frac{1}{(m-1)!}\Omega^{(0)}$
and $\Omega^{(m-1)}= \sum_{\ell_1=1}^m \mathcal{J}_{\ell_1}$.  In Appendix~\ref{app:EC} we give a mathematical interpretation 
of these forms in terms of $U(1)$-equivariant cohomology of $S^{2m-1}$.
The following formula for the change in 
$\Omega^{(r)}$ under a gauge transformation is the essential ingredient in our construction of the full gauged WZ action.

\textit{Claim:} Under a gauge transformation $b_{\ell}\to e^{i \xi}b_{\ell}$ we have 
$\Omega^{(r)} \to \Omega^{(r)} + \delta_{\xi} \Omega^{(r)}$ with
\beq
	\delta_{\xi}\Omega^{(r)}= \frac{1}{2}d\Omega^{(r+1)}\wedge d\xi\ . \label{eq:important}
\eeq

\textit{Proof:} Using Eqs.~\eqref{eq:J-K-var} we can show
\begin{widetext}
\beqa
	\delta_{\xi}\Omega^{(r)} &=& \sum_{\ell_1,\dots,\ell_{m-r}=1}^m (n_{2\ell_1-1}^2 + n_{2\ell_1}^2) \mathcal{K}_{\ell_2}\wedge \dots \wedge \mathcal{K}_{\ell_{m-r}}\wedge d\xi \label{eq:Omega-var}  \\
&+& \sum_{s=2}^{m-r} \sum_{\ell_1,\dots,\ell_{m-r}=1}^m \mathcal{J}_{\ell_1}\wedge \mathcal{K}_{\ell_2}\wedge \dots \wedge \overline{\mathcal{K}_{\ell_s}} \wedge \dots \wedge \mathcal{K}_{\ell_{m-r}}\wedge (n_{2\ell_s-1}dn_{2\ell_s-1} + n_{2\ell_s}dn_{2\ell_s})\wedge d\xi\ , \nnb
\eeqa
\end{widetext}
where the overline again means to omit that term from the wedge product. Next we use the two properties
\begin{subequations}
\label{eq:constraints-even}
\beqa
	\sum_{\ell=1}^m (n_{2\ell-1}^2 + n_{2\ell}^2) &=& 1 \\
	\sum_{\ell=1}^m  (n_{2\ell-1}dn_{2\ell-1} + n_{2\ell}dn_{2\ell}) &=& 0\ ,
\eeqa
\end{subequations}
which follow from the fact that $\mb{n}$ is a unit vector field with $2m$ components, to find that
\beq
	\delta_{\xi}\Omega^{(r)} = \sum_{\ell_2,\dots,\ell_{m-r}=1}^m  \mathcal{K}_{\ell_2}\wedge \dots \wedge \mathcal{K}_{\ell_{m-r}}\wedge d\xi\ ,
\eeq
or after re-indexing,
\beq
	\delta_{\xi}\Omega^{(r)} = \sum_{\ell_1,\dots,\ell_{m-(r+1)}=1}^m  \mathcal{K}_{\ell_1}\wedge \dots \wedge \mathcal{K}_{\ell_{m-(r+1)}}\wedge d\xi\ .
\eeq
So in fact, only the term in the first line of Eq.~\eqref{eq:Omega-var} has contributed.
Next we write $\mathcal{K}_{\ell_1} = \frac{1}{2}d\mathcal{J}_{\ell_1}$ and use the fact that $\mathcal{K}_{\ell}$ is closed
to find
\beqa
	\delta_{\xi}\Omega^{(r)} &=& \frac{1}{2}\sum_{\ell_1,\dots,\ell_{m-(r+1)}=1}^m  d\mathcal{J}_{\ell_1}\wedge\mathcal{K}_{\ell_2} \wedge \dots \wedge \mathcal{K}_{\ell_{m-(r+1)}}\wedge d\xi \nnb \\
	&=& \frac{1}{2} d\Omega^{(r+1)}\wedge d\xi\ ,
\eeqa
which completes the proof. $\blacksquare$

With Eq.~\eqref{eq:important} in hand we can now construct the properly gauged action step by step. We go through the first
few steps explicitly, and then write down the final answer. To start, the change of the WZ term under a gauge transformation
is
\begin{align}
	\delta_{\xi} S_{WZ}[\mb{n}] =& \frac{2\pi k}{\mathcal{A}_{2m-1}}\frac{1}{(m-1)!}\int_{\mathcal{B}} \delta_{\xi}\Omega^{(0)} \nnb  \\
	=& \frac{2\pi k}{\mathcal{A}_{2m-1}}\frac{1}{(m-1)!}\frac{1}{2}\int_{\mathcal{B}} d\Omega^{(1)}\wedge d\xi \nnb \\
	=& \frac{2\pi k}{\mathcal{A}_{2m-1}}\frac{1}{(m-1)!}\frac{1}{2}\int_{\mathbb{R}^{d-1,1}} \Omega^{(1)}\wedge d\xi \ . 
\end{align}
So the first counterterm we should add is
\beq
	S^{(1)}_{ct}[\mb{n},A]= -\frac{2\pi k}{\mathcal{A}_{2m-1}}\frac{1}{(m-1)!}\frac{1}{2}\int_{\mathbb{R}^{d-1,1}} \Omega^{(1)}\wedge A\ .
\eeq
The part of the action containing the WZ term is now
\beq
	S'_{WZ,gauged}[\mb{n},A]= S_{WZ}[\mb{n}] + S^{(1)}_{ct}[\mb{n},A]\ ,
\eeq
and under a gauge transformation we find
\begin{align}
	\delta_{\xi} S'_{WZ,gauged}[\mb{n},A] &= \nnb \\
  -\frac{2\pi k}{\mathcal{A}_{2m-1}}&\frac{1}{(m-1)!}\frac{1}{2}\int_{\mathbb{R}^{d-1,1}} \delta_{\xi}\Omega^{(1)}\wedge A \ ,
\end{align}
which becomes
\begin{align}
\delta_{\xi} S'_{WZ,gauged}[\mb{n},A] &= \nnb \\
	 -\frac{2\pi k}{\mathcal{A}_{2m-1}}\frac{1}{(m-1)!}&\frac{1}{2^2}\int_{\mathbb{R}^{d-1,1}} d\Omega^{(2)}\wedge d\xi \wedge A\ .
\end{align}
Now we note that 
\beq
	d \left(\Omega^{(2)}\wedge d\xi \wedge A \right) = d\Omega^{(2)}\wedge d\xi \wedge A + \Omega^{(2)}\wedge d\xi \wedge F\ ,
\eeq
and we use this to do an integration by parts. Neglecting boundary terms (in general we neglect all terms coming from
the boundaries of the physical boundary spacetime), we now have
\beq
	\delta_{\xi} S'_{WZ,gauged}[\mb{n},A] = \frac{2\pi k}{\mathcal{A}_{2m-1}}\frac{1}{(m-1)!}\frac{1}{2^2}\int_{\mathbb{R}^{d-1,1}} \Omega^{(2)}\wedge d\xi \wedge F\ .
\eeq
Therefore we should choose the second counterterm to be
\beq
	S^{(2)}_{ct}[\mb{n},A]= -\frac{2\pi k}{\mathcal{A}_{2m-1}}\frac{1}{(m-1)!}\frac{1}{2^2}\int_{\mathbb{R}^{d-1,1}} \Omega^{(2)}\wedge A \wedge F\ ,
\eeq
and the total gauged action is now
\beq
	S''_{WZ,gauged}[\mb{n},A]= S_{WZ}[\mb{n}] + S^{(1)}_{ct}[\mb{n},A] + S^{(2)}_{ct}[\mb{n},A]\ .
\eeq

At this point the pattern is clear. After iterating this procedure we find that a total of $m-1$ counterterms are needed to 
construct a gauged WZ action which satisfies Eq.~\eqref{eq:gauge-trans}. 
The $r^{th}$ counterterm (for $r=1,\dots,m-1$) is given by
\beq
	S^{(r)}_{ct}[\mb{n},A]= -\frac{2\pi k}{\mathcal{A}_{2m-1}}\frac{1}{(m-1)!}\frac{1}{2^r}\int_{\mathbb{R}^{d-1,1}} \Omega^{(r)}\wedge A \wedge F^{r-1}\ , \label{eq:BIQH-cts}
\eeq
where $F^{r-1}$ is shorthand for the wedge product of $F$ with itself $r-1$ times. The total gauged action is then
\beq
	S_{WZ,gauged}[\mb{n},A]= S_{WZ}[\mb{n}] + \sum_{r=1}^{m-1} S^{(r)}_{ct}[\mb{n},A]\ .
\eeq
In Appendix~\ref{app:EC} we discuss this gauged WZ action from the point of view of $U(1)$-equivariant cohomology over
the sphere $S^{2m-1}$.

When we look at the change of the full action $S_{WZ,gauged}[\mb{n},A]$ under a gauge transformation we find that it is not
completely gauge-invariant. In other words, the $U(1)$ symmetry of the boundary theory of the BIQH state 
is anomalous, as we expect on physical grounds. The anomaly is controlled only by the final counterterm $S^{(m-1)}_{ct}[\mb{n},A]$, 
since all other contributions cancel by construction. Under a gauge transformation we have
\begin{align}
	\delta_{\xi} S_{WZ,gauged}[\mb{n},A] =& \nnb \\
 -\frac{2\pi k}{\mathcal{A}_{2m-1}}\frac{1}{(m-1)!}&\frac{1}{2^{m-1}}\int_{\mathbb{R}^{d-1,1}} \delta_{\xi}\Omega^{(m-1)}\wedge A \wedge F^{m-2}\ .
\end{align}
Now we use $\delta_{\xi}\Omega^{(m-1)} = d\xi$, the formula $\mathcal{A}_{2m-1}= \frac{2\pi^m}{(m-1)!}$,
and integrate by parts to arrive at the final formula
\beq
	\delta_{\xi} S_{WZ,gauged}[\mb{n},A]= k \int_{\mathbb{R}^{d-1,1}} \xi \left(\frac{F}{2\pi}\right)^{m-1}\ , \label{eq:WZ-anomaly}
\eeq
or in terms of the boundary spacetime dimension $d$,
\beq
	\delta_{\xi} S_{WZ,gauged}[\mb{n},A]= k \int_{\mathbb{R}^{d-1,1}} \xi \left(\frac{F}{2\pi}\right)^{\frac{d}{2}}\ . \label{eq:NLSM-anomaly}
\eeq

\subsection{Chern-Simons effective action for bulk electromagnetic response}

We now use the result of the previous subsection to understand the bulk electromagnetic response of
BIQH states in all odd spacetime dimensions. As we discussed in the Introduction, a Quantum Hall state in 
$2m-1$ dimensions is characterized by the presence of a CS term  
in the effective action $S_{eff}[A]$ for the electromagnetic field $A$. Recall that on $(2m-1)$-dimensional
spacetime the CS term takes the form
\beq
	S_{CS}[A]= \frac{N_{2m-1}}{(2\pi)^{m-1} m!}\int_{\mathcal{M}} A\wedge (dA)^{m-1}\ .
\eeq
Now it is well known that under a gauge transformation $A\to A+d\xi$ the CS action changes by a boundary term,
\beq
	\delta_{\xi} S_{CS}[A]= \frac{N_{2m-1}}{m!}\int_{\pd\mathcal{M}} \xi \left(\frac{F}{2\pi}\right)^{m-1} \ . \label{eq:CS-var}
\eeq
We can then deduce the coefficient $N_{2m-1}$ for the bulk response of BIQH states by matching the variation of the
bulk CS effective action for $A$ with the anomaly of the boundary theory of the BIQH state (the $O(2m)$ NLSM with WZ term)
which we calculated in the previous subsection. The gauge transformation of the bulk CS term must cancel the anomaly of 
the boundary theory in order for the entire system (bulk plus boundary) to be gauge-invariant. This is exactly the concept of 
anomaly inflow\cite{callan1985anomalies} which we mentioned in the 
introduction. Comparing Eq.~\eqref{eq:CS-var} to Eq.~\eqref{eq:WZ-anomaly} for the $U(1)$ anomaly of the $O(2m)$ theory 
with WZ term, we deduce that the coefficient $N_{2m-1}$ must be given by 
\beq
	N_{2m-1}= -(m!)k, \ k\in\mathbb{Z}\ , \label{eq:CS-level}
\eeq
in order to cancel the anomaly of the boundary theory. Therefore we find that the level $N_{2m-1}$ of the CS effective
action for BIQH states in $2m-1$ spacetime dimensions is quantized in units of $m!$. 
This answer agrees with the known cases for three and five spacetime 
dimensions and gives a prediction for all odd dimensions beyond those. In Sec.~\ref{sec:gauge-invariance}
we discuss this peculiar quantization of the CS level from a mathematical point of view by studying the
transformation of the CS term under large $U(1)$ gauge transformations on general Euclidean manifolds (including
manifolds which do not admit a spin structure).

We also remark here that based on the form of the CS response for the BIQH state in $2m-1$ dimensions,
we can conclude that the chiral anomaly of the boundary theory of the BIQH state is $m!$ times larger than the chiral
anomaly of the boundary theory for a fermionic SPT phase in $2m-1$ dimensions with a bulk CS response at level one. 
So, for example, the anomaly of the boundary theory is twice as large when the bulk is three-dimensional ($m=2$ case) and
six times as large when the bulk is five-dimensional ($m=3$).

\subsection{A derivation of the response from the bulk physics}

To close this section we present an alternative derivation of the response of the BIQH state. This derivation uses only bulk 
properties of the BIQH state, which should be contrasted with our derivation using gauged WZ actions which was based
on the anomaly of the boundary theory. Recall again that the bulk of the BIQH state is described by an $O(2m)$ NLSM with theta term 
and theta angle $\theta=2\pi k$ (so we have a theta term and not a WZ term in the bulk description).
The main reason for including this alternative derivation is that it
provides a clear physical reason for the appearance of the $m!$ factor in the response.
The derivation in this subsection uses only the dimensional reduction properties of the
CS response action for the external field, and the theta term of the NLSM,  which we now review. 

We start by considering the CS response action at level $N$ in $2m-1$ dimensions,
\beq
	S_{CS}[A]= \frac{N}{(2\pi)^{m-1} m!}\int_{\mathbb{R}^{D,1}} A\wedge (dA)^{m-1}\ , \label{eq:CS}
\eeq
where $D$ is the spatial dimension so that $D+1 = 2m-1$. Let $\mb{x}=(x^1,\dots,x^D)$ be the spatial coordinates.
Now suppose we thread a delta function of $2\pi$ flux at a point $\mb{x}_0$ in the $(x^{D-1},x^D)$ 
plane (i.e., $x_0^j = 0$, $j= 1,\dots,D-2$). Concretely, we set 
\beqa
	F_{x^{D-1}x^D} &=& 2\pi \delta(x^{D-1}-x^{D-1}_0)\delta(x^{D}-x^{D}_0)\ ,
\eeqa
and we assume that  $F_{x^j x^{D-1}}= F_{x^jx^D}= 0$ 
$\forall j=1,\dots,D-2$, and that $F_{x^j x^k}$ is independent of $(x^{D-1},x^D)$ for $j,k=1,\dots,D-2$.
Then, for this configuration, the CS response action reduces to
\beq
	S_{CS}[A] \to  \frac{N}{(2\pi)^{m-2} (m-1)!}\int_{\mathbb{R}^{D-2,1}} \tilde{A}\wedge (d\tilde{A})^{m-2}\ . \label{eq:CS-dim-red}
\eeq
The key point is that it reduces to a CS term \emph{at the same level} $N$ on the $(D-2)$-dimensional space located at the point 
$\mb{x}_0$ in the $(x^{D-1},x^D)$ plane. 

Now that we know what happens in the CS response action when we thread a $2\pi$ delta function flux of $F$ in a particular plane,
let us also see what happens in the NLSM description of the BIQH phase when this flux is inserted. In the NLSM description, the
$m$ bosons $b_{\ell}$ are all charged under the $U(1)$ symmetry. Therefore, threading a 
$2\pi$ delta function flux at the point $\mb{x}_0$ in the $(x^{D-1},x^D)$ plane will cause all of the bosons
$b_{\ell}$ to have a vortex configuration in that plane around the point $\mb{x}_0$. By a vortex configuration we just mean
that the phases of the complex numbers $b_{\ell}$ all wind by $2\pi$ as one encircles the point $\mb{x}_0$ in the 
$(x^{D-1},x^D)$ plane. So we conclude that threading a $2\pi$ delta function flux of $F$ will create $m$ vortex excitations
in the $O(2m)$ NLSM which describes the bulk of the BIQH. 

On the other hand, we are going to show that
if a \emph{single boson} $b_{\ell}$ for some $\ell$
has a vortex configuration at a point $\mb{x}_0$ in the $(x^{D-1},x^D)$ plane, then the $O(2m)$ NLSM
action with $\theta=2\pi k$ reduces to an $O(2m-2)$ NLSM with $\theta=2\pi k$ living on the $(D-2)$-dimensional space at 
$\mb{x}_0$. So if we have a vortex in one boson only, then the NLSM theory for the BIQH state in $2m-1$ dimensions
reduces to the NLSM theory for the BIQH state in $2m-3$ dimensions (inside the vortex core) and \emph{with the same theta angle}. 

We now prove the assertion in the previous paragraph that a vortex in one boson $b_{\ell}$ in the $O(2m)$ NLSM traps an 
$O(2m-2)$ NLSM with the same theta angle inside the vortex core. To do this we consider an explicit vortex ansatz for the 
NLSM field in which the last boson $b_{m}= n_{2m-1}+ in_{2m}$ takes on a vortex configuration. 
To set up the notation let $(r,\phi)$ be polar coordinates for the $(x^{D-1},x^D)$ plane, and let $\mb{y}= (x^1,\dots,x^D)$
be the coordinates for the remaining directions of space. 
Then our vortex ansatz has the form
\beq
	\mb{n}(t,\mb{x})= \{\sin(f(r))\mb{N}(t,\mb{y}),\cos(f(r))\mb{m}(\phi)\}\ .
\eeq
where $\mb{N}(t,\mb{y})$ is a $(2m-2)$-component unit vector field depending only on $t$ and $\mb{y}$, and 
$\mb{m}(\phi)= (\cos(\phi),\sin(\phi))$ represents the vortex configuration of the last two components of $\mb{n}$. The
function $f(r)$ is assumed to satisfy the boundary conditions
\beqa
	f(0) &=& \frac{\pi}{2} \\
	\lim_{r\to\infty} f(r) &=& 0\ ,
\eeqa
which means that the field $\mb{N}(t,\mb{y})$ lives in the core of the vortex. This vortex ansatz is equivalent to the
$q=1$, $n_q=1$, case of the more general defect configurations for NLSMs considered in Appendix~\ref{app:dim-red-NLSM}. 
Using the dimensional reduction formula from Eq.~\eqref{eq:dim-red-theta} of Appendix~\ref{app:dim-red-NLSM} we immediately 
derive that on this configuration the theta term of the $O(2m)$ NLSM reduces to
\beqa
	S_{\theta}[\mb{n}] &=& \frac{\theta}{\mathcal{A}_{2m}}\int_{\mathbb{R}^{D,1}}\mb{n}^*\omega_{2m}\nnb \\
	&\to& \frac{\theta}{\mathcal{A}_{2m-2}}\int_{\mathbb{R}^{D-2,1}}\mb{N}^*\omega_{2m-2}\ .
\eeqa
This is the theta term for the $O(2m-2)$ NLSM with field $\mb{N}$ living in the vortex core, and we see that the theta angle
is the same as for the original $O(2m)$ NLSM. This proves our claim from the previous paragraph.

From the discussion above we see that threading a $2\pi$ flux of $F$ in the $O(2m)$ NLSM theory will produce
$m$ copies of the $O(2m-2)$ theory, since the $2\pi$ flux creates a vortex in all $m$ species of bosons, and a vortex in just
one species produces one copy of the $O(2m-2)$ NLSM with theta term. 
We should mention a technical point that the $m$ vortices cannot all be localized at a point and should spread or separate slightly in space after we thread the $2\pi$ flux. This
is because the amplitude $|b_{\ell}|$ should vanish at the core of a vortex in the phase of $b_{\ell}$, but the NLSM constraint
$\sum_{\ell} |b_{\ell}|^2$ does not allow the amplitudes $|b_{\ell}|$ for all $\ell$ to simultaneously vanish at a particular point. However, this subtlety does
not effect the basic physical point which is that threading the $2\pi$ flux of $F$ produces $m$ vortices (at nearly the same point),
each of which carries a copy of the lower dimensional BIQH state.

Let us denote the CS level for the response of the $O(2m)$ NLSM with $\theta=2\pi k$ in $2m-1$
dimensions by $N_{2m-1}.$ From what we have just learned, and from Eq.~\eqref{eq:CS-dim-red} for the reduction of the CS term
after threading $2\pi$ flux, we find that the CS levels for the response of the NLSMs in dimensions $2m-1$ and 
$2m-3= 2(m-1)-1$  must obey the recursion relation
\beq
	N_{2m-1} = m N_{2m-3}\ .
\eeq
We can now iterate this equation to generate
\beq
	N_{2m-1} = (m!)N_1\ .
\eeq
This equation gives the electromagnetic response of the $O(2m)$ NLSM with $\theta=2\pi k$ in terms of the response 
of the $O(2)$ NLSM in one dimension with $\theta=2\pi k$. 
In Appendix~\ref{app:O2} we directly calculate $N_1$ for the $O(2)$ NLSM (in the limit of large coupling $g$) and
show that $N_1 = -k$ in that case. This then implies that
\beq
	N_{2m-1}= -(m!) k\ ,
\eeq
and this agrees (in magnitude and in sign) with our boundary calculation using gauged WZ actions. 
Thus, the dimensional reduction approach employed in this
subsection gives a clear physical picture for the $m!$ factor in the response, and crucially depends on the fact that all the bosons $b_{\ell}$ carry a $U(1)$ charge.

\section{General Gauge invariance argument for the BIQH response and comparison with the fermionic case}
\label{sec:gauge-invariance}

In this section we show that the factor of $m!$ in the BIQH response derived in Sec.~\ref{sec:BIQH} 
can be understood by studying large $U(1)$ gauge transformations of the CS action on general (closed, compact) 
Euclidean manifolds which do not necessarily admit a spin structure. Physically, we require the \emph{exponential} 
of the CS term to be gauge-invariant, since this object is part of the partition function of a short-range entangled (gapped)
phase coupled to the external field $A$.  In such phases, since the ground state is always unique, one can always safely 
integrate out the matter field and obtain a gauge-invariant action. In contrast, if we do the same thing for a topologically 
ordered state, for example a Laughlin state, we will indeed get a non-gauge-invariant response theory. This is because the 
calculation to arrive at a response theory is only perturbatively defined around a single ground state.

The level $N_{2m-1}$ of the CS term must be quantized for the exponential of the
CS term to be gauge-invariant, but we find that the required quantization of $N_{2m-1}$ is 
different depending on whether
or not the Euclidean manifold admits a spin structure. Bosonic theories may be formulated on any generic manifold, but
the Dirac equation cannot be formulated properly on a manifold which does not admit  a spin structure, and so we cannot
place fermions on these manifolds. In particular we find that the CS action will be gauge-invariant on a generic manifold
if the level $N_{2m-1}$ is quantized in integer multiples of $m!$, which agrees with our direct calculation for the NLSM theory
from Sec.~\ref{sec:BIQH}. For the fermionic case we use the Atiyah-Singer index theorem for the twisted Dirac 
complex\cite{EGH} 
to  show that the CS response action will not, in general, be $U(1)$ gauge-invariant unless suitable gravitational terms are also 
included in the response action. We also discuss an explicit example of how these gravitational terms can 
contribute to the response of a fermionic SPT phase with $U(1)$ symmetry. Furthermore, using these examples, we compare 
the quantization of FIQH and BIQH states, as well as another type of bosonic SPT state with non-trivial topological electromagnetic-
gravitational response.

\subsection{Gauge invariance argument for bosonic and fermionic states}

In Euclidean spacetime the CS term takes the form
\beq
	S_{CS}[A]= -i\frac{N_{2m-1}}{(2\pi)^{m-1} m!}\int_{\mathcal{M}} A\wedge F^{m-1}\ .
\eeq
Here $\mathcal{M}$ is
a $(2m-1)$-dimensional closed, compact manifold, 
and for the moment let us assume that $N_{2m-1}$ is some number, not necessarily an integer.
A more careful way to define the CS term is to consider an extension of the field configuration $A$ into a $2m$-dimensional 
manifold $\mathcal{B}$ such that $\pd\mathcal{B}=\mathcal{M}$ (this type of analysis of CS terms dates back at least to 
Ref.~\onlinecite{alvarez1985topological}). Let $\tilde{A}$ denote this extension.
Then the CS term is more properly written as
\beq
	S_{CS}[A]= -i\frac{N_{2m-1}}{(2\pi)^{m-1} m!}\int_{\mathcal{B}} \tilde{F}^{m}\ ,
\eeq
where $\tilde{F}=d\tilde{A}$. In this formulation, a large $U(1)$ gauge transformation of the action can be understood as
a change of the extension of $A$ into the larger space $\mathcal{B}$. 
Suppose $\tilde{A}^{(1)}$ and $\tilde{A}^{(2)}$ are two different extensions of $A$. In order for
the CS term to be well-defined, we require that the difference 
\beq
	-i\frac{N_{2m-1}}{(2\pi)^{m-1} m!}\int_{\mathcal{B}} (\tilde{F}^{(1)})^{m} - \left( -i\frac{N_{2m-1}}{(2\pi)^{m-1} m!}\int_{\mathcal{B}} (\tilde{F}^{(2)})^{m} \right)
\eeq
be an integer multiple of $2\pi i$ so that the exponential of the difference of the two Euclidean actions is equal to one.
This is equivalent to the requirement that the exponential of the CS term be invariant under a large $U(1)$ gauge 
transformation. 
This difference can in turn be written as the integral of the field strength $F$ of a gauge field in $2m$ dimensions over the closed manifold
$2m$-dimensional manifold $X$ constructed by gluing $\mathcal{B}$ to another copy of $\mathcal{B}$ (with the opposite orientation) along their
boundary (which is the original lower-dimensional manifold $\mathcal{M}$). 
So the requirement for a well-defined CS term is to check that 
\beq
	I[A]= -i\frac{N_{2m-1}}{(2\pi)^{m-1} m!}\int_{X} F^{m}\ , \label{eq:chern-character-app}
\eeq
is equal to $2\pi k$ for some integer $k$, where $X$ is a $2m$-dimensional closed, compact manifold, and $F$ is now the field 
strength of a gauge field $A$ living in $2m$ dimensions.

We must also make one crucial assumption about the configuration of $F$ on X, which is that $F$ should be chosen to satisfy 
the Dirac quantization condition
\beq
	\int_{\mathcal{C}}\frac{F}{2\pi} \in \mathbb{Z}\ ,\label{eq:Dirac-condition}
\eeq
where $\mathcal{C}$ is any non-trivial two-cycle on $X$ 
(i.e., an element of the second homology group $H_2(X,\mathbb{R})$). 
This requirement tells us how a general background field $F$ on $X$ can be expanded in terms of the elements of the second 
cohomology group $H^2(X,\mathbb{R})$ of $X$ (more precisely, we expand $F$ in terms of elements of the second de Rham
cohomology group $H^2_{dR}(X)$, which is in turn isomorphic to $H^2(X,\mathbb{R})$ by de Rham's theorem).

If we enforce the Dirac quantization condition of Eq.~\eqref{eq:Dirac-condition}, then on a generic closed, compact Euclidean 
manifold $X$ we have
\beq
	\int_{X} \left(\frac{F}{2\pi}\right)^{m} \in \mathbb{Z}\ .
\eeq
Briefly, this comes from the fact that (assuming the Dirac quantization condition) $\frac{F}{2\pi}$ is the first Chern class $c_1$
of a complex line bundle over $X$. The integral over $X$ of its $m^{th}$ power $(c_1)^m$ is then one of the Chern numbers of this
complex line bundle, and is therefore an integer\cite{milnor}.
Note that here we also need to assume that $X$ is orientable. 
From this result we deduce that the (exponential of the) CS term will be invariant under large $U(1)$ gauge transformations on 
any Euclidean manifold provided that
\beq
	N_{2m-1}= (m!) k\ , \ k\in\mathbb{Z}\ 
\eeq
which agrees with our result from Sec.~\ref{sec:BIQH} derived using the NLSM description of the BIQH state.
In Appendix~\ref{app:CP} we show that the minimum value with $\int_{X} \left(\frac{F}{2\pi}\right)^{m} =1$ can be 
achieved for $X=C\mathbb{P}^m$ if we thread $2\pi$ flux of $F$ through the non-trivial
two-cycle on $C\mathbb{P}^m$.

We can also compare this result with the result for FIQH phases
with $U(1)$ symmetry in the same dimension. In any odd dimension, we can consider the massive Dirac fermion as a model for a FIQH state with the global $U(1)$ 
symmetry associated to charge conservation. The Lagrangian of a massive Dirac fermion on 
flat, $(2m-1)$-dimensional Minkowski spacetime takes the form
\begin{align}
\mathcal{L}_\text{Dirac} [\psi, A] = \overline{\psi} (i \slashed{\pd} -\slashed{A} - M) \psi\ ,
\end{align}
where $\gamma^\mu$, $\mu=0,\dots,2m-2$, are the standard Gamma matrices satisfying 
$\{\gamma^\mu, \gamma^\nu\} = 2\eta^{\mu\nu}$ with $\eta^{\mu\nu} = \text{diag}(1,-1,-1,...,-1)$, 
$\overline{\psi}= \psi^{\dg}\gamma^0$, and $M>0$ is 
the mass of the Dirac fermion. We also used the Feynman slash notation $\slashed{\pd} \equiv \gamma^{\mu}\pd_{\mu}$, 
etc. Here we have also coupled the fermion $\psi$ to the background $U(1)$ gauge field (electromagnetic field) $A_{\mu}$.
After integrating out the massive Dirac fermion, we arrive at a topological response theory given by the CS theory at level one:
\begin{align}
S_{\text{Dirac}}[A] = -i\frac{1}{(2\pi)^{m-1}m!} \int_{\mathcal{M}} A \wedge F^{m-1}\ ,
\label{Dirac_Gauge_Response}
\end{align}
where in this case the spacetime manifold $\mathcal{M}$ is just $(2m-1)$-dimensional Minkowski spacetime.
 In deriving this response theory we have employed a Pauli-Villars regularization procedure (see Ref.~\onlinecite{Redlich}
or the more recent discussion in Ref.~\onlinecite{witten2015fermion}) 
such that integrating out a Dirac fermion with a negative mass $M$ does not produce any topological term (i.e., a CS term with level zero).
Also, we have omitted all the non-topological terms, for example the Maxwell term, 
from the final response action. Since a single massive Dirac fermion gives rise to a CS term for $A$ at level one, we have  
the result that
\begin{align}
N_{2m-1} \in \mathbb{Z}
\end{align}
for general $U(1)$ fermionic SPT phases in $2m-1$ dimensions. 

 However, as we know from the discussion of the CS term earlier in this section, on a generic manifold $\mathcal{M}$ the 
CS term will not be invariant under large $U(1)$ gauge transformations unless the level $N_{2m-1}$ is an integer multiple of $m!$.
Thus, one might naively conclude that the response action for the FIQH state on a generic manifold $\mathcal{M}$ is not invariant under
large $U(1)$ gauge transformations. Of course, this is not the case. The resolution of this problem is to recall that on a curved manifold
$\mathcal{M}$ a Dirac fermion also has non-trivial gravitational and (when coupled to the gauge field $A$) mixed gauge and gravitational
responses. The gravitational part of the response comes from the coupling of the Dirac fermion to the metric $g_{\mu\nu}$ of the
curved spacetime $\mathcal{M}$. The response action for the FIQH state (as modeled by the massive Dirac fermion) 
will include these additional terms.
%However, if we follow the more careful definition of the CS term as we did earlier in this section (i.e., paying attention to the 
%behavior of the CS term under large $U(1)$ gauge transformations), then we would naively conclude that 
%$S_{\text{Dirac}}[A]$ is not gauge-invariant. 
% The same argument applies to the CS term with a generic integer level 
%$N_{2m-1}$. This is due to the fact that when $N_{2m-1}$ is not a multiple of $m!$, the integral in 
%Eq.~\eqref{eq:chern-character-app} does not evaluate to $2\pi i\times$(integer). 
%Even requiring that the manifolds $\mathcal{M}$, $\mathcal{B},$ and $X$  be spin manifolds, 
%which is natural for Dirac fermions, does not completely resolve the issue. Indeed, $C\mathbb{P}^m$
%for $m$ odd \emph{is} a spin manifold, but the integral in Eq.~\eqref{eq:chern-character-app} for $X=C\mathbb{P}^m$ 
%does not evaluate to $2\pi i\times$(integer), as we show in Appendix~\ref{app:CP}.
%We argue that the natural solution is to consider 
%coupling the FIQH state not only to the background $U(1)$ gauge field $A$ but also to the background gravitational field
%$g_{\mu\nu}$. 
The effective action for a massive Dirac fermion on a $(2m-1)$-dimensional closed, compact manifold $\mathcal{M}$ can be written in the
form\cite{alvarez1985anomalies}
\begin{align}
S_{\text{FIQH}}[A , g] = 2\pi i \int_{\mathcal{B}} \text{ch}(\tilde{F}) \wedge \hat{A}(\mathcal{B})\ ,
\label{fSPT_Gauge_Gravity_Response}
\end{align}
where $\partial\mathcal{B} = \mathcal{M}$, $\text{ch}(\tilde{F}) = e^{\frac{\tilde{F}}{2\pi}}$ is the Chern character of the 
extended field strength $\tilde{F}$ and $ \hat{A}(\mathcal{B})$ is the \emph{A-roof genus} (or Dirac genus) on 
$\mathcal{B}$. Since we are focusing on fermionic phases here, we should only consider spin manifolds $\mathcal{M}$ and $\mathcal{B}$. The A-
roof genus $\hat{A}(\mathcal{B})$ can be expressed in terms of the Pontryagin classes 
$p_i (\mathcal{B})$ of $\mathcal{B}$ as\cite{alvarez1985structure},
\begin{align}
\hat{A}(\mathcal{B}) = 1 -\frac{1}{24} p_1 + \frac{1}{5760}(7p_1^2 -4p_2) + ...,
\end{align}
with
\begin{align}
& p_1 = -\frac{1}{8\pi^2} \Tr \tilde{\mathcal{R}}^2, \\
& p_2 = -\frac{1}{64\pi^4} \Tr \tilde{\mathcal{R}}^4 + \frac{1}{128\pi^4} \left( \Tr \tilde{\mathcal{R}}^2 \right)^2.
\end{align}
Here, $\tilde{R}$ is the $2m\times 2m$ matrix of two-forms (curvature two-form) on $\mathcal{B}$:
\begin{align}
\tilde{R}_\mu^\nu = \frac{1}{2} \tilde{R}_{\alpha\beta\mu} {}^\nu dx^\alpha \wedge dx^\beta
\end{align}
which depends on the Riemann curvature tensor $\tilde{R}_{\alpha\beta\mu}{}^{\nu}$ in the extended space $\mathcal{B}$.
In Eq.~\eqref{fSPT_Gauge_Gravity_Response} it is understood that the integral is only over the terms of (differential form) degree $2m$ in the 
product $\text{ch}(\tilde{F}) \wedge \hat{A}(\mathcal{B})$ on $\mathcal{B}$. 
It is easy to see that when we only consider the electromagnetic response in $S_{\text{FIQH}}[A , g] $ (e.g., by  
setting all $p_i$ to 0 on $\mathcal{B}$), it recovers the response theory Eq.~\eqref{Dirac_Gauge_Response} of the massive Dirac 
fermion in $2m-1$ dimensions. More importantly, the response theory $S_{\text{FIQH}}[A , g] $ is fully gauge-invariant. 
This is because on any closed, compact $2m$-dimensional spin manifold $X$, the Atiyah-Singer index theorem for
the twisted Dirac complex (see, for example, Ref.~\onlinecite{EGH}) states that 
\begin{align}
\int_{X} \text{ch}(\tilde{F}) \wedge \hat{A}(X) = \text{index}(\slashed{D}) \in \mathbb{Z}\ ,
\end{align}
where $\text{index}(\slashed{D})$ is the index (the difference between the number of positive and negative 
chirality zero modes) of the Dirac operator on $X$, and is necessarily an integer. Although we originally derived
Eq.~\eqref{fSPT_Gauge_Gravity_Response} by using the theory of a massive Dirac fermion on the curved manifold
$\mathcal{M}$ as a model for the FIQH state, we argue that due to the requirement of large $U(1)$ gauge invariance,  
Eq.~\eqref{fSPT_Gauge_Gravity_Response} is the minimal (or ``level 1") non-trivial gauge and 
gravitational response theory of any putative FIQH phase with $U(1)$ symmetry in $(2m-1)$ dimensions.

%Therefore, we claim that Eq.~\eqref{fSPT_Gauge_Gravity_Response} is the minimal (or ``level 1") non-trivial gauge and 
%gravitational response theory of a FIQH phase with $U(1)$ symmetry in $(2m-1)$ dimensions. 

There is one more subtlety here. When $m$ is even  (i.e., when
the spacetime dimension is $4k-1$ with $k\in\mathbb{Z}$), the object  $\text{ch}(\tilde{F}) \wedge \hat{A}(\mathcal{B})$ contains a purely gravitational term that comes from $\hat{A}(\mathcal{B})$ alone. Such a term itself can be well-defined (the index theorem for the untwisted Dirac complex guarantees that
it integrates to an integer on a closed, compact spin manifold) and can capture the non-trivial gravitational response of certain short-range entangled states even without the inclusion of a global $U(1)$ symmetry. For example, for $m=2$ the purely gravitational term is 
given by $-\frac{1}{24}p_1$ on $\mathcal{B}$, which is equivalent to the three-dimensional 
gravitational Chern-Simons term on $\mathcal{M}.$ This term is tied to the chiral central charge. 
Hence, we can separately consider the purely 
gravitational term $\hat{A}(\mathcal{B})$ and the rest of the terms $\left[ \text{ch}(\tilde{F}) \wedge \hat{A}(\mathcal{B}) - \hat{A}(\mathcal{B})\right]$ in Eq.~\eqref{fSPT_Gauge_Gravity_Response}.

In general, we can consider the FIQH phase at level $N_{2m-1}\in \mathbb{Z}$, whose topological response theory 
(minus the purely gravitational term) is given by
\begin{align}
S'_{\text{FIQH}}[A , g] = 2\pi iN_{2m-1} \int_{\mathcal{B}} \left[ \text{ch}(\tilde{F}) \wedge \hat{A}(\mathcal{B}) - \hat{A}(\mathcal{B}) \right]\ .
\label{Eq:FIQH_Action_General}
\end{align}
$S'_{\text{FIQH}}[A , g]$ naturally contains both a term capturing the electromagnetic response of the FIQH state and other terms that describe various different types of mixed gauge-gravitational response. The coexistence of all these terms is enforced by the properties of spin manifolds and the Atiyah-Singer index theorem, and reflects the fermionic nature of the FIQH phase.
This combination also informs us that we should \emph{not} use each of the terms to independently classify fermionic SPTs with $U(1)$ symmetry. For bosonic systems, we can, in principle, separately study each single term in $S'_{\text{FIQH}}[A , g] $ by itself, and use each of them to characterize a different class of bosonic SPTs. However, just like the quantization of the level of the $U(1)$ CS term, we expect gauge invariance to enforce a larger quantization unit of the ``level" when we isolate a single term as a bosonic response theory, as opposed to the case where that term appears in the full combination $S'_{\text{FIQH}}[A , g]$ as a part of a fermionic theory. The difference in the quantization unit of the ``level" between fermionic and bosonic systems will also lead to very different behaviors under dimensional reduction, the details of which will be elaborated using examples. In Sec.~\ref{Sec:Example_FIQH_response}, we provide an example of the electromagnetic and gravitational response theory of FIQH states in five dimensions. In Sec. \ref{Sec:Comparison_Quantization_DReduction}, we compare the example fermionic response theory with five-dimensional bosonic theories, including the BIQH state and another type of bosonic $U(1)$ SPT state with non-trivial mixed electromagnetic and gravitational response.

\subsection{An example of electromagnetic and gravitational response theories of FIQH states and their dimensional reduction}
\label{Sec:Example_FIQH_response}

In this section, we restrict our discussion to the topological response theory of a five-dimensional FIQH phase, and
we study its dimensional reduction to the response theory of a FIQH state in three dimensions.  
We start with the response theory of the FIQH phase at level $N_5=1$ on a 
five-dimensional spin manifold $\mathcal{M}^5$:
\begin{align}
S_{\text{FIQH}}[A , g] & = 2\pi i \int_{\mathcal{B}^6} \left[ \frac{1}{6} \left(\frac{\tilde{F}}{2\pi}\right)^3 -\frac{p_1}{24} \frac{\tilde{F}}{2\pi}  \right]\ ,
\label{fSPT_4+1D_Response}
\end{align}
where $\mathcal{B}^6$ is a six-dimensional spin manifold such that $\mathcal{M}^5 = \partial \mathcal{B}^6$. We first consider its dimensional reduction to the response theory of a FIQH state in three dimensions.  
In order to do so, we take the spacetime manifold to be
$\mathcal{M}^5 = S^2 \times \mathcal{M}^3$ 
where $\mathcal{M}^3$ is a closed, compact three-dimensional manifold, and $S^2$ is a two-sphere. 
In this case, it is natural to consider the bounding space 
$\mathcal{B}^6 = S^2 \times \mathcal{B}^4$ where $\mathcal{B}^4$ is a four-dimensional spin manifold such that 
$\mathcal{M}^3 = \partial \mathcal{B}^4$. Also, we consider the configuration with $2\pi$ flux of $\tilde{F}$ piercing
the $S^2$ part. The response theory is then reduced to 
\begin{align}
S_{\text{FIQH}}[A , g] \Big|_{S^2 \times \mathcal{M}^3} & = 2\pi i \int_{\mathcal{B}^4} \left( \frac{\tilde{F}^2}{8\pi^2} -\frac{p_1}{24} \right) \nonumber 
\\
& = i \int_{\mathcal{M}^3} \Bigg[ \frac{A\wedge F}{4\pi} \nnb \\
 & - \frac{1}{24} \frac{1}{4\pi} \Tr \left( \omega \wedge d \omega + \frac{2}{3} \omega \wedge \omega \wedge \omega \right)\Bigg]\ ,
\end{align}
where $\omega$ is the $SO(1,2)$ spin connection on $\mathcal{M}^3$. The first term describes the standard Integer
Quantum Hall effect in three dimensions with unit Hall conductance. The second term, which is the gravitational Chern-Simons 
term, captures the gravitational response of a three-dimensional chiral state with chiral central charge $c=1$. On the other 
hand, we can directly consider a five-dimensional massive Dirac fermion as a model of a five-dimensional FIQH state at level one 
on this background. When put on the manifold 
$S^2\times \mathcal{M}^3$ with $2\pi$ flux of $F$ inside the $S^2$ part, the five-dimensional 
massive Dirac fermion effectively reduces to a three-dimensional massive Dirac fermion on $\mathcal{M}^3$ at low energies 
when the linear size of the $S^2$ part is small compared to the length scale set by the Dirac fermion mass $M$. The $U(1)$ and 
gravitational response of the three-dimensional FIQH state is indeed given by the dimensionally-reduced response theory 
$S_{\text{FIQH}}[A , g] \Big|_{S^2 \times \mathcal{M}^3}$.

Finally, let us also remark here that the response theory Eq.~\eqref{fSPT_4+1D_Response} for the five-dimensional FIQH state
can also be used to derive the electromagnetic and gravitational responses of a topological superconductor in four dimensions using a dimensional 
reduction procedure\cite{QiWittenZhang2013}.

\subsection{Comparing bosonic and fermionic systems: quantization and dimensional reduction}
\label{Sec:Comparison_Quantization_DReduction}

As we have discussed, we can consider each term of $S_{\text{FIQH}}[A, g]$ separately as a topological response theory for 
bosonic $U(1)$ SPTs in five dimensions:
\begin{align}
S_{\text{BIQH}}[A] & = 2\pi i N_5 \int_{\mathcal{B}^6} \frac{1}{6} \left(\frac{\tilde{F}}{2\pi}\right)^3\ ,
\\
S_{\text{BSPT}}[A,g] & = -2\pi i N_5' \int_{\mathcal{B}^6} \frac{p_1}{24}\wedge \frac{\tilde{F}}{2\pi}\ .
\label{Eq:BSPT_4+1_respsonse}
\end{align}
$S_{\text{BIQH}}[A]$ is the response theory of a five-dimensional BIQH state, and requires a quantization of level as
 $N_5 \in 6\mathbb{Z}$ as we showed in this section and in Sec.~\ref{sec:BIQH}. $S_{\text{BSPT}}[A,g]$ characterizes an independent class of bosonic SPT states in five dimensions without a requirement of $U(1)$ symmetry\cite{wang2015GaugeGravity}. Similar to the BIQH and FIQH cases, gauge invariance requires $ N_5' \int_{X^6}\frac{p_1}{24} \wedge\frac{\tilde{F}}{2\pi} \in \mathbb{Z}$ on any closed six-dimensional manifold $X^6$. Since $p_1$ and $\frac{\tilde{F}}{2\pi}$ are both cohomology classes of $X^6$ with integer coefficients, gauge invariance then enforces the quantization
$N_5' \in 24\mathbb{Z}$. We would like to point out that previously Ref.~\onlinecite{wang2015GaugeGravity} considered only closed six-dimensional manifolds that can be decomposed into products of two and four-dimensional manifolds, and concluded that $N_5' \in 8\mathbb{Z}$. However, when we take into account 
more general six-dimensional manifolds, for example $C\mathbb{P}^3$, we arrive at the stronger quantization condition $N_5' \in 24\mathbb{Z}$. \footnote{When we consider $C\mathbb{P}^3$ with the $U(1)$ gauge field given by its fundamental line bundle, we find that $N_5' \int_{C\mathbb{P}^3}\frac{p_1}{24} \frac{\tilde{F}}{2\pi} = N_5'/6$. Combining with the result $N_5' \in 8\mathbb{Z}$ from Ref. \onlinecite{wang2015GaugeGravity}, we can conclude that the gauge invariance argument requires $N_5' \in 24\mathbb{Z}$. On the other hand, since $p_1$ and $\frac{\tilde{F}}{2\pi}$ are both cohomology classes with integer coefficients, any $N_5' \in 24\mathbb{Z}$ will satisfy the gauge invariance requirement. }
As seen here, for both of the bosonic theories $S_{\text{BIQH}}[A]$ and $S_{\text{BSPT}}[A,g]$, the quantization units of their levels are larger than when these two terms appear together in the fermionic theory $S_{\text{FIQH}}[A,g]$ in 
Eq.~\eqref{fSPT_4+1D_Response}.  

Now let us consider a similar dimensional reduction of both $S_{\text{BIQH}}[A_\mu]$ and $S_{\text{BSPT}}[A_\mu,g]$ to 
three dimensions, as we did in the fermion case. Now the five-dimensional spacetime manifold $\mathcal{M}^5$ is taken to be the product 
$S^2 \times \mathcal{M}^3$ with $\mathcal{M}^3$ a three-dimensional manifold. Again, we consider the configuration with $2\pi$ flux of $\tilde{F}$ piercing the $S^2$ part. The dimensionally reduced theories are given by
\begin{align}
S_{\text{BIQH}}[A] \Big|_{S^2 \times \mathcal{M}^3}  & =  i2\pi \frac{N_5}{2}  \int_{\mathcal{M}^3} \frac{A\wedge F}{(2\pi)^2},
\nonumber \\
 & 
\\
S_{\text{BSPT}}[A,g] \Big|_{S^2 \times \mathcal{M}^3}  & = \nnb \\
 -i 2\pi \frac{N'_5}{24}  &\int_{\mathcal{M}^3} \frac{1}{4\pi} \Tr \left( \omega \wedge d \omega + \frac{2}{3} \omega \wedge \omega \wedge \omega \right).
\end{align}
For the BIQH state, due to the bosonic quantization $N_5 \in 6 \mathbb{Z}$, we notice that the most fundamental 
three-dimensional BIQH state (with CS level $N_3=2$) cannot be realized from such a dimensional reduction from a five-dimensional BIQH state. From our analysis of the CS level of the BIQH state, it should be generally true that there are certain lower-dimensional BIQH states that cannot be realized from the dimensional reduction of higher-dimensional BIQH states. In fact, this phenomenon is not restricted to BIQH states.
 For the bosonic SPT states described by Eq.~\eqref{Eq:BSPT_4+1_respsonse}, due to the quantization $N_5' \in 24 \mathbb{Z}$, the action $S_{\text{BSPT}}[A,g] \Big|_{S^2 \times \mathcal{M}^3}$ only captures chiral bosonic states with chiral central charge $c \in 24 \mathbb{Z}$. The $E_8$ state in $(2+1)$ dimensions, which has chiral central charge $c=8$, is absent in this dimensional reduction picture. This is in strong contrast with the fermionic theory studied in Sec.~\ref{Sec:Example_FIQH_response}, in which case lower-dimensional response theories of FIQH at any level can be obtained from dimensionally reducing higher-dimensional FIQH states.

\section{Electromagnetic response of BTI states in all even dimensions}
\label{sec:BTI}

In this section we construct the gauged WZ action for the boundary of BTI states in all even dimensions. Again,
the action that we construct satisfies the gauging principle of
Eq.~\eqref{eq:gauge-trans}. Unlike the BIQH case, however, the gauged boundary action that we find for BTI states \emph{is}
completely gauge-invariant.  
 From the form of the gauged action for the boundary of the BTI, we find that if the NLSM
field on the boundary condenses in such a way that the $\mathbb{Z}_2$ symmetry of the BTI is broken, but the $U(1)$ symmetry
remains intact, then the boundary of the BTI can exhibit a $\mathbb{Z}_2$ symmetry-breaking Quantum Hall response 
(recall from Sec.~\ref{sec:background} that the BTI phase also 
has a $\mathbb{Z}_2$ symmetry such that the total symmetry group is $U(1)\rtimes\mathbb{Z}_2$)
\footnote{Our result can also be applied to systems with a symmetry of the
form $U(1)\times \mathbb{Z}_2$, but only in the case that the $U(1)$ symmetry rotates the maximal number of components
of $n_a$ as in the $U(1)\rtimes\mathbb{Z}_2$ cases considered in this paper. For example, according to 
Ref.~\onlinecite{CenkeClass1} bosonic SPT phases in four dimensions with $U(1)\times \mathbb{Z}^T_2$ symmetry have
a $(\mathbb{Z}_2)^3$ classification. However, only one of the three root phases is described by an $O(5)$ NLSM with symmetry
assignment that rotates four out of the five components of $\mb{n}$\cite{VS2013}, so this is the only case in which our 
technique can be applied directly. For the other cases one must use the more general methods of 
Ref.~\onlinecite{HullSpence2} to gauge the $U(1)$ symmetry.}. We
find that the boundary Quantum Hall response is characterized by a CS level $N_{2m-1}$ which is quantized in units of 
$\frac{m!}{2}$, i.e., the minimal boundary Quantum Hall response is half that of the minimal BIQH state that can be realized intrinsically
in the same spacetime dimension. This boundary response implies a bulk response of the form of 
Eq.~\eqref{eq:chern-character} with the parameter $\Theta_{2m}$ quantized as 
$\Theta_{2m}= 2\pi\left(\frac{m!}{2}\right)$.

In Appendix~\ref{app:EC} we re-interpret the gauged action constructed in this section in terms of $U(1)$-equivariant cohomology
of the sphere $S^{2m}$. There we show that the problem of constructing a gauged WZ action for the boundary of the BTI
phase in $2m$ dimensions is equivalent to the problem of constructing an equivariant extension of $\omega_{2m}$, the volume
form for $S^{2m}$, and we explicitly construct such an extension. The fact that an extension exists
is mathematically equivalent to the result in this section that the gauged WZ action for the boundary of the BTI is completely
gauge-invariant. We also show that the forms $\Phi^{(r)}$ which appear later in this section in the counterterms of 
Eq.~\eqref{eq:r-ct-BTI} are exactly the same forms which are needed for the construction of the equivariant extension of 
$\omega_{2m}$.

We now construct the gauged WZ action for the boundary of BTI states. Recall that
in the BTI case we define the integer $m$ via
$2m+1=d+2$, so that the SPT phases we study live in $2m$ spacetime dimensions and have a 
$2m-1$ dimensional boundary (the bulk spacetime dimension was defined to be $d+1$). 
We again make use of the forms $\mathcal{J}_{\ell}$ and $\mathcal{K}_{\ell}$, $\ell= 1,\dots,m$ defined in 
Eqs.~\eqref{eq:J-K-def}. Now, however, the NLSM field has the extra component $n_{2m+1}$, 
so the relations of Eq.~\eqref{eq:constraints-even} are replaced with
\begin{subequations}
\label{eq:constraints-odd}
\beqa
	\sum_{\ell=1}^m (n_{2\ell-1}^2 + n_{2\ell}^2) &=& 1 - n_{2m+1}^2\\
	\sum_{\ell=1}^m  (n_{2\ell-1}dn_{2\ell-1} + n_{2\ell}dn_{2\ell}) &=& - n_{2m+1} dn_{2m+1} \ .
\eeqa
\end{subequations}
In this case the WZ term takes the form
\beq
	S_{WZ}[\mb{n}]= \frac{2\pi k}{\mathcal{A}_{2m}}\int_{\mathcal{B}} \omega_{2m}\ ,
\eeq
where $\mathcal{B}= [0,1]\times\mathbb{R}^{d-1,1}$ is the extended boundary spacetime.

For the BTI case it is convenient to define the forms $\Phi^{(r)}$ for $r=0,1,\dots,m-1$ as
\beq
	\Phi^{(r)}= \sum_{\ell_1,\dots,\ell_{m-r}=1}^m \mathcal{K}_{\ell_1}\wedge \cdots\wedge \mathcal{K}_{\ell_{m-r}}\ , \label{eq:phi-forms}
\eeq
and in addition we define $\Phi^{(m)}= 1$, so that $\Phi^{(r)}$ is defined for all $r=0,1,\dots,m$.
Also, note that all of these forms are closed since each $\mathcal{K}_{\ell}$ is closed.
Just as in the BIQH case, the essential ingredient in the construction of the gauged WZ action is a formula for how these forms 
change under a gauge transformation. 

\textit{Claim:} Under a gauge transformation $b_{\ell}\to e^{i \xi}b_{\ell}$ we have 
$\Phi^{(r)} \to \Phi^{(r)} + \delta_{\xi} \Phi^{(r)}$ with
\beq
	\delta_{\xi}\Phi^{(r)}= - (m-r)n_{2m+1}dn_{2m+1}\wedge \Phi^{(r+1)}\wedge d\xi\ . \label{eq:gauge-trans-odd}
\eeq

\textit{Proof:} Using the symmetry of the summand of $\Phi^{(r)}$ under the exchange of any two of the indices
$\ell_1,\dots,\ell_{m-r},$ we first find that
\begin{widetext}
\beqa
	\delta_{\xi}\Phi^{(r)}= (m-r)\sum_{\ell_1,\dots,\ell_{m-r}=1}^m (n_{2\ell_1-1}dn_{2\ell_1 -1} + n_{2\ell_1}dn_{2\ell_1})\wedge d\xi\wedge \mathcal{K}_{\ell_2}\wedge \cdots\wedge \mathcal{K}_{\ell_{m-r}} \ .
\eeqa
\end{widetext}
Now we can move $d\xi$ all the way to the right by commuting it past the two-forms $K_{\ell_2},\dots,K_{\ell_{m-r}}$. This
gives
\beq
	\delta_{\xi}\Phi^{(r)}= (m-r)\sum_{\ell_1=1}^m  (n_{2\ell_1-1}dn_{2\ell_1 -1} + n_{2\ell_1}dn_{2\ell_1}) \wedge \Phi^{(r+1)} \wedge d\xi\ ,
\eeq
where we used the fact that 
\beq
	\Phi^{(r+1)}= \sum_{\ell_2,\dots,\ell_{m-r}=1}^m \mathcal{K}_{\ell_2}\wedge \cdots\wedge \mathcal{K}_{\ell_{m-r}}\ .
\eeq
Finally we can do the sum over $\ell_1$ using the second relation of Eqs.~\eqref{eq:constraints-odd}, and this gives the final 
formula of Eq.~\eqref{eq:gauge-trans-odd}. $\blacksquare$

In terms of the form $\Phi^{(0)}$ we can write the volume form on $S^{2m}$ as
\begin{align}
	\omega_{2m} = \frac{1}{(m-1)!}\Bigg[\sum_{\ell_1,\dots,\ell_m=1}^m &\mathcal{J}_{\ell_1}\wedge\mathcal{K}_{\ell_2}\wedge\cdots\wedge \mathcal{K}_{\ell_m}\wedge dn_{2m+1} \nnb \\
&+\ \frac{n_{2m+1}}{m}\Phi^{(0)}  \Bigg]\ .
\end{align}
The last term in this expression is just the term
\beq
	n_{2m+1} dn_1\wedge dn_2 \wedge\cdots \wedge dn_{2m-1}\wedge dn_{2m}\ ,
\eeq
but re-written using the formula
\beq
	dn_1\wedge dn_2 \wedge\cdots \wedge dn_{2m-1}\wedge dn_{2m}= \frac{1}{m!}\sum_{\ell_1,\dots,\ell_m=1}^m  \mathcal{K}_{\ell_1}\wedge \cdots\wedge \mathcal{K}_{\ell_{m}}\ .
\eeq

We are now in a position to construct the properly gauged action step by step as in Section \ref{sec:BIQH}  on the BIQH system.
We demonstrate the first few steps in the construction and then write down the final answer. To start we have
\beqa
	\delta_{\xi} \omega_{2m} &=& -\frac{1}{(m-1)!} dn_{2m+1}\wedge \Phi^{(1)}\wedge d\xi \nnb \\
	&=& -\frac{1}{(m-1)!} d \left(n_{2m+1} \Phi^{(1)}\wedge d\xi \right)\ .
\eeqa
This is computed using Eq.~\eqref{eq:gauge-trans-odd} for the case $r=0$ combined with the formula
\begin{align}
	\delta_{\xi}\left( \sum_{\ell_1,\dots,\ell_m=1}^m \mathcal{J}_{\ell_1}\wedge\mathcal{K}_{\ell_2}\wedge\cdots\wedge \mathcal{K}_{\ell_m}\wedge dn_{2m+1} \right) =& \nnb \\
 -(1-n_{2m+1}^2)dn_{2m+1}\wedge\Phi^{(1)}\wedge& d\xi\ ,
\end{align}
which is easily proven using Eq.~\eqref{eq:J-K-var} and Eq.~\eqref{eq:constraints-odd}. Then we have
\beq
	\delta_{\xi} S_{WZ}[\mb{n}]=  -\frac{2\pi k}{\mathcal{A}_{2m}}\frac{1}{(m-1)!}\int_{\mathbb{R}^{d-1,1}} n_{2m+1} \Phi^{(1)}\wedge d\xi\ .
\eeq
We therefore choose the first counterterm to be 
\beq
	S^{(1)}_{ct}[\mb{n},A]= \frac{2\pi k}{\mathcal{A}_{2m}}\frac{1}{(m-1)!}\int_{\mathbb{R}^{d-1,1}} n_{2m+1} \Phi^{(1)}\wedge A\ .
\eeq
The total gauged WZ action is now
\beq
	S'_{gauged,WZ}[\mb{n},A]=  S_{WZ}[\mb{n}] + S^{(1)}_{ct}[\mb{n},A] \ ,
\eeq
and under a gauge transformation we find
\begin{align}
	\delta_{\xi} S'_{gauged,WZ}[\mb{n},A]=& \nnb \\
 -\frac{2\pi k}{\mathcal{A}_{2m}}\frac{1}{(m-2)!}&\int_{\mathbb{R}^{d-1,1}} n^2_{2m+1} dn_{2m+1}\wedge \Phi^{(2)}\wedge d\xi \wedge A\ .
\end{align}
Next we integrate by parts using the formula
\begin{align}
	d \left( \frac{1}{3} n^3_{2m+1}\Phi^{(2)}\wedge d\xi\wedge A  \right) =& \nnb \\
 n^2_{2m+1}dn_{2m+1}\wedge \Phi^{(2)}\wedge d\xi &\wedge A - \frac{1}{3} n^3_{2m+1}\Phi^{(2)}\wedge d\xi\wedge F\ ,
\end{align}
to find (neglecting boundary terms)
\begin{align}
	\delta_{\xi} &S'_{gauged,WZ}[\mb{n},A] =\nnb \\
 &-\frac{2\pi k}{\mathcal{A}_{2m}}\frac{1}{(m-2)!}\frac{1}{3}\int_{\mathbb{R}^{d-1,1}} n^3_{2m+1} \Phi^{(2)}\wedge d\xi \wedge F\ .
\end{align}
We should then take the second counterterm to be
\beq
	S^{(2)}_{ct}[\mb{n},A]= \frac{2\pi k}{\mathcal{A}_{2m}}\frac{1}{(m-2)!}\frac{1}{3}\int_{\mathbb{R}^{d-1,1}} n^3_{2m+1} \Phi^{(2)}\wedge A \wedge F\ .
\eeq

To see the full structure of the counterterms it is necessary to go one step further. At this point the total gauged action is
\beq
	S''_{gauged,WZ}[\mb{n},A]=  S_{WZ}[\mb{n}] + S^{(1)}_{ct}[\mb{n},A] + S^{(2)}_{ct}[\mb{n},A]\ , 
\eeq
and under a gauge transformation we have
\begin{align}
	\delta_{\xi}& S''_{gauged,WZ}[\mb{n},A]= \nnb \\
-\frac{2\pi k}{\mathcal{A}_{2m}}&\frac{1}{(m-3)!}\frac{1}{3}\int_{\mathbb{R}^{d-1,1}} n^4_{2m+1} dn_{2m+1}\wedge \Phi^{(3)}\wedge d\xi \wedge A \wedge F\ .
\end{align}
We again integrate by parts and show
\begin{align}
	\delta_{\xi} S''_{gauged,WZ}[\mb{n},A]&= \nnb \\
-\frac{2\pi k}{\mathcal{A}_{2m}}\frac{1}{(m-3)!}&\frac{1}{5\cdot 3}\int_{\mathbb{R}^{d-1,1}} n^5_{2m+1} \Phi^{(3)}\wedge d\xi \wedge F^2\ .
\end{align}
Note that the denominator contains the \emph{double factorial} $5!!= 5\cdot 3= 5\cdot 3\cdot 1$. In general, we find
 that all of the counterterms contain a double factorial. Then
the third counterterm takes the form
\beq
	S^{(3)}_{ct}[\mb{n},A]= \frac{2\pi k}{\mathcal{A}_{2m}}\frac{1}{(m-3)!}\frac{1}{5!!}\int_{\mathbb{R}^{d-1,1}} n^5_{2m+1} \Phi^{(3)}\wedge A \wedge F^2\ .
\eeq

At this point the pattern is clear. Continuing with this procedure we find that a total of $m$ counterterms are needed
to construct a gauged boundary action which satisfies Eq.~\eqref{eq:gauge-trans}, and the final gauged action is 
\emph{completely gauge-invariant}. It takes the form
\beq
	S_{WZ,gauged}[\mb{n},A]= S_{WZ}[\mb{n}] + \sum_{r=1}^m S^{(r)}_{ct}[\mb{n},A]\ , \label{eq:BTI-bdy}
\eeq
where the $r^{th}$ counterterm is
\begin{align}
	S^{(r)}_{ct}[\mb{n},A]&= \nnb \\
 \frac{2\pi k}{\mathcal{A}_{2m}}\frac{1}{(m-r)!}&\frac{1}{(2r-1)!!}\int_{\mathbb{R}^{d-1,1}}(n_{2m+1})^{2r-1}\ \Phi^{(r)}\wedge A \wedge F^{r-1}\ , \label{eq:r-ct-BTI}
\end{align}
where $(2r-1)!!$ is the double factorial,
\beq
	(2r-1)!!= (2r-1)(2r-3)\cdots(3)(1)\ .
\eeq
The final counterterm is just
\beq
	S^{(m)}_{ct}[\mb{n},A]= \frac{2\pi k}{\mathcal{A}_{2m}}\frac{1}{(2m-1)!!}\int_{\mathbb{R}^{d-1,1}}(n_{2m+1})^{2m-1}\  A \wedge F^{m-1}\ ,
\eeq
and its change under a gauge transformation comes only from the transformation of $A$ (the last component $n_{2m+1}$
of the NLSM field does not transform under $U(1)$). This explains why the final gauged action is completely gauge-invariant: the
change due to the transformation of $A$ in the last term cancels the transformation from the previous counterterm in the action,
and there are no further changes in the last term which remain to be canceled.

Now let us show that the boundary of a BTI phase exhibits a $\mathbb{Z}_2$ symmetry breaking response when the field $n_a$ condenses
in such a way that it preserves the $U(1)$ symmetry, but breaks the $\mathbb{Z}_2$ symmetry. The only possible way for
$n_a$ to condense and fulfill these requirements is to have
\begin{subequations}
\label{eq:BTI-condense}
\beqa
	n_{2m+1} &=& \pm 1, \\
	n_a&=& 0,\ \forall\  a\neq 2m+1\ .
\eeqa
\end{subequations}
In this case, all terms in $S_{WZ,gauged}[\mb{n},A]$ vanish except for the final counterterm ($r=m$), which gives the 
boundary electromagnetic response,
\beq
	S_{eff,bdy}[A]= \pm\frac{2\pi k}{\mathcal{A}_{2m}}\frac{1}{(2m-1)!!}\int_{\mathbb{R}^{d-1,1}} A \wedge F^{m-1}\ ,
\eeq
where we used $0!=1$ and $\Phi^{(m)}= 1$. Now we use the formulas
\beq
	\mathcal{A}_{2m} = \frac{2\pi^m \sqrt{\pi}}{\Gamma(m+\tfrac{1}{2})}\ ,
\eeq
and 
\beq
	(2m-1)!!= \frac{2^m}{\sqrt{\pi}}\Gamma(m+\tfrac{1}{2})\ ,
\eeq
to find
\beq
	S_{eff,bdy}[A]= \pm\frac{1}{2}\frac{k}{(2\pi)^{m-1}}\int_{\mathbb{R}^{d-1,1}} A \wedge F^{m-1}\ .
\eeq
Comparing to Eq.~\eqref{eq:CS-term}, we see that this is a CS response with level
\beq
	N_{2m-1}= \pm \left(\frac{m!}{2}\right) k\ ,
\eeq 
which is exactly \emph{half} the response of the BIQH state which appears intrinsically in the same spacetime dimension 
(which we calculated in Section \ref{sec:BIQH}). As we discussed in Sec.~\ref{sec:background}, this boundary CS
response is equivalent to a bulk electromagnetic response of the form of Eq.~\eqref{eq:chern-character} with response parameter
\beq
	\Theta_{2m}= 2\pi  \left(\frac{m!}{2}\right) k\ .
\eeq
However, we should recall from the discussion in Sec.~\ref{sec:background} that the BTI phase with $k=2$ is smoothly connected to 
the phase with $k=0$. More generally the BTI phase with $\theta=2\pi k$ is smoothly connected to the phase with
$\theta=2\pi (k\pm 2)$. This means that the single non-trivial BTI phase is represented by the choice $k=1$.

Finally, we note that the boundary of the BTI can be driven into the $\mathbb{Z}_2$ symmetry breaking phase without explicitly
breaking the $\mathbb{Z}_2$ symmetry. This can be done by adding a term $\mu\ n_{2m+1}^2$ to the Lagrangian. This
term is invariant under the full $U(1)\rtimes\mathbb{Z}_2$ symmetry of the BTI but, for $\mu >0$ and sufficiently
large, will drive the system into a phase in which the $\mathbb{Z}_2$ symmetry is spontaneously broken and 
$n_a= \pm \delta_{a,2m+1}$ (i.e., $n_{2m+1}=\pm 1$ and $n_a=0$ for $a\neq 2m+1$).

\section{Applications}
\label{sec:applications}

In this section we explore several applications of the results obtained so far. We start with the observation that 
the gauged boundary action for the BIQH state in $2m-1$ spacetime dimensions can be used as building block to construct 
 a bosonic analogue of a Weyl, or chiral, semi-metal in \emph{any} even dimension. We refer to this state
as a bosonic chiral semi-metal (BCSM). We write down an effective theory for this state in any even dimension $d$, compute 
its electromagnetic response, and compare this response with
the response of an ordinary fermionic chiral semi-metal. This construction represents a generalization to higher even dimensions  
of the work in Ref. \onlinecite{lapa2016bosonic} that constructed a bosonic analogue of a \emph{Dirac} semi-metal in three 
dimensions.

As a second application, we show that the boundary theory of the BTI exhibits a bosonic analogue of the parity anomaly of a single
Dirac fermion in odd dimensions. As we discuss below, this is closely related to the fact (derived in 
Sec.~\ref{sec:BTI}) that the boundary theory of the BTI can exhibit a half-quantized BIQH state when the
$\mathbb{Z}_2$ symmetry of the BTI is broken \emph{spontaneously} at the boundary. This situation is clearly 
analogous to the time-reversal symmetry-breaking half-quantized Integer Quantum Hall state which appears on the 
surface of the familiar electron topological insulator\cite{kanehasan}. 
This leads us to argue that the boundary theory for a BTI state in $2m$ dimensions cannot
exist intrinsically in $2m-1$ dimensions without breaking (partially or fully) the symmetry of the BTI state.

Finally, we perform a detailed study of $\mathbb{Z}_2$ symmetry-breaking domain walls on the boundary of BTI states.
We use a dimensional reduction formula for NLSMs with WZ term, analogous to the dimensional reduction formula 
for theta terms that we derive in Appendix~\ref{app:dim-red-NLSM}, to show that a $\mathbb{Z}_2$ symmetry-breaking
domain wall on the boundary of a BTI state in $2m$ dimensions hosts a lower-dimensional theory which is identical to the 
boundary theory of the BIQH state in $2m-1$ dimensions. We show that the $U(1)$ anomaly of the theory on the domain wall is
exactly canceled by an inflow of charge from the two $\mathbb{Z}_2$ breaking regions on either side of the domain wall. This
calculation is an important consistency check for our results on the response of BIQH and BTI states, and also provides 
a clear example of the phenomenon of anomaly inflow in the context of bosonic SPT phases.

\subsection{Bosonic analogue of a Weyl semi-metal in any even dimension}

In this section we describe how a bosonic analogue of a Weyl semi-metal can be constructed in any even spacetime dimension
$d$ using two copies of an $O(d+2)$ NLSM with Wess-Zumino (WZ) term. Before discussing the bosonic analogue, let
us first review the basic construction of a Weyl (or more generally a chiral) semi-metal of fermions in any even dimension
$d$. Note that our construction here still assumes a point-like structure of the Fermi surface even in higher dimensions, as opposed to
the recent construction in Ref.~\onlinecite{lian2016five} using Weyl sheets in six spacetime dimensions.
 We consider a Dirac fermion $\Psi$ in $d$ dimensions. To write down an action for a Dirac fermion we need the 
gamma matrices $\gamma^{\mu}$, $\mu= 0,\dots,d-1,$ which obey the Clifford algebra 
$\{\gamma^{\mu},\gamma^{\nu}\}= 2\eta^{\mu\nu}$ (and we choose $\eta= \text{diag}(1,-1,\dots,-1)$).
When $d$ is even we have an extra element 
$\overline{\gamma}$ of the Clifford algebra which anti-commutes with the other gamma matrices and can be chosen to 
satisfy $\overline{\gamma}^{\dg}= \overline{\gamma}$ and $\overline{\gamma}^2= \mathbb{I}$ 
($\overline{\gamma}$ is the higher-dimensional analog of $\gamma^5$ in $d=4$). 
In the basis (known as the Weyl basis in $d=4$) in which $\overline{\gamma}$ takes the block diagonal form 
\beq
	\overline{\gamma}= \begin{pmatrix}
	\mathbb{I} & 0 \\ 
	 0 & - \mathbb{I}
\end{pmatrix}\ ,
\eeq
the fermion $\Psi$ breaks up into chiral and anti-chiral parts as
\beq
	\Psi= (\Psi_{+},\Psi_{-})^T\ .
\eeq

Now a minimal, two-node chiral (or Weyl) semi-metal (CSM) in $d$ dimensions is described at low energies by chiral fermions
$\Psi_{\pm}$ separated in momentum by $2 \mb{B}$ and in energy by $2 B_t$, where $\mb{B}= (B_1,\dots,B_{d-1})$
should be thought of as a vector in a $(d-1)$-dimensional momentum space (or Brillouin zone). 
We assume here that the components $B_{\mu}$ ($\mu=0,\dots,d-1$, $B_0=B_t$) 
are constant, although the results below are expected to hold approximately if the components $B_{\mu}$ are slowly varying
functions of $x^{\mu}$.  In addition, both chiral fermions carry charge $e$ of an external
$U(1)$ gauge field $A_{\mu}$. Using the extra gamma matrix $\overline{\gamma}$, an action capturing
this low-energy physics takes the form
\beq
	S_{CSM}[\Psi,A,B] = \int d^d x\ i\overline{\Psi}(\slashed{\pd} - i e \slashed{A} - i\slashed{B}\overline{\gamma})\Psi\ , \label{eq:CSM}
\eeq
where $\overline{\Psi}= \Psi^{\dg}\gamma^0$ and we used the Feynman slash notation 
$\slashed{\pd}= \gamma^{\mu}\pd_{\mu}$, etc. In addition, we have assumed that the separation of $\Psi_{\pm}$
in momentum and energy is symmetric about $B_{\mu}=0$, so that $\Psi_{\pm}$ is located at $\pm B_{\mu}$ in 
momentum/energy space. We also note here that in this low-energy description, the chiral fermion fields
$\Psi_{\pm}$ couple only to the linear combinations $e A_{\mu} \pm B_{\mu}$ of the vector fields $A_{\mu}$ and $B_{\mu}$.
This feature will be important later in our construction of a bosonic analogue of the CSM.

The quasi-topological part of the electromagnetic response of the CSM follows directly from the 
\emph{axial anomaly} of a Dirac fermion in $d$ dimensions\cite{zyuzin2012}.
This is because this response is generated by attempting to 
remove the coupling to $B_{\mu}$ from the action via the chiral rotation
\beq
	\Psi \to e^{i\xi \overline{\gamma}}\Psi\ , \label{eq:chiral-trans}
\eeq
with the parameter $\xi$ chosen as
\beq
	\xi= B_{\mu}x^{\mu}\ .
\eeq
This transformation removes the coupling to $B_{\mu}$ from the action. The physical interpretation of this transformation is 
that it moves the two cones of the chiral semi-metal to the origin of the Brillouin zone. 
However, the path integral measure is not
invariant under this transformation. Instead, the change in the path integral 
measure generates a new term in the action of the form (``f" stands for fermionic)
\beq
	S^{(f)}_{eff}[A,B]= -\frac{2}{\left(\frac{d}{2}\right)!}\left(\frac{e}{2\pi}\right)^{\frac{d}{2}}\int_{\mathbb{R}^{d-1,1}}\xi\  (F)^{\frac{d}{2}}\ . \label{eq:CSM-resp-0}
\eeq
Noting that $d\xi= B_{\mu}dx^{\mu} \equiv B$ (for constant $B_{\mu}$), and integrating by parts gives the 
final form of the chiral semi-metal response
\beq
	S^{(f)}_{eff}[A,B]= \frac{2}{\left(\frac{d}{2}\right)!}\left(\frac{e}{2\pi}\right)^{\frac{d}{2}} \int_{\mathbb{R}^{d-1,1}} B\wedge A\wedge (F)^{\frac{d}{2}-1}\ . \label{eq:CSM-resp}
\eeq
It is also interesting to consider the form Eq.~\eqref{eq:CSM-resp-0} of the semi-metal response (before integrating by parts),
as it has the form of the ``Chern character" terms discussed earlier in the paper, but with a spacetime-dependent
angle $\xi= B_{\mu}x^{\mu}$ appearing in the integrand.

So under the chiral transformation of Eq.~\eqref{eq:chiral-trans}, the CSM action of Eq.~\eqref{eq:CSM} transforms
as
\beq
	S_{CSM}[\Psi,A,B] \to S_{CSM}[\Psi,A,0] + S^{(f)}_{eff}[A,B]\ ,
\eeq
where we again emphasize that the term $S^{(f)}_{eff}[A,B]$ was generated by the change in the path integral measure
under the chiral transformation of Eq.~\eqref{eq:chiral-trans}. Thus, we can say that the electromagnetic
response of the CSM with non-zero separation vector $B_{\mu}$ differs from the response of a CSM with separation 
vector $B_{\mu}=0$ (i.e., a system where the two chiral parts of the Dirac fermion sit at the same point in momentum
space) by the term $S^{(f)}_{eff}[A,B]$ from Eq.~\eqref{eq:CSM-resp}.
For $d=2$ and $d=4$ the responses are
\beq
	S^{(f)}_{eff}[A,B] = \frac{e}{\pi}\int_{\mathbb{R}^{1,1}} B\wedge A\ ,
\eeq
and
\beq
	S^{(f)}_{eff}[A,B] = \frac{e^2}{4\pi^2}\int_{\mathbb{R}^{3,1}} B\wedge A\wedge F\ ,
\eeq
respectively. We see that the general expression of Eq.~\eqref{eq:CSM-resp} agrees with the known expressions in low 
dimensions~\cite{zyuzin2012,ramamurthy2015}. 

Having reviewed the basic properties of fermionic chiral semi-metals, we
are now ready to describe our construction of a bosonic analogue of a CSM (BCSM). To motivate our construction we
note that the low-energy theory of the CSM has (at least) two essential properties which we use as a guide to construct
the BCSM model. The first property is that the CSM model is constructed from two building blocks, namely the chiral fermion
theories with fields $\Psi_{\pm}$, such that each building block \emph{on its own} would have an anomaly in the $U(1)$ 
symmetry which sends $\Psi_{s} \to e^{i\xi_s}\Psi_{s},\ s=\pm$. 
The second property (already noted above) is that the two building blocks 
$\Psi_{\pm}$ couple only to the linear combinations $e A_{\mu}\pm B_{\mu}$ of vector fields. 
This property, combined with the
axial anomaly of the Dirac fermion, is responsible for the form of the CSM response shown 
in Eq.~\eqref{eq:CSM-resp}. We now describe the construction of a bosonic theory with very similar properties. 

Our low-energy theory for a BCSM in $d$ dimensions ($d$ even) consists of two copies of the 
$O(d+2)$ NLSM with WZ term, i.e., two copies of the boundary theory of the BIQH state in $d+1$ dimensions. To understand 
this system we briefly recall a few facts from Sec.~\ref{sec:BIQH} about the boundary theory of the BIQH state. The
boundary of the BIQH state in $2m-1$ dimensions is described by an $O(2m)$ NLSM with WZ term. Here the dimension 
$d$ is related to $m$ by $d= 2m-2$ as we are constructing a model using the boundary theory for the BIQH state.
Finally, recall that under a $U(1)$ transformation the NLSM field transforms as in Eq.~\eqref{eq:U1} (in units where
the boson charge $e$ is set to $1$). We showed that the properly
gauged boundary action had an anomaly in this $U(1)$ symmetry, 
with the anomaly given explicitly by Eq.~\eqref{eq:NLSM-anomaly}.

To construct an effective theory for a bosonic semi-metal in $d$ dimensions we use two copies of the boundary
theory of the BIQH state. 
We label the fields of the two copies by $\mb{n}_{\pm}$, or $b_{\ell,\pm}$ when written in terms of bosons, and
we take the two copies to have opposite level on their WZ term, $k_{\pm}= \pm k$. Finally, in the effective theory we model 
the separation of the two copies in momentum/energy space by coupling the fields $b_{\ell,\pm}$ to the linear combinations
$A_{\mu}\pm B_{\mu}$ of the external $U(1)$ gauge field $A_{\mu}$ and the momentum/energy shift field $B_{\mu}$.
Then our action for the BCSM theory takes the form
\begin{align}
	\tilde{S}_{BCSM}[\mb{n}_{+},&\mb{n}_{-},A,B]= \nnb \\
 &S_{gauged}[\mb{n}_{+},A+B] + S_{gauged}[\mb{n}_{-},A-B]\ , \label{eq:BCSM-no-ct}
\end{align}
where $S_{gauged}[\mb{n},A]$ is the properly gauged action for one $O(d+2)$ NLSM with WZ term and coupled
to the external field $A$ (as constructed in Sec.~\ref{sec:BIQH}). We put a tilde on 
$\tilde{S}_{BCSM}[\mb{n}_{+},\mb{n}_{-},A,B]$ because, as we now discuss, this action has an inconsistency and must be 
modified.

Suppose that the
vector field $B_{\mu}$, which is a constant in the context of the chiral semi-metal, instead had a non-trivial spacetime
dependence, i.e., $dB \neq 0$. In this case the action in Eq.~\eqref{eq:BCSM-no-ct} is \emph{not} invariant under
the $U(1)$ gauge transformation $b_{\ell,\pm} \to e^{i\theta}b_{\ell,\pm}$, $A\to A+d\theta$. Instead, under
this transformation one can show that the change in the action of  Eq.~\eqref{eq:BCSM-no-ct} is
\begin{widetext}
\beq
	\delta_{\theta}\tilde{S}_{BCSM}[\mb{n}_{+},\mb{n}_{-},A,B] =-\frac{k}{(2\pi)^{m-1}}\sum_{p=0}^{m-1}\binom{m-1}{p}[1 + (-1)^{m-p}] \int_{\mathbb{R}^{d-1,1}}  d\theta\wedge (dA)^p \wedge B \wedge (dB)^{m-2-p} \ .
\eeq
where $2m-1=d+1$. This equation requires some explanation. To compute it we used the relation Eq.~\eqref{eq:WZ-anomaly} 
for the $U(1)$ anomaly for each gauged WZ theory in Eq.~\eqref{eq:BCSM-no-ct} (but coupled to the combinations
of fields $A\pm B$ instead of $A$ alone), then expanded the powers $(dA\pm dB)^{m-1}$ using the binomial expansion, and
finally performed an integration by parts to move one derivative off of $B$ and onto $\theta$.

So in the presence of a spacetime-dependent $B_{\mu}$, our putative semi-metal model is not invariant under $U(1)$ gauge
transformations. To remedy this we modify the action by adding the counterterm 
\beq
	S_{ct}[A,B]= \frac{k}{(2\pi)^{m-1}}\sum_{p=0}^{m-1}\binom{m-1}{p}[1 + (-1)^{m-p}] \int_{\mathbb{R}^{d-1,1}}  A\wedge (dA)^p \wedge B \wedge (dB)^{m-2-p}\ .
\eeq
\end{widetext}
The change in this counterterm under $A\to A+d\theta$ exactly compensates for the change in Eq.~\eqref{eq:BCSM-no-ct}
under the $U(1)$ gauge transformation, and so the modified BCSM action
\begin{align}
	S_{BCSM}[\mb{n}_{+},\mb{n}_{-},A,B] &= \nnb \\
\tilde{S}_{BCSM}[\mb{n}_{+}&,\mb{n}_{-},A,B] + S_{ct}[A,B]\ , \label{eq:BCSM-ct}
\end{align}
is completely gauge-invariant even in the presence of a spacetime-dependent $B_{\mu}$. The counterterm
$S_{ct}[A,B]$ is the analogue in our bosonic theory of the \emph{Bardeen counterterm} that one adds to the theory of a
Dirac fermion coupled to vector and axial vector gauge fields to ensure conservation of the vector current in the quantum
theory\cite{bardeen1969anomalous}. Since this counterterm is absolutely necessary for the more general case of a 
spacetime-dependent $B_{\mu}$, we argue that one should include it even in the simple semi-metal setting in which we
take $B_{\mu}$ to be a constant. If we now restrict to the case of a constant $B_{\mu}$, then only the 
$p=m-2$ term in the counterterm survives, and the counterterm reduces to
\beq
	S_{ct}[A,B] \to -\frac{2k}{(2\pi)^{m-1}}(m-1)\int_{\mathbb{R}^{d-1,1}}  B\wedge A\wedge (dA)^{m-2}\ ,
\eeq
where we used $\binom{m-1}{m-2}=m-1$.

To compute the response of the modified BCSM theory in Eq.~\eqref{eq:BCSM-ct}, we attempt to remove the coupling to 
$B$ from the action via the chiral transformation
\beq
	b_{\ell,\pm} \to e^{\pm i\xi}b_{\ell,\pm}\ ,
\eeq
where $\xi= B_{\mu}x^{\mu}$ as in the fermionic case. Note that this transformation takes the opposite sign for the
two copies of the NLSM theory: this is the analogue in the bosonic theory of the chiral transformation of 
Eq.~\eqref{eq:chiral-trans} that we performed in the fermionic case. Using the $U(1)$ anomaly for the boundary
theory of the BIQH state from Eq.~\eqref{eq:NLSM-anomaly}, we find that under this transformation the original effective
action for the BCSM state transforms as
\begin{align}
	\tilde{S}_{BCSM}[\mb{n}_{+},&\mb{n}_{-},A,B] \to \nnb \\
 &\tilde{S}_{BCSM}[\mb{n}_{+},\mb{n}_{-},A,0] + \tilde{S}^{(b)}_{eff}[A,B]\ ,
\end{align}
where 
\beq
	 \tilde{S}^{(b)}_{eff}[A,B]= -\frac{2k}{(2\pi)^{m-1}} \int_{\mathbb{R}^{d-1,1}} B\wedge A\wedge (dA)^{m-2}\ . \label{eq:BCSM-resp}
\eeq
However, this is not the end of the story as the full
action for the BCSM state contains the counterterm $S_{ct}[A,B]$. When we combine Eq.~\eqref{eq:BCSM-resp} with
the counterterm (neglecting those parts of the counterterm containing $dB$), then we obtain the final expression for the
response of the BCSM,
\beq
	S^{(b)}_{eff}[A,B] = -2k m\left(\frac{e}{2\pi}\right)^{m-1} \int_{\mathbb{R}^{d-1,1}} B\wedge A\wedge (dA)^{m-2}\ ,
\eeq
or in terms of $d$, 
\begin{align}
	S^{(b)}_{eff}[A,B] &= \nnb \\
 -2k&\left(\frac{d}{2}+1\right)\left(\frac{e}{2\pi}\right)^{\frac{d}{2}} \int_{\mathbb{R}^{d-1,1}} B\wedge A\wedge (dA)^{\frac{d}{2}-1}\ , \label{eq:BCSM-EM-final}
\end{align}
where we have restored the charge $e$ of the bosons. This equation is the final form of the response of our BCSM model.

If we set $k=1$ and compare the BCSM response in Eq.~\eqref{eq:BCSM-EM-final} to the fermionic CSM response in
 Eq.~\eqref{eq:CSM-resp}, then 
we see that the response of the BCSM in $d$ dimensions is larger by a factor of $\left(\frac{d}{2}+1\right)!$. To understand
this number recall that our BCSM model in $d$ dimensions is constructed from two copies of the boundary state for a BIQH state 
in $d+1$ dimensions. Setting $d+1=2m-1$, we see that $\left(\frac{d}{2}+1\right)!= m!$, so we find that the
coefficients for the response of the bosonic and fermionic semi-metals in $d$ dimensions differ by exactly the same factor
we found in Sec.~\ref{sec:BIQH} for the coefficients for the response of BIQH and FIQH states in one dimension higher.

We can also see from Eq.~\eqref{eq:BCSM-EM-final} that \emph{at the level of the electromagnetic response}, the BCSM
theory at level $k$ is equivalent to $k$ copies of the BCSM theory at level $1$. However, as a quantum field theory we certainly expect the
theory at level $k$ to be distinct from $k$ copies of the theory at level $1$. This can be seen very clearly in the case where $d=2$. In this
case the BCSM model consists of two copies of an $O(4)$ NLSM with WZ terms at levels $k$ and $-k$, respectively. The $O(4)$ NLSM
can be reformulated in terms of a $2\times 2$ $SU(2)$ matrix field, 
and so (when the coupling constant for the NLSM takes on a particular value), the
$O(4)$ NLSM with WZ term at level $k$ is equivalent to the $SU(2)_k$ Wess-Zumino-Witten conformal field theory. Now the $SU(2)_k$
theory is distinct from $k$ copies of the $SU(2)_1$ theory (this can be seen by comparing central charges), and so we conclude that even in
the simplest case of two dimensions, the BCSM model at level $k$ is distinct (as a quantum field theory) from $k$ copies of the 
BCSM model at level 1. However,
it is entirely possible that $k$ copies of the BCSM model at level $1$ could flow under the Renormalization Group to the BCSM model at level $k$.
In the simple $d=2$ case discussed in this paragraph this flow is allowed by Zamolodchikov's c-theorem\cite{zamolodchikov1986}.

\subsection{Bosonic analogue of the parity anomaly on the boundary of BTI phases}

In this subsection we show that the half-quantized BIQH on the BTI boundary, which we derived
in Sec.~\ref{sec:BTI}, represents a bosonic
analogue of the parity anomaly\cite{Redlich,niemi1983axial,alvarez1985anomalies,witten2015fermion,witten2016parity}, which is an anomaly that is usually associated to massless Dirac fermions in odd dimensions. To start, we give a brief review of the parity anomaly in the fermionic case before 
explaining the bosonic analogue. 

To understand the parity anomaly for Dirac fermions in odd dimensions, 
consider a theory of a single massless Dirac fermion $\Psi$ with $U(1)$ symmetry in $2m-1$ dimensions. We can couple
$\Psi$ to an external electromagnetic field $A$ and then integrate out $\Psi$ to obtain the partition function
\beq
	Z[A]= \int [D\Psi][D\overline{\Psi}]e^{iS[\Psi,A]}\ ,
\eeq
and the effective action for the external field $A$,
\beq
	S_{eff}[A]= -i\ln(Z[A])\ .
\eeq
The action $S[\Psi,A]$ (with $\Psi$ a massless fermion) has an additional discrete symmetry, which is 
time-reversal symmetry
when the spacetime dimension equals $3$ mod $4$, or charge-conjugation (particle-hole) symmetry\cite{elitzur1986origins} when 
the spacetime dimension equals $1$ mod $4$
(in Euclidean spacetime the discrete symmetry in any dimension can be chosen to be reflection of a single coordinate).
However, when one proceeds to calculate the effective action $S_{eff}[A]$, one finds that there is no choice of regularization
procedure which yields an action $S_{eff}[A]$ which has this extra discrete symmetry and is also gauge-invariant. In other words,
one can choose to preserve either the discrete symmetry, or gauge invariance, but not both. For example, when Pauli-Villars 
regularization is used to compute $S_{eff}[A],$ the mass terms for the regulator fermions explicitly break the discrete symmetry,
and this results in the appearance of a term in $S_{eff}[A]$ which also breaks the discrete symmetry. 

At this point it helps to look at a specific example. We choose the case of a massless Dirac fermion 
$\Psi$ in three spacetime dimensions with $U(1)$ symmetry and $\mathbb{Z}_2^T$ time-reversal symmetry,
which was the case originally studied in Refs.~\onlinecite{Redlich,niemi1983axial}. This case also applies
directly to the study of the surface of the familiar electron topological insulator in four dimensions.
 Because
of the $U(1)$ symmetry, $\Psi$ can be coupled to the external field $A$. 
To discuss the transformation of $\Psi$ under time-reversal, it is convenient\cite{son2015composite} 
to choose the gamma matrices in the ``mostly minus" metric to be $\gamma^0= \sigma^z$, $\gamma^1= i\sigma^y$ and 
$\gamma^2= -i\sigma^x$, where $\sigma^a$, $a=x,y,z,$ are the three Pauli matrices
(and recall that a single Dirac fermion in three dimensions has two components). 
With this choice, the time-reversal transformation of $\Psi$ takes the form
\beq
	\mathbb{Z}_2^T:\ \Psi(t,\mb{x}) \to i\sigma^y\Psi(-t,\mb{x})\ ,
\eeq
while the components $A_{\mu}$ of $A$ transform as 
\beqa
	\mathbb{Z}_2^T:\ A_{0}(t,\mb{x}) &\to&  A_{0}(-t,\mb{x}) \\
		A_{i}(t,\mb{x}) &\to&  -A_{i}(-t,\mb{x})\ ,\ i=1,2\ .
\eeqa

In the absence of a mass term for $\Psi$ the
action $S[\Psi,A]$ for $\Psi$ minimally coupled to $A$ has  time-reversal symmetry in addition to the $U(1)$ symmetry. 
However, when Pauli-Villars regularization is used to compute $S_{eff}[A]$, one finds that $S_{eff}[A]$ contains the 
time-reversal-breaking CS term for $A$\footnote{In a more precise treatment Pauli-Villars 
regularization leads to an effective action which is proportional to the Atiyah-Potodi-Singer \emph{eta invariant} of the Dirac 
operator\cite{alvarez1985anomalies}. In certain cases the expression in terms of the
eta invariant can then be replaced with the simpler expression in terms of a CS term with half-quantized level. However, this more 
precise treatment using the eta invariant still gives an effective action that breaks the time-reversal symmetry of the original 
Lagrangian for $\Psi$ and $A$.}. In addition, the level of this CS term is equal to $\pm \frac{1}{2}$, which is half
of the minimum Hall conductance possible for free fermions in three dimensions (i.e., the CS term with level $\pm \frac{1}{2}$
is like a half-quantized Integer Quantum Hall state of fermions). 
One can think of the parity anomaly as a quantum version of the spontaneous breaking of a discrete symmetry. Indeed, the
value of the induced CS term in $S_{eff}[A]$ is determined by the sign of the mass of the Pauli-Villars regulator fermion, and this
choice of sign is arbitrary in the same way that the choice of a particular point on the vacuum manifold of a ``Mexican hat" 
potential is arbitrary. 

To demonstrate that a bosonic analogue of the parity anomaly occurs in the boundary of a BTI phase, we first need to
discuss the symmetries of the BTI theory coupled to $A$. As we discussed in Sec.~\ref{sec:background}, the
NLSM field $n_a$ transforms under the $\mathbb{Z}^T_2$ or $\mathbb{Z}^C_2$ symmetry of the BTI as shown in 
Eq.~\eqref{eq:TC-trans}. Recall that in the case where the $\mathbb{Z}_2$ symmetry is time-reversal, we also need to send 
$t\to -t$ in the argument of the NLSM field $n_a$ and in the action. 
For a spacetime of dimension $d$ (which in our convention is the dimension of the boundary of the SPT 
phase) the field $A$ transforms under time-reversal and charge-conjugation as
\begin{align}
	\mathbb{Z}_2^T:\ A_{0}(t,\mb{x}) &\to  A_{0}(-t,\mb{x}) \\
		A_{i}(t,\mb{x}) &\to  -A_{i}(-t,\mb{x})\ ,\ i=1,\dots,d-1\ ,
\end{align}
and
\beq
	\mathbb{Z}_2^C:\ A_{\mu}(t,\mb{x}) \to - A_{\mu}(t,\mb{x})\ ,\ \forall\ \mu\ ,
\eeq
where $\mb{x}= (x^1,\dots,x^{d-1})$ denotes the spatial coordinates.

The gauged WZ action of Eq.~\eqref{eq:BTI-bdy} for the boundary of the BTI
phase has the $\mathbb{Z}^C_2$ (for $m$ odd) or $\mathbb{Z}^T_2$ (for $m$ even) symmetry of the BTI, in addition to the
invariance under $U(1)$ gauge transformations. To see this we simply note that the counterterms from Eq.~\eqref{eq:r-ct-BTI} 
transform under these two $\mathbb{Z}_2$ symmetries as
\beq
	\mathbb{Z}^T_2:\ S^{(r)}_{ct}[\mb{n},A] \to (-1)^{m}S^{(r)}_{ct}[\mb{n},A] \ ,
\eeq
and
\beq
	\mathbb{Z}^C_2:\ S^{(r)}_{ct}[\mb{n},A] \to (-1)^{m+1}S^{(r)}_{ct}[\mb{n},A]\ .
\eeq
So the gauged WZ action for the BTI boundary has $\mathbb{Z}^T_2$ symmetry for $m$ even and $\mathbb{Z}^C_2$
symmetry for $m$ odd. 

Now, although the gauged WZ action for the BTI boundary has the $\mathbb{Z}_2$ symmetry, we 
have seen in Sec.~\ref{sec:BTI} that it is possible to add the symmetry-allowed term $\mu\ n_{2m+1}^2$ to the 
Lagrangian and drive the boundary of the BTI into a phase in which the $\mathbb{Z}_2$ symmetry is \emph{spontaneously}
 broken by
the condensate $n_a= \pm \delta_{a,2m+1}$. In addition, we showed that when the field $n_a$ condenses in this way it
leads to a CS term in the effective action for $A$ on the boundary of the BTI phase. The CS term in $2m-1$ dimensions
breaks $\mathbb{Z}^T_2$ symmetry for $m$ even, and $\mathbb{Z}^C_2$ symmetry for $m$ odd, so the effective
action for $A$ does not have the $\mathbb{Z}_2$ symmetry of the BTI phase. We also saw
that the CS level turned out to be quantized in \emph{half-integer} multiples of $m!$. 

Since the CS term in the effective action for the boundary breaks the $\mathbb{Z}_2$  symmetry of the BTI phase, and since
the boundary also exhibits a ``half" BIQH response, we conclude that the boundary theory of the BTI phase exhibits an
anomaly in the $\mathbb{Z}_2$ symmetry which is almost an exact analogue of the parity anomaly of a Dirac fermion in odd
dimensions.

There is, however, one important difference between the bosonic analogue of the parity anomaly discussed here and the original
parity anomaly for Dirac fermions. The difference is the fact that in the bosonic case the spontaneous breaking of the
$\mathbb{Z}_2$ symmetry is a classical effect, whereas in the original fermionic case the $\mathbb{Z}_2$ symmetry is broken 
spontaneously only at the quantum level (by the choice of the sign of the mass of the regulator fermions). One likely explanation
for this difference is as follows. Since the NLSM description of the bosonic SPT phase is an effective field theory description, i.e.,
it does not involve the microscopic degrees of freedom in the SPT phase, it is entirely possible that 
the quantum anomaly of any putative microscopic description of the SPT phase is already captured at the classical level in the 
effective NLSM description of the phase. 
This is, in fact, exactly the way in which quantum anomalies of fermionic theories are captured at the classical level in 
effective descriptions of those fermionic theories in terms of gauged WZ actions\cite{witten1983global,manes1985}. 
In addition, we have already seen in this paper that the pertubative $U(1)$ anomaly on the boundary of BIQH
states is completely captured at the classical level in the gauged WZ description of the BIQH boundary.
For this reason we believe that the bosonic analogue of the parity anomaly discussed here is a bona-fide quantum effect 
that occurs in the boundary theory of a BTI phase, and that this anomaly would appear as a true quantum anomaly in a more 
microscopic description of the boundary of the BTI. We are therefore led to argue that, due to this anomaly, the boundary
theory of a $2m$-dimensional BTI phase cannot be realized intrinsically in $2m-1$ dimensions without breaking either
the $U(1)$ or the $\mathbb{Z}_2$ symmetry of the BTI phase.

Finally, let us describe one more way of understanding the bosonic analogue of the parity anomaly in
the specific case of the boundary theory of the four-dimensional BTI. As we know, the boundary theory is an
$O(5)$ NLSM with WZ term in three dimensions. Let us investigate what happens in this theory when we thread a
$2\pi$ delta function flux of the gauge field at a point in space. This method of analysis in known to expose the
parity anomaly in the theory of a single massless Dirac fermion in three dimensions 
(see, for example, Ref.~\onlinecite{karch2016particle}) and so it should work in this case as well.
For simplicity we consider a deformation of the
$O(5)$ theory in which we set $n_5=0$ (this deformation preserves the $U(1)$ and time-reversal symmetry). In this case
the WZ term at level $k$ reduces to a theta term for the four component field $(n_1,\dots,n_4)$ with the theta
angle equal to $\theta=\pi k$. In particular, for $k=1$ (which represents the non-trivial BTI phase in four dimensions)
the WZ term with level $k=1$ reduces to a theta term with theta angle $\theta=\pi$. 

Threading a $2\pi$ delta function flux at a point $\mb{x}_0$ in space will cause the phase of both bosons $b_1= n_1 + in_2$ 
and $b_2= n_3+in_4$ to wind by $2\pi$ about $\mb{x}_0$, i.e., there is a vortex centered at $\mb{x}_0$ in the phase of 
both bosons. In Appendix B of Ref.~\onlinecite{lapa2016bosonic},
two of us performed a detailed study of vortex configurations of a \emph{single} boson 
$b_1$ or $b_2$ in the $O(4)$ NLSM with theta 
term and with $\theta=\pi$. In particular we quantized global fluctuations of the theory over such a vortex background
and showed that the ground state of these fluctuations was doubly degenerate, with the two degenerate states having
charges $\pm\frac{1}{2}$. This analysis confirmed the arguments of Ref.~\onlinecite{VS2013} that a vortex in a single boson
$b_1$ or $b_2$ should carry charge $\pm\frac{1}{2}$. 
As stated above, threading a $2\pi$ flux of $A_{\mu}$ at $\mb{x}_0$ induces a vortex in \emph{both}
$b_1$ and $b_2$ at that point. This composite object has an integer charge and so is naively gauge-invariant, however,
this composite object can actually be shown to be a fermion\cite{SenthilFisher,VS2013,lapa2016bosonic}. This fact
clearly shows the anomaly in theory, as there should be no local fermionic particle with integer charge in a system made 
up of bosons of unit charge. The existence of a fermion with unit charge in the boundary theory of the BTI 
can also be inferred from the presence of a CS term at level $N_3=1$ in the response action for the BTI boundary
using an argument from Ref.~\onlinecite{SenthilLevin}.

\subsection{$\mathbb{Z}_2$ symmetry-breaking domain walls on the boundary of BTI}

We close this section by analyzing the physics of a domain wall between two opposite
$\mathbb{Z}_2$ symmetry-breaking regions on the boundary of a BTI state in $2m$ dimensions. 
In particular, we derive the low-energy theory that
exists on the domain wall, and we show that this theory has a $U(1)$ anomaly which is exactly canceled by the contributions of 
the CS response actions for the $\mathbb{Z}_2$ symmetry-breaking regions on either side of the domain wall. We find
that the theory which lives on the domain wall is identical to the boundary theory for the BIQH phase in $2m-1$ dimensions, and
so this demonstration of anomaly cancellation for domain wall configurations on the BTI boundary provides a nice consistency
check between our gauged actions for BIQH and BTI surfaces. 

To start, recall from Sec.~\ref{sec:BTI} that the boundary of a BTI phase in $2m$ dimensions, 
which is described by an $O(2m+1)$ NLSM with WZ
term at level $k$, can exhibit a $\mathbb{Z}_2$ symmetry-breaking response of the form ($d=2m-1$ is again the boundary
dimension)
\beq
	S_{eff}[A]= \pm \frac{1}{2} \frac{k}{(2\pi)^{m-1}}\int_{\mathbb{R}^{d-1,1}} A\wedge F^{m-1}\ ,
\eeq
when the NLSM field $n_a$ condenses as in Eq.~\eqref{eq:BTI-condense}, i.e., $n_{2m+1} = \pm 1$ and all other
components of $\mb{n}$ equal to zero. As we discussed earlier, this particular condensation pattern is the only one which preserves 
the $U(1)$ symmetry of the BTI phase while breaking the $\mathbb{Z}_2$ symmetry. 

We now consider a domain wall configuration between opposite $\mathbb{Z}_2$ breaking regions on the boundary. Let
$(x^0,\dots,x^{d-1})$ be the boundary spacetime coordinates. We study a configuration of the system in which 
$n_{2m+1}$ condenses as $n_{2m+1}=1$ in the region $x^{d-1}>0,$ and as $n_{2m+1}=-1$ in the region $x^{d-1}<0.$ Hence,
the domain wall is in the $x^{d-1}$ direction. Then on the two sides of the domain wall the electromagnetic response is
given by
\beq
	S_{eff,+}[A] = \frac{1}{2} \frac{k}{(2\pi)^{m-1}}\int_{\mathbb{H}^{d-1,1}_{+}} A\wedge F^{m-1}\ ,
\eeq
and 
\beq
	S_{eff,-}[A] = -\frac{1}{2} \frac{k}{(2\pi)^{m-1}}\int_{\mathbb{H}^{d-1,1}_{-}} A\wedge F^{m-1}\ ,
\eeq
respectively, where $\mathbb{H}^{d-1,1}_{+}$ denotes the half space $\{x \in \mathbb{R}^{d-1,1} |\ x^{d-1} >0\}$, and
similarly for $\mathbb{H}^{d-1,1}_{-}$. If we perform a gauge transformation then the change in the total effective
action is
\beq
	\delta_{\xi}S_{eff,+}[A] + \delta_{\xi}S_{eff,-}[A] = \frac{k}{(2\pi)^{m-1}}\int_{\mathbb{R}^{d-2,1}} \xi F^{m-1}\ , \label{eq:DW-anomaly}
\eeq
where the integration is over the domain wall which is just the space $\mathbb{R}^{d-2,1}$ sitting at $x^{d-1}=0$. Note 
also that the contributions from $S_{eff,\pm}[A]$ add instead of subtract due to the fact that the domain wall is on the 
opposite side of the two regions (the domain
wall lies to the right of the region $\mathbb{H}^{d-1,1}_{+}$ and to the left of the region $\mathbb{H}^{d-1,1}_{-}$,
so when we integrate the total derivative the boundary terms coming from each integral appear with opposite signs).

Next we derive the theory which lives on the domain wall and show that it has a $U(1)$ anomaly which precisely cancels the
gauge transformation from Eq.~\eqref{eq:DW-anomaly}. To derive this theory we analyze the BTI surface theory in the 
presence of a domain wall in $n_{2m+1}$. Concretely, we assume that the $O(2m+1)$ NLSM field takes on the domain wall
configuration,
\beq
	\mb{n}= \{\sin(f(x^{d-1}))\mb{N}(x^0,\dots,x^{d-2}),\cos(f(x^{d-1}))\}\ ,
\eeq
where $\mb{N}$ is a $2m$-component unit vector which depends only on the coordinates $(x^0,\dots,x^{d-2})$ on the
domain wall, and where $f(x^{d-1})$ is a function with the limiting behavior
\beqa
	\lim_{x^{d-1}\to \infty} f(x^{d-1}) &=& 0 \\
	\lim_{x^{d-1}\to -\infty} f(x^{d-1}) &=& \pi\ .
\eeqa
This guarantees that $\mb{n}$ asymptotes to a configuration with $n_{2m+1}= \pm 1$ as $x^{d-1}\to \pm\infty$.
To solve for the theory which lives on the domain wall, we evaluate the $O(2m+1)$ NLSM action (with WZ term) on this 
configuration. Evaluating the kinetic term of the NLSM on the domain wall configuration is simple, provided
that we assume the function $f(x^{d-1})$ is sufficiently well-behaved so that the integration over $x^{d-1}$ gives a finite 
answer. We therefore focus our attention on the WZ term since any anomalous behavior of the domain wall theory
should come from this term. The WZ term involves an extension $\tilde{\mb{n}}$ of the field $\mb{n}$ into a fictitious extra 
direction with coordinate $s\in[0,1]$, and so we need to decide how to extend our domain wall configuration into this extra
direction. Here we assume the extension takes the form
\beq
	\tilde{\mb{n}}= \{\sin(f(x^{d-1}))\tilde{\mb{N}}(s,x^0,\dots,x^{d-2}),\cos(f(x^{d-1}))\}\ , \label{eq:DW-extended}
\eeq
so that all of the $s$-dependence of the extension is in the extended $2m$-component field $\tilde{\mb{N}}$, while the
function $f(x^{d-1})$ still depends only on $x^{d-1}$.

We now examine how the WZ term of the $O(2m+1)$ NLSM reduces on the extended domain wall configuration $\tilde{\mb{n}}$
of Eq.~\eqref{eq:DW-extended}. The analysis is similar (but not identical) to that in Appendix~\ref{app:dim-red-NLSM} for the
dimensional reduction of theta terms in NLSMs on defect configurations of the NLSM field. Recall that the WZ term takes the form
\beq
	S_{WZ}[\mb{n}]= \frac{2\pi k}{\mathcal{A}_{2m}} \int_{[0,1]\times\mathbb{R}^{d-1,1}}\tilde{\mb{n}}^*\omega_{2m},
\eeq
where $\omega_{2m}$ is the volume form for the sphere $S^{2m},$ and $[0,1]\times\mathbb{R}^{d-1,1}$
is the extended spacetime (which we called $\mathcal{B}$ before). Using the relations
\beq
	dn_a = \sin(f)dN_a + \cos(f) N_a df\ ,\ a=1,\dots,2m \ ,
\eeq
and
\beq
	dn_{2m+1} = -\sin(f) df\ ,
\eeq
one can show that on this configuration the volume form $\omega_{2m}$ for $\mb{n}$ reduces to
\beq
	\omega_{2m} \to [\sin(f)]^{2m-1} df\wedge \omega_{2m-1}\ ,
\eeq
where 
\beq
	\omega_{2m-1}= \sum_{a=1}^{2m} (-1)^{a-1} N_a dN_1 \wedge \cdots \wedge \overline{dN_a} \wedge \cdots \wedge dN_{2m}\ ,
\eeq
is the volume form for $N_a$. Since the WZ term involves the pullback $\tilde{\mb{n}}^*\omega_{2m}$ of the volume form
to the extended spacetime, we find that the WZ term reduces as
\begin{widetext}
\beqa
	S_{WZ}[\mb{n}] &\to& \frac{2\pi k}{\mathcal{A}_{2m}}\int_{-\infty}^{\infty} dx^{d-1}\ f'(x^{d-1}) [\sin(f(x^{d-1}))]^{2m-1} \int_{[0,1]\times\mathbb{R}^{d-2,1}}\tilde{\mb{N}}^*\omega_{2m-1} \nnb \\
	&=& \frac{2\pi k}{\mathcal{A}_{2m}}\left(-\frac{\sqrt{\pi}\ \Gamma(m)}{\Gamma(m+\frac{1}{2})}\right)\int_{[0,1]\times\mathbb{R}^{d-2,1}}\tilde{\mb{N}}^*\omega_{2m-1} \nnb \\
	&=& -\frac{2\pi k}{\mathcal{A}_{2m-1}} \int_{[0,1]\times\mathbb{R}^{d-2,1}}\tilde{\mb{N}}^*\omega_{2m-1}\ .
\eeqa
\end{widetext}

We see that the theory localized on the domain wall is an $O(2m)$ NLSM for the field $\mb{N}$, with a WZ term at level $-k$.
This theory also appears at the boundary theory of the BIQH phase in $2m-1$ dimensions, as discussed in Sec.~\ref{sec:BIQH}. 
The extra minus sign on the level of the domain wall 
theory, as compared with the level of the boundary theory of the BTI, is very important. 
Indeed, from our previous formula Eq.~\eqref{eq:WZ-anomaly} for the $U(1)$ anomaly of the $O(2m)$ NLSM with WZ term we see 
that, under a gauge transformation, the theory on the domain wall transforms as
\beq
	\delta_{\xi}S_{DW}[\mb{N},A] =  -\frac{k}{(2\pi)^{m-1}}\int_{\mathbb{R}^{d-2,1}} \xi F^{m-1}\ .
\eeq
This exactly cancels Eq.~\eqref{eq:DW-anomaly}, which was the contribution flowing into the domain wall from the 
$\mathbb{Z}_2$  breaking regions on either side, and so this calculation gives a nice example of anomaly 
inflow at the domain walls on the boundary of SPT phases. It also provides an important consistency check of the gauged 
WZ actions calculated in this paper for the boundaries of BIQH and BTI phases (since it relates the calculation
of the BTI boundary CS response to the calculation of the anomaly of the BIQH boundary theory).

\section{Conclusion}
\label{sec:conclusion}

In this paper we calculated the electromagnetic response of bosonic SPT phases with $U(1)$ symmetry in all spacetime 
dimensions. In particular, we focused our attention on BIQH phases in odd dimensions and BTI phases in even dimensions.
To calculate the response of these phases we used the NLSM description of bosonic SPT phases and the tool of gauged WZ 
actions. This enabled us to compute the 
coefficients $N_{2m-1}$ and $\Theta_{2m}$ which determine, via Eqs.~\eqref{eq:CS-term} 
and \eqref{eq:chern-character}, the electromagnetic response of BIQH and BTI states in all odd and even spacetime 
dimensions, 
respectively. We found that for BIQH states the coefficient $N_{2m-1}$ was quantized in units of $m!$, and that the 
non-trivial BTI state in $2m$ dimensions has $\Theta_{2m}= 2\pi \left(\frac{m!}{2}\right)$. This response for the BTI is equivalent to a 
$\mathbb{Z}_2$ symmetry-breaking Quantum Hall state on the boundary of the BTI with CS level equal to 
$\frac{m!}{2}$, which is exactly half the response of the BIQH state which can be realized intrinsically in the same spacetime 
dimension. 
In Sec.~\ref{sec:gauge-invariance} we showed that the value of $m!$ for the CS level can be understood
by studying the transformation of the CS term under large $U(1)$ gauge transformations on general Euclidean manifolds which 
may or may not admit a spin structure. In that section we also applied this gauge invariance argument to study the 
electromagnetic \emph{and} gravitational responses of fermionic SPT phases with $U(1)$ symmetry in odd spacetime 
dimensions.

We then used our gauged WZ actions for the boundaries of the BIQH and BTI phases to further investigate
the physics of BIQH and BTI states, and to construct other interesting states. In particular, we showed how two copies of the
BIQH boundary theory could be used to construct an effective theory for a bosonic analogue of a Weyl, or chiral, semi-metal 
(a ``bosonic chiral semi-metal" or BCSM state) in any even spacetime dimension. We also showed that the boundary of the BTI state exhibits a bosonic analogue of the parity anomaly of a Dirac fermion in odd dimensions, and
we used this fact to argue that the boundary theory of the BTI in $2m$ dimensions cannot be realized intrinsically in 
$2m-1$ dimensions while preserving the symmetry of the BTI state. We also explored the phenomenon of anomaly inflow
at $\mathbb{Z}_2$ symmetry-breaking domain walls on the boundaries of BTI states.

From a technical point of view one of the most interesting results of the paper is our explicit construction of gauged WZ actions 
for $O(2m)$ and $O(2m+1)$ NLSMs, and with the gauge group $U(1)$. This construction 
allowed us to overcome the difficulties associated with calculating the electromagnetic response of bosonic SPT phases from
their NLSM description. In addition, as we reviewed in Appendix~\ref{app:EC}, 
the problem of constructing a gauged WZ action is equivalent to the 
mathematical problem of constructing equivariant extensions of the volume form on the target space of the NLSM. Then
from a mathematical point of view we can say that we have succeeding in constructing an equivariant extension of the
volume form $\omega_{2m}$ on $S^{2m}$ (this is the BTI case), whereas in the case of $S^{2m-1}$ we have constructed
an extension of $\omega_{2m-1}$ which is almost, but not quite, equivariantly closed (this is the BIQH case). The fact that
we could not construct an equivariant extension of $\omega_{2m-1}$ is mathematically equivalent to the statement that the
boundary theory of the BIQH phase has a perturbative anomaly in the $U(1)$ symmetry, as we expect based on physical arguments.

Our work in this paper opens up many possible directions for future investigations. 
In particular, it would be nice to have a physical argument
along the lines of the one in Ref.~\onlinecite{SenthilLevin} for why the CS level for the BIQH phase is quantized in units of $m!$ in 
$2m-1$ dimensions. Perhaps one can find a physical argument for this quantization by studying generalized braiding processes of 
extended excitations in gapped bosonic phases in higher dimensions, but we do not have any concrete suggestions as to which 
excitations and braiding processes might be relevant. Another possible direction would be to apply the general
method of gauging WZ actions from Ref.~\onlinecite{HullSpence2} to compute the ``electromagnetic" response of SPT
phases with the symmetry of a non-Abelian Lie group $G$. From a mathematical point of view it would also be interesting
to investigate whether the theory of $G$-equivariant cohomology over an appropriate target manifold could be used to 
classify SPT phases with the symmetry of a Lie group $G$. 
Finally, it would be interesting to use the bosonic analogue of the parity anomaly discussed in this paper
as a guide to investigate possible dual descriptions of the boundary of BTI phases in all dimensions, 
analogous to the recent investigations into the dual description of the boundary of the electron topological insulator and
BTI in four spacetime 
dimensions\cite{son2015composite,wang2015dual,metlitski2016particle,mross2015,seiberg2016gapped,seiberg2016duality,karch2016particle,
xu2015,lapa2016bosonic}.

\acknowledgements

MFL would like to thank Michael Stone for introducing him to the literature on the connection between gauged Wess-Zumino
actions and equivariant cohomology, and for numerous helpful discussions on the content of the paper.
CMJ would like to thank Qiaochu Yuan and Xiao-Liang Qi for helpful discussions. 
The authors also thank Shinsei Ryu for helpful comments on the
first version of the paper.
MFL and TLH acknowledge support from the ONR YIP Award N00014-15-1-2383. 
CMJ's research was in part completed in Stanford University under the support of the David and Lucile Packard Foundation. CMJ's research at KITP is supported by the NSF Grant No. NSF PHY-1125915 and a fellowship from the Gordon and Betty Moore Foundation (Grant 4304). PY was supported in part by the NSF through grant DMR 1408713 at the University of Illinois. This work was also done in part at the Perimeter Institute and supported by the Government of Canada through Industry Canada and by the Province of Ontario through the Ministry of Research and Innovation (PY).
MFL, PY, and TLH gratefully acknowledge the support of the Institute for Condensed Matter Theory at UIUC.

\appendix

\section{Equivariant Cohomology Intepretation of Gauged Wess-Zumino Actions}
\label{app:EC}

In this Appendix we review the connection between the theory of gauged WZ actions and equivariant cohomology.
This allows us to give a concrete mathematical interpretation of the form of the gauged WZ actions for the boundary theories
of BIQH and BTI states that we derived in Sec.~\ref{sec:BIQH} and Sec.~\ref{sec:BTI} of this paper. 
Briefly, equivariant cohomology can be thought of as a generalization of de Rham cohomology to the case where a continuous 
group $G$ acts on the manifold. In the cases of interest in this paper the group $G$ is just the group $U(1)$ representing the 
charge-conservation symmetry of the SPT phases we study (i.e., the BIQH and BTI states), and this group acts on the target 
space of the NLSM via the rotations shown in Eq.~\eqref{eq:U1}. The connection between gauged WZ actions and equivariant
cohomology has been explored in Refs.~\onlinecite{papadopoulos,WittenHolo,figueroa1,figueroa2}. The connection was first
discussed explicitly by Witten in Ref.~\onlinecite{WittenHolo} for the case of two spacetime dimensions. Later, 
Figueroa-O'Farrill and Stanciu\cite{figueroa1,figueroa2} considered NLSMs with a generic target space and
in any spacetime dimension, and they gave an explanation of the results of Ref.~\onlinecite{HullSpence2} in terms of 
equivariant cohomology. In addition, Wu\cite{wu1993cohomological} considered the equivalent mathematical problem of 
finding obstructions to the \emph{equivariant extension} (to be defined below) 
of closed differential forms which are invariant under a group action.
The result of these papers is that the problem of constructing a gauge-invariant WZ action 
is equivalent to the problem of constructing
an equivariant extension of the volume form on the target manifold of the NLSM. We now give a brief review of 
equivariant cohomology and the connection to gauged WZ actions in the case where $G=U(1)$, and then we apply this knowledge 
to give a mathematical interpretation of the counterterms of Eq.~\eqref{eq:BIQH-cts} and Eq.~\eqref{eq:r-ct-BTI} which appear 
in the gauged WZ actions constructed in this paper.

\subsection{Equivariant cohomology}

To introduce equivariant cohomology we first need to recall some basic facts about calculus on manifolds. For a $D$-dimensional
manifold $\mathcal{M}$ a vector field $\underline{V}$ in the coordinate patch with coordinates $y=(y^1,\dots,y^D)$ can be 
expanded as
\beq
	\underline{V}= V^a \frac{\pd}{\pd y^a}\ .
\eeq
The partial derivatives $\frac{\pd}{\pd y^a}$ provide a basis for the tangent space $T_y\mathcal{M}$ 
of $\mathcal{M}$ at the point $y$, and a general vector field $\underline{V}$ is a section of the tangent bundle $T\mathcal{M}$ 
of $\mathcal{M}$. The differential forms $dy^a$ provide a basis which is dual to the basis provided by
$\frac{\pd}{\pd y^a}$, i.e., the $dy^a$ form a basis for the cotangent space $T^*_y\mathcal{M}$ at the point $y$. 
A general differential $p$-form $\al$ is a section of the bundle whose fiber over the point $y$ is 
$\bigwedge^p (T^*_y\mathcal{M})$, the $p^{th}$ exterior power of $T^*_y\mathcal{M}$.

Now for any vector field $\underline{V}$ and $p$-form 
$\al= \frac{1}{p!}\al_{b_1\cdots b_p}dy^{b_1}\wedge\cdots\wedge dy^{b_p}$ we can define the operator 
$i_{\underline{V}}$, called \emph{interior multiplication} by $\underline{V}$, by
\beq
	i_{\underline{V}}\al=  \frac{1}{(p-1)!}V^a \al_{a b_2\cdots b_p}dy^{b_2}\wedge\cdots\wedge dy^{b_p}\ .
\eeq
So $i_{\underline{V}}$ takes a $p$-form to a $(p-1)$-form. 
For later use we also note that applying the interior multiplication twice gives zero,
$i_{\underline{V}}^2= 0$, and that $i_{\underline{V}} f = 0$ for any function (zero form) on $\mathcal{M}$.
The Lie derivative $\mathcal{L}_{\underline{V}}$ of any differential form $\al$ along the vector field $\underline{V}$ is then
given by \emph{Cartan's formula},
\beq
	\mathcal{L}_{\underline{V}}\al = d(i_{\underline{V}}\al) + i_{\underline{V}}(d\al)\ ,
\eeq
or simply
\beq
	\mathcal{L}_{\underline{V}}= d i_{\underline{V}} + i_{\underline{V}} d\ , \label{eq:cartan}
\eeq
in operator form.

We are now ready to introduce $U(1)$-equivariant cohomology over $\mathcal{M}$. To start, we pick some
vector field $\underline{V}$ which generates a $U(1)$ action, or circle action, on the manifold. 
This can be understood concretely in terms of the flow generated by $\underline{V}$ as follows. First, recall that a
vector field $\underline{V}$ generates a flow on the manifold via the set of differential equations
\beq
	\frac{d y^a(t)}{dt}= V^a(y^1,\dots,y^D)\ ,\ a=1,\dots,D\ .
\eeq
The condition that $\underline{V}$ generate a $U(1)$ action on the manifold means that this flow 
carries each point on $\mathcal{M}$ along a closed path, and each point takes the same amount of ``time" $t$ to return
to its initial position. Now define the modified exterior derivative
\beq
	\tilde{d}= d - i_{\underline{V}}\ .
\eeq
Note that $\tilde{d}$ takes a $p$-form to a linear combination of a $(p+1)$-form and a $(p-1)$-form. If we compute the square
of $\tilde{d}$ then we find that
\beq
	\tilde{d}^2= -\mathcal{L}_{\underline{V}}\ , 
\eeq
which means that $\tilde{d}^2= 0$ on the subspace of forms which have a vanishing Lie derivative along $\underline{V}$. It is 
therefore possible to define the cohomology of the operator $\tilde{d}$ in this subspace of differential forms in the same way
that one defines the ordinary de Rham cohomology of the exterior derivative $d$. 

Given this structure one can then try to understand what kinds of objects are closed under the action of $\tilde{d}$. 
From the definition of
$\tilde{d}$ it is clear that a differential form of a definite degree will not, in general, be closed under the action of $\tilde{d}$.
Instead, an equivariantly closed ``form" $\al$ is actually a formal linear combination of differential forms of different degrees, 
i.e., a section of the bundle whose fiber over the point $y$ is the exterior algebra 
$\bigwedge(T^*_y\mathcal{M})= \bigoplus_{r=0}^D\ \bigwedge^r (T^*_y\mathcal{M})$.
For the purposes of this paper we are interested in the case where $\al$ is a sum of a form of degree $D$ (the highest
possible degree form on the manifold), and several other forms whose parity (even or odd) is the same as that of the
form of degree $D$. In this case we can expand $\al$ as
\beq
	\al = \sum_{r=0}^{D'} \al^{(D-2r)} \ ,
\eeq
where $\al^{(D-2r)}$ is a differential form of degree $D-2r$ and 
\beq
	D'= \begin{cases}
	\frac{D}{2} & D=\ \text{even} \\
	\frac{D-1}{2} & D=\ \text{odd}\ .
\end{cases}
\eeq
The condition $\tilde{d}\al = 0$ then implies that the forms $\al^{(D-2r)}$ obey the set of equations
\begin{subequations}
\label{eq:equivariant-eqns}
\begin{align}
	  i_{\underline{V}}\al^{(D-2r)} &= d\al^{(D-2r-2)} \ ,\ r=0,\dots,D'-1 \\
	 i_{\underline{V}}\al^{(D-2D')} &= 0\ .
\end{align}
\end{subequations}
In these equations the second line is trivially satisfied in the case that $D$ is even, since in that case $D-2D'= 0$ and
so $\al^{(D-2D')}$ is just a function. The relation $d\al^{(D)}= 0$ is also
trivially satisfied since $\al^{(D)}$ is a highest-degree form on $\mathcal{M}$, and so we have not included it in the set of 
equations for the forms that make up $\al$. The form $\al$ constructed in this way is known as an \emph{equivariant 
extension} of the form $\al^{(D)}$. We now move on and discuss the connection between these ideas and the theory
of gauged WZ actions. 

\subsection{The connection to gauged WZ actions}

To understand the connection between equivariant cohomology and gauged WZ actions, consider a general NLSM with
$D$-dimensional target space $\mathcal{M}$ (a closed, compact manifold). 
We denote the NLSM field by $\mbs{\phi}= (\phi^1,\dots,\phi^D)$, so $\mbs{\phi}$ labels a point on 
$\mathcal{M}$. We formulate this NLSM on a spacetime manifold $\pd \mathcal{B}$ of dimension $D-1$, where $\mathcal{B}$
is an extended spacetime of dimension $D$. So the NLSM field $\mbs{\phi}$ is a map from $\pd\mathcal{B}$ to 
$\mathcal{M}$. Finally, let $\al^{(D)}(\mbs{\phi})$ be the volume form on the target space $\mathcal{M}$. Then a WZ
term for this NLSM takes the form (we absorb any constant factors needed for consistency of the WZ term into the definition
of $\al^{(D)}$)
\beq
	S_{WZ}[\mb{\phi}]= \int_{\mathcal{B}}\tilde{\mbs{\phi}}^*\al^{(D)}\ ,
\eeq
where $\tilde{\mbs{\phi}}$ is an extension of $\mbs{\phi}$ into $\mathcal{B}$ and $\tilde{\mbs{\phi}}^*\al^{(D)}$
again denotes the pullback of $\al^{(D)}$ to $\mathcal{B}$ via the map $\tilde{\mbs{\phi}}$. In what follows we again
omit the pullback symbols $\mbs{\phi}^*$ and $\tilde{\mbs{\phi}}^*$ for notational simplicity. 

Now we suppose that the NLSM has a $U(1)$ symmetry and we attempt to probe this symmetry by coupling the system
to the external field $A$. The transformation of the field $\mbs{\phi}$ under the $U(1)$ symmetry is generated by 
a vector field $\underline{V}$, i.e., under an infinitesimal gauge transformation the field $\mbs{\phi}$ transforms
as 
\beq
	\phi^a \to \phi^a + \xi V^a \ ,
\eeq
where $\xi$ is a small gauge transformation parameter. More generally, a differential $p$-form 
$\beta= \frac{1}{p!}\beta_{a_1\cdots a_p} d\phi^{a_1}\wedge\cdots\wedge d\phi^{a_p}$ on $\mathcal{M}$ transforms under
a small gauge transformation as 
\beq
	\beta \to \beta + \mathcal{L}_{\xi \underline{V}}\beta\ , \label{eq:gauge-trans-lie}
\eeq
where $\mathcal{L}_{\xi \underline{V}}$ is the Lie derivative along the ``small" vector field $\xi\underline{V}$. We can now use
this more general geometric formulation to try and gauge the WZ term. 
We should mention that in the case of a $U(1)$ symmetry it suffices to study the 
change in the action under infinitesimal gauge transformations since there is only one gauge transformation parameter $\xi$
(as opposed to the non-Abelian case where there are several parameters $\xi_{J}$ with $J$ indexing the generators of the
Lie group).

Under a small gauge transformation the WZ term transforms as
\beqa
	\delta_{\xi}S_{WZ}[\mb{\phi}] &=&  \int_{\mathcal{B}}\mathcal{L}_{\xi \underline{V}}\al^{(D)} \nnb \\
	&=&  \int_{\pd\mathcal{B}}\xi (i_{\underline{V}}\al^{(D)})\ ,
\eeqa
where we used the Lie derivative formula Eq.~\eqref{eq:gauge-trans-lie}, the fact
that $d \al^{(D)}=0$, and the property $i_{\xi \underline{V}}= \xi i_{\underline{V}}$ of the interior multiplication. This
change can be canceled by a term
\beq
	S^{(1)}_{ct}[\mb{\phi},A]= \int_{\pd\mathcal{B}} A \wedge \al^{(D-2)}\ ,
\eeq
where $\al^{(D-2)}$ is some $(D-2)$-form, provided that $\al^{(D-2)}$ satisfies the equation
\beq
	i_{\underline{V}}\al^{(D)}= d\al^{(D-2)}\ . \label{eq:descent-eq1}
\eeq
To see this, consider the change in $S^{(1)}_{ct}[\mb{\phi},A]$ when $A \to A + d\xi$. We find a term
\beq
	\int_{\pd\mathcal{B}} d\xi \wedge \al^{(D-2)} = -\int_{\pd\mathcal{B}} \xi  d\al^{(D-2)}\ ,
\eeq
where we performed an integration by parts and ignored boundary terms (since $\pd\mathcal{B}$ has no boundary). At this point
the candidate gauged WZ action takes the form
\beq
	S'_{WZ,gauged}[\mb{\phi},A]=  S_{WZ}[\mb{\phi}]  + S^{(1)}_{ct}[\mb{\phi},A]\ .
\eeq	
Now under a small gauge transformation we find
\beq
	\delta_{\xi}S'_{WZ,gauged}[\mb{\phi},A] = \int_{\pd\mathcal{B}} A \wedge (\mathcal{L}_{\xi\underline{V}}\al^{(D-2)})\ ,
\eeq
which can be reduced to 
\beq
	\delta_{\xi}S'_{WZ,gauged}[\mb{\phi},A] = \int_{\pd\mathcal{B}}\xi F\wedge (i_{\underline{V}}\al^{(D-2)})\ ,
\eeq
with the help of Eq.~\eqref{eq:descent-eq1}, the property $i_{\underline{V}}^2=0$, and an integration by parts.
This change can then be canceled by a term
\beq
	S^{(2)}_{ct}[\mb{\phi},A]= \int_{\pd\mathcal{B}} A \wedge F\wedge \al^{(D-4)}\ ,
\eeq
where $\al^{(D-4)}$ is some $(D-4)$-form that satisfies the equation
\beq
	i_{\underline{V}}\al^{(D-2)}= d\al^{(D-4)}\ , \label{eq:descent-eq2}
\eeq
and so on.

Proceeding in this way we find that a gauge-invariant WZ term can be constructed if and only if there exist differential
forms $\al^{(D-2r)}$, $r=1,\dots,D'$, such that together with the volume form $\al^{(D)}$ they satisfy 
Eqs.~\eqref{eq:equivariant-eqns}. Thus, we find that the problem of constructing a gauge-invariant WZ action is exactly the 
same as the problem of constructing an equivariant extension of the volume form $\al^{(D)}$ on the target space manifold 
$\mathcal{M}$. We now use this information to re-interpret the gauged WZ actions for the boundary theories of the BIQH and 
BTI phases.

\subsection{Application to BIQH and BTI boundary theories}

In the BIQH and BTI cases the vector field $\underline{V}$ which generates the $U(1)$ gauge transformations is
\beq
	\underline{V}= \sum_{\ell=1}^m \left( -n^{2\ell}\frac{\pd}{\pd n^{2\ell-1}} + n^{2\ell-1}\frac{\pd}{\pd n^{2\ell}} \right)\ .
\eeq
Now the NLSMs which describe the boundary of the BIQH and BTI have target spaces $S^{2m-1}$ and $S^{2m}$, respectively. 
We now consider the mathematical problem of constructing equivariant extensions of the volume forms $\omega_{2m-1}$
and $\omega_{2m}$ for these two manifolds. In the BTI case we will see that an equivariant extension of $\omega_{2m}$
exists, and we will give an explicit formula for it. On the other hand, in the BIQH case we will attempt to construct an equivariant
extension of $\omega_{2m-1}$, but we will find that it is not quite closed under the action of $\tilde{d}=d-i_{\underline{V}}$.
This gives a mathematical interpretation of the $U(1)$ anomaly that we found for the 
boundary theory of the BIQH phase. 

We start with the BTI case. Recall from our study of the boundary theory of the BTI that the construction of the gauged WZ 
action involved the forms $\Phi^{(r)}$ from Eq.~\eqref{eq:phi-forms}. If we apply interior multiplication by $\underline{V}$
to these forms we find
\beq
	i_{\underline{V}}\Phi^{(r)}= (m-r) n_{2m+1} dn_{2m+1} \wedge \Phi^{(r+1)}\ , \label{eq:int-mult-phi}
\eeq
which bears a close resemblance to Eq.~\eqref{eq:gauge-trans-odd}. In addition, for the volume form 
$\omega_{2m}$ we have
\beq
	i_{\underline{V}}\omega_{2m}= \frac{1}{(m-1)!} d \left(n_{2m+1} \Phi^{(1)} \right)\ . \label{eq:int-mult-omega-2m}
\eeq
We now use these relations to construct an equivariant extension of $\omega_{2m}$, i.e., a solution of 
Eqs.~\eqref{eq:equivariant-eqns} with $\al^{(D)}= \omega_{2m}$ (so $D=2m$). 
To start we need a form $\al^{(2m-2)}$ which satisfies 
\beq
	i_{\underline{V}}\omega_{2m} = d\al^{(2m-2)}\ ,
\eeq
and from Eq.~\eqref{eq:int-mult-omega-2m} the answer is obviously 
\beq
	\al^{(2m-2)}= \frac{1}{(m-1)!} n_{2m+1} \Phi^{(1)}\ .
\eeq
Next we need a form $\al^{(2m-4)}$ such that 
\beq
	i_{\underline{V}}\al^{(2m-2)} = d\al^{(2m-4)}\ ,
\eeq
and Eq.~\eqref{eq:int-mult-phi} tells us exactly how to find such a form. 
Proceeding in this way we eventually find that an equivariant
extension of $\omega_{2m}$ exists and is given explicitly by
\begin{align}
	\tilde{\omega}_{2m} = \omega_{2m} + \sum_{r=1}^m \frac{1}{(m-r)!(2r-1)!!}(n_{2m+1})^{2r-1}\Phi^{(r)}\ .
\end{align}

The terms appearing in the equivariantly closed form $\tilde{\omega}_{2m}$ are exactly the same as the terms which appear
multiplying the factors $A\wedge F^{r-1}$ in the counterterms of Eq.~\eqref{eq:r-ct-BTI} for the gauged action of the BTI 
boundary. So our construction of a gauged WZ action for the BTI boundary is equivalent to the construction of an 
equivariant extension of the volume form $\omega_{2m}$ on $S^{2m}$. 

Moving on to the BIQH phase, we recall that in the BIQH case the construction of the gauged WZ action involved the forms
$\Omega^{(r)}$ defined in Eq.~\eqref{eq:omega-forms}. Applying the interior multiplication by $\underline{V}$ to
these forms gives
\beq
	i_{\underline{V}}\Omega^{(r)} = \frac{1}{2}d\Omega^{(r+1)}\ , \label{eq:int-mult-BIQH}
\eeq
which bears a close resemblance to Eq.~\eqref{eq:important}. We also saw that the volume form $\omega_{2m-1}$ for 
$S^{2m-1}$ could be written in terms of the $\Omega^{(r)}$ as $\omega_{2m-1}= \frac{1}{(m-1)!}\Omega^{(0)}$. 
Using this fact, and Eq.~\eqref{eq:int-mult-BIQH}, we can then attempt to construct an equivariant extension of
$\omega_{2m-1}$, using the same procedure as in the BTI case. In this way we find a candidate for an equivariant
extension of $\omega_{2m-1}$, which is given explicitly by
\beq
	\tilde{\omega}_{2m-1}= \omega_{2m-1} + \frac{1}{(m-1)!}\sum_{r=1}^{m-1} \frac{1}{2^r}\Omega^{(r)}\ .
\eeq
However, this object is not quite closed under the action of $\tilde{d}=d-i_{\underline{V}}$, and instead we find that
\beq
	\tilde{d}\tilde{\omega}_{2m-1}= -\frac{1}{(m-1)!}\frac{1}{2^{m-1}}\ .
\eeq
In fact, what has happened is that the second line of Eqs.~\eqref{eq:equivariant-eqns} fails to hold in this case. 
This failure of $\tilde{\omega}_{2m-1}$ to be equivariantly
 closed is the mathematical reason for why the BIQH boundary action is not
gauge-invariant, but instead has a pertubative anomaly in the $U(1)$ symmetry.

It turns out that there is a simple mathematical explanation for why an equivariant extension of $\omega_{2m-1}$ does
not exist in this case~\footnote{This explanation was pointed out to us by Michael Stone and we thank him for sharing it with us.}. 
For the $U(1)$ symmetry
considered in this paper (see Eq.~\eqref{eq:U1}) the action of the group $U(1)$ on $S^{2m-1}$ is free, i.e., only the identity element 
of $U(1)$ leaves all the points in $S^{2m-1}$ fixed. In this case the $U(1)$-equivariant cohomology of $S^{2m-1}$ is equal to
the ordinary de Rham cohomology of the quotient manifold $S^{2m-1}/U(1)$ (see, for example, Ref.~\onlinecite{szabo1996}). 
Now for the specific $U(1)$ symmetry we have chosen the quotient is just 
$S^{2m-1}/U(1) = C\mathbb{P}^{m-1}$, and we know that the cohomology ring of $C\mathbb{P}^{m-1}$ is generated
by the K\"{a}hler two-form $K$ (which we will meet in Appendix~\ref{app:CP}). This means
 that only the even-dimensional cohomology groups of $C\mathbb{P}^{m-1}$ are non-trivial. 
On the other hand, the volume form of $S^{2m-1}$ is a $(2m-1)$-form, i.e.,
a form of \emph{odd} degree. Since the $U(1)$-equivariant cohomology of $S^{2m-1}$ is equivalent to the ordinary cohomology 
of $C\mathbb{P}^{m-1}$, we conclude that an equivariant extension of $\omega_{2m-1}$ does not exist for this $U(1)$ 
symmetry (if such an extension did exist, then it would imply the existence of a non-trivial closed form of odd degree on 
$C\mathbb{P}^{m-1}$, but no such form exists).

\section{Chern character on $C\mathbb{P}^m$}
\label{app:CP}

In this Appendix we compute the integral 
\beq
	\int_{X} \left(\frac{F}{2\pi}\right)^{m} \label{eq:integral-appendix}
\eeq
for the specific case of $X=C\mathbb{P}^m$ (complex projective space with $m$ complex dimensions). When the
field strength $F$ satisfies the Dirac quantization condition of Eq.~\eqref{eq:Dirac-condition}
in Sec.~\ref{sec:gauge-invariance} we find that the integral can be equal to one. 
This answer is already well-known, but it provides a nice example of the need for the peculiar quantization of the CS 
level on generic manifolds, as we discussed in Sec.~\ref{sec:gauge-invariance}.

To compute the integral in Eq.~\eqref{eq:integral-appendix} we are going to need some
background information about the complex projective space $C\mathbb{P}^m$. We choose to follow the discussion in 
Ref.~\onlinecite{EGH}. Note that in this section we depart from previous notation and use an overline $\bar{z}$, 
and not a star, to denote the complex conjugate of a complex number $z$.
For $C\mathbb{P}^m$ the second Betti number is $b_2= \text{dim}[H_2(X,\mathbb{R})]= 1$, meaning that
$C\mathbb{P}^m$ has a single non-trivial two-cycle. This two-cycle, 
which we call $\mathcal{C}$, is essentially a copy of $C\mathbb{P}^1$. To understand this two-cycle, and the element of 
$H^2(X,\mathbb{R})$ which is dual to it, first introduce the K\"{a}hler form $K$ on $C\mathbb{P}^m$,
\beq
	K = \frac{i}{2}g_{ab}\ dz^a \wedge d\bar{z}^b\ , \label{eq:kah}
\eeq
where 
\beq
	g_{ab}= \frac{1}{\mathcal{D}^2}\left[\mathcal{D}\delta_{ab} - \bar{z}_a z_b \right]\ ,
\eeq
and
\beq
	\mathcal{D}= 1+ \sum_{c=1}^m z_c \bar{z}_c\ .
\eeq
Here $z_a$, $a=1,\dots,m$,
are complex coordinates which each take values on the whole complex plane $\mathbb{C}$. The indices of $z_a$ can be raised 
and lowered with $\delta_{ab}$ and $\delta^{ab}$, and as usual there is an implied sum over any index which appears once in 
a lower position and once in an upper position in any expression. The quantity $g_{ab}$ is known as the
Fubini-Study metric and it satisfies $\bar{g}_{ab}=g_{ba}$. 
In addition we have $dK= 0$, so $K$ is closed. 
That $K$ is closed follows immediately from the fact that it can be written as
\beq
	K= \frac{i}{2}\pd \overline{\pd} \ln(\mathcal{D})\ ,
\eeq
where $\pd\equiv \pd_{z^a}dz^a$, $\overline{\pd}\equiv \pd_{\overline{z}^a}d\overline{z}^a$ are the Dolbeault operators (on any K\"{a}hler manifold one
has $K= \frac{i}{2}\pd \overline{\pd}\rho$, where the function $\rho$ is called the K\"{a}hler potential). 
Since the exterior derivative decomposes as $d= \pd + \overline{\pd}$, and since the Dolbeault operators satisfy 
$\pd^2 = \overline{\pd}^2 = \{\pd,\overline{\pd}\}=0$, we immediately see that $dK= 0$. 
 Hence, the K\"{a}hler form is closed. However, it is not exact, and we will use it in order to write down non-trivial
configurations of $F$ on $C\mathbb{P}^m$.

The K\"{a}hler form $K$ is a representative of the non-trivial element of $H^2(X,\mathbb{R})$. In the coordinate patch
that we have chosen (in which $K$ takes the form shown in Eq.~\eqref{eq:kah})
we can take the non-trivial two-cycle $\mathcal{C}$ to be any \emph{one} of the $m$ complex planes whose coordinates
are $z_a$. For example let us take $\mathcal{C}$ to be the $z_1$ plane. In that plane (with all other $z_a=0$) we have
\beq
	K \to \left(\frac{i}{2}\right)\frac{dz^1\wedge d\bar{z}^1}{(1+ z_1\bar{z}_1)^2}\ .
\eeq
If we introduce the real coordinates $x_1$ and $x_2$ by $z_1= x_1 + i x_2$ then we have
$dz^1\wedge d\bar{z}^1= - 2 i dx^1 \wedge dx^2$, and integrating $K$ over the $(x_1,x_2)$ plane gives
\beq
	\int_{\mathcal{C}}K= \int \frac{ dx^1\wedge dx^2}{(1 + x_1^2 + x_2^2)^2}= \pi\ . 
\eeq
We learn from this that a normalized form with unit flux through $\mathcal{C}$ is $\frac{K}{\pi}$, so we should set
$\frac{F}{2\pi}=\frac{K}{\pi}$ or just
\beq
	F= 2K\ ,
\eeq
in order to satisfy the Dirac quantization condition of Eq.~\eqref{eq:Dirac-condition}.

Now in order to compute the integral in Eq.~\eqref{eq:integral-appendix} we need to do the integral
\beq
	\int_{C\mathbb{P}^m} K^m\ ,
\eeq
so we need to compute the wedge product of $K$ with itself $m$ times. We have
\begin{widetext}
\beq
	K^m= \left(\frac{i}{2}\right)^m \frac{1}{\mathcal{D}^{2m}} dz^{a_1}\wedge d\bar{z}^{b_1}\wedge \cdots \wedge dz^{a_m}\wedge d\bar{z}^{b_m}\prod_{r=1}^m \left[\mathcal{D}\delta_{a_r b_r} - \bar{z}_{a_r} z_{b_r} \right]\ .
\eeq
\end{widetext}
To simplify this, first note that
\begin{align}
	dz^{a_1}\wedge d\bar{z}^{b_1}\wedge& \cdots \wedge dz^{a_m}\wedge d\bar{z}^{b_m} = \nnb \\
&\ep^{a_1\cdots a_m}\ep^{b_1\cdots b_m} dz^{1}\wedge d\bar{z}^{1}\wedge \cdots \wedge dz^{m}\wedge d\bar{z}^{m}\ .
\end{align}
Now we have to contract $\ep^{a_1\cdots a_m}\ep^{b_1\cdots b_m}$ with the product
\beq
	\prod_{r=1}^m \left[\mathcal{D}\delta_{a_r b_r} - \bar{z}_{a_r} z_{b_r} \right]\ .
\eeq
When expanded out this product contains $2^m$ terms. However, most of these terms contain two or more factors of
$\bar{z}_{a_r} z_{b_r}$, for example a term might contain two of them such as $\bar{z}_{a_1} z_{b_1}\bar{z}_{a_2} z_{b_2}$. All 
such terms with two or more factors of $\bar{z}_{a_r} z_{b_r}$ will vanish when contracted with 
$\ep^{a_1\cdots a_m}\ep^{b_1\cdots b_m}$ because of the anti-symmetry of the Levi-Civita symbol, so we only have
to worry about terms with zero or one factor of $\bar{z}_{a_r} z_{b_r}$. The term with no factors of $\bar{z}_{a_r} z_{b_r}$
is 
\beq
	\mathcal{D}^m\prod_{r=1}^m \delta_{a_r b_r}\ ,
\eeq
and we have
\beq
	\ep^{a_1\cdots a_m}\ep^{b_1\cdots b_m}\mathcal{D}^m\prod_{r=1}^m \delta_{a_r b_r} = m! \mathcal{D}^m\ .
\eeq
Then there are $m$ terms which each have a single factor of $\bar{z}_{a_r} z_{b_r}$. The first such term is
\beq
	-\bar{z}_{a_1} z_{b_1} \mathcal{D}^{m-1}\prod_{r=2}^m \delta_{a_r b_r}\ ,
\eeq
and we find
\begin{align}
	-\ep^{a_1\cdots a_m}\ep^{b_1\cdots b_m}\bar{z}_{a_1} z_{b_1} \mathcal{D}^{m-1}&\prod_{r=2}^m \delta_{a_r b_r} = \nnb \\
&-\mathcal{D}^{m-1} (m-1)! \sum_{c=1}^m z_c \bar{z}_c\ .
\end{align}
So all together we find that (recalling that there are $m$ terms with one factor of $\bar{z}_{a_r} z_{b_r}$ and they all
give an identical contribution)
\beq
	\ep^{a_1\cdots a_m}\ep^{b_1\cdots b_m}\prod_{r=1}^m \left[\mathcal{D}\delta_{a_r b_r} - \bar{z}_{a_r} z_{b_r} \right] = m! \mathcal{D}^{m-1}\ ,
\eeq
where we used $ (\mathcal{D} - \sum_{c=1}^m z_c\bar{z}_c ) = 1$.
We finally obtain
\beq
	K^m= m! \left(\frac{i}{2}\right)^m \frac{ dz^{1}\wedge d\bar{z}^{1}\wedge \cdots \wedge dz^{m}\wedge d\bar{z}^{m}}{\mathcal{D}^{m+1}}\ .
\eeq

To do the integral over $C\mathbb{P}^m$ we now introduce $2m$ real coordinates $x_j$, $j=1,\dots,2m$, defined by
$z_j= x_{2j-1}+ix_{2j}$. Let $r^2= \sum_{j=1}^{2m} x_j^2$. The integral becomes 
\beqa
	\int_{C\mathbb{P}^m} K^m &=& m! \int d^{2m}x \frac{1}{(1+r^2)^{m+1}} \nnb \\
	&=& m!\ \mathcal{A}_{2m-1} \int_0^{\infty} dr\ \frac{r^{2m-1}}{(1+r^2)^{m+1}} \nnb \\
	&=& m!\ \mathcal{A}_{2m-1}\  \frac{1}{2 m} \nnb \\
	&=& \pi^m\ ,
\eeqa
where we used spherical coordinates on $\mathbb{R}^{2m}$ to do the integral.
Setting $F= 2K$, we then find that
\beq
	\int_{C\mathbb{P}^m} \left(\frac{F}{2\pi}\right)^{m} = 1\ .
\eeq

\section{From BIQH to BTI states via dimensional reduction}
\label{app:dim-red}

In this Appendix we discuss a dimensional reduction procedure which allows one to generate a BTI state in $2m-2$ dimensions
from a BIQH state in $2m-1$ dimensions. The procedure is carried out at the level of the effective action $S_{eff}[A]$ and
is similar, but not equivalent to, the procedure used in Ref.~\onlinecite{QHZ2008} to obtain the time-reversal invariant topological
insulator in four dimensions from an Integer Quantum Hall state of fermions in five dimensions. 

To start, we imagine separately gauging the $U(1)$ symmetry associated with each species of ``boson" 
$b_{\ell}$ ($\ell=1,\dots,m$) in the NLSM description of the BIQH state in $2m-1$ dimensions. 
That is, we consider an $O(2m)$ NLSM describing a BIQH state, and we study this 
state with a $U(1)^m$ symmetry which acts on the bosons as
\beq
	b_{\ell} \to e^{i\xi_{\ell}} b_{\ell}\ ,\ \ell=1,\dots,m\ ,
\eeq
where $\xi_{\ell}$ are a set of $m$ independent gauge transformation parameters. We then couple this system to
$m$ $U(1)$ gauge fields $A^{(\ell)}_{\mu}$. 

In this paper we have not calculated the response of the $O(2m)$ NLSM when this $U(1)^m$ subgroup is gauged. However,
from our results in this paper we can make an argument for what the general form should be. The effective
response action $S_{eff}[A^{(1)},\dots,A^{(m)}]$ should have at least two properties: (i) it should reduce to a 
CS response with level $N_{2m-1}= m!$ for the gauge field $A$ if we set $A^{(\ell)}= A$ $\forall \ell$, and (ii) it should
be invariant under any permutation of the labels $\ell$ of the different gauge fields. The second property follows from the fact 
that the action for the $O(2m)$ NLSM is invariant under any permutation of the labels $\ell$ of the bosons $b_{\ell}$.
This fact is not completely obvious, and so we prove it now.

The $O(2m)$ NLSM with theta term or WZ term is invariant under the action of the
\emph{alternating} group $A_{2m}$ of even signature permutations of the labels $a= 1,\dots,2m$ of the components
$n_a$ of the NLSM field $\mb{n}$. Now the permutations of the labels $\ell=1,\dots,m$ of the bosons $b_{\ell}$ consist of
two transpositions in the \emph{symmetric} group $S_{2m}$. This is because a permutation which swaps $\ell$ with $\ell'$ must
swap $n_{2\ell-1}$ with $n_{2\ell'-1}$ and $n_{2\ell}$ with $n_{2\ell'}$. Since the signature of a permutation $\sigma$ in the 
symmetric group is given by $\text{sgn}(\sigma)=(-1)^{N_T}$, with $N_T$ the number of transpositions in $\sigma$, 
it immediately follows that the permutations of the boson labels $\ell$ are contained within the group $A_{2m}$. 
This proves property (ii).

Using properties (i) and (ii) we can now argue that the response action for a gauged $U(1)^m$ symmetry must take the form
\begin{align}
	S_{eff}[A^{(1)},\dots,A^{(m)}] &= \nnb\\
 \frac{k}{(2\pi)^{m-1}m!}&\int_{\mathcal{M}} \Big(A^{(1)}\wedge dA^{(2)}\wedge \dots \wedge dA^{(m)} \nnb \\
&\ \ \ \ \ \ \ \ \ + \text{permutations}\Big)\ .
\end{align}
If $\mathcal{M}$ has no boundary then we can integrate by parts and write this simply as
\begin{align}
	S_{eff}[A^{(1)},\dots,A^{(m)}] &= \nnb \\
	 \frac{k}{(2\pi)^{m-1}}&\int_{\mathcal{M}}A^{(m)}\wedge dA^{(1)}\wedge \dots \wedge dA^{(m-1)} \ ,
\end{align}
or we could choose some other ordering of the gauge fields $A^{(\ell)}$. 
We now describe the dimensional reduction procedure which allows one to derive the
electromagnetic response for the BTI state in $2m-2$ dimensions from this response action for the BIQH state in $2m-1$ 
dimensions. For concreteness we work on flat spacetime with spatial coordinates $x^{j}$, $j=1,\dots,2m-2$.

To obtain the BTI state from the higher dimensional BIQH state we first compactify the space by wrapping the $x^{2m-2}$
direction into a circle, which turns the space $\mathbb{R}^{2m-2}$ into the ``cylinder" $\mathbb{R}^{2m-3}\times S^1$.
We then thread a $\pi$ flux of the gauge field $A^{(m)}$ through the hole in the cylinder, and finally shrink the circumference
of the cylinder to zero. This leaves us with a response action for a phase in $2m-2$ spacetime dimensions. Mathematically, this
procedure assumes the following configuration of gauge fields $A^{(\ell)}$: (i) $A^{(\ell)}$ 
$\ell=1,\dots,m-1$, are independent of $x^{2m-2}$ and have their last component $A^{(\ell)}_{\mu=2m-2}$ equal to zero,
(ii) the components $A^{(m)}_{\mu}$, $\mu=0,\dots,2m-3$, of the $m^{th}$ gauge field are equal to zero, 
and (iii)
the last component of $A^{(m)}$ satisfies $\int dx^{2m-2}\ A^{(m)}_{2m-2}= \pi$. 

Under these assumptions the effective
action for the BIQH phase with gauged $U(1)^m$ symmetry reduces as
\begin{align}
	S_{eff}[A^{(1)},\dots,A^{(m)}] &\to \nnb \\
 \frac{\pi k}{(2\pi)^{m-1}}&\int_{\mathbb{R}^{2m-3,1}}dA^{(1)}\wedge \dots \wedge dA^{(m-1)}\ .
\end{align}
If we now break the remaining $U(1)^{m-1}$ symmetry of this phase down to $U(1)$ by setting $A^{(\ell)}=A$ for 
$\ell=1,\dots,m-1$ then we obtain (with $F=dA$)
\beq
	S_{eff}[A^{(1)},\dots,A^{(m)}] \to \frac{\pi k}{(2\pi)^{m-1}}\int_{\mathbb{R}^{2m-3,1}}F^{m-1}\ ,
\eeq
which is a response action of the form of Eq.~\eqref{eq:chern-character} with response parameter 
$\Theta_{m-1}= \pi (m-1)! k$. This is exactly the response action for a BTI phase in $2m-2$ dimensions, so the dimensional
reduction procedure described here does allow one to obtain the BTI response action from the BIQH response action in one
higher dimension. 

The main difference between the dimensional reduction procedure shown here and the procedure used in 
Ref.~\onlinecite{QHZ2008} is that in our procedure we only thread $\pi$ flux of \emph{one} flavor $\ell$ of gauge field 
$A^{(\ell)}$ through hole in the cylinder. On the other hand, the procedure in Ref.~\onlinecite{QHZ2008} (in which there
is only a single $U(1)$ gauge field $A$) is equivalent to threading $\pi$ flux of \emph{all} the gauge fields $A^{(\ell)}$. 
This second method \emph{does not} give the proper quantization of the parameter $\Theta_{m-1}$
for the BTI phase in one lower dimension. The answer turns out to be too large by a factor of $m$. The physical 
reason for the different dimensional reduction procedure needed to go from BIQH to BTI states can be seen from our 
alternative calculation in Sec.~\ref{sec:BIQH} of the
BIQH response. There we showed that threading $2\pi$ flux of the external gauge field $A$ generates a vortex in 
all $m$ bosons $b_{\ell}$, and so it creates $m$ excitations. This explains why the more familiar dimensional reduction
procedure of threading $\pi$ flux for $A$ gives an answer which is $m$ times too large.

\section{Dimensional reduction formulas for theta terms in nonlinear sigma models}
\label{app:dim-red-NLSM}

In this section we derive a general dimensional reduction formula for theta terms of $O(D+2)$ NLSMs in $D+1$ dimensions. 
The formula shows how the theta term of the NLSM can reduce to a theta term for a lower-dimensional NLSM when evaluated
on a ``defect configuration" of the NLSM field. The formula we derive applies to any spacetime dimension $D+1,$ and defects of any 
codimension, with the simplest cases being vortices and hedgehog defects. The physical content of the dimensional reduction 
formula can be summarized in the following way: a topological defect of (spatial) co-dimension $(q+1)$ in an $O(D+2)$ NLSM
with theta term can trap an $O(D-q+1)$ NLSM with theta term in its core. The theta angle of the lower-dimensional NLSM
is related to that of the original NLSM in a simple way which we calculate below. 
We use a special case of this formula in the last subsection of Sec.~\ref{sec:BIQH} to study vortices in the NLSM description of the 
BIQH state, but the general result presented in this Appendix should be very useful for working with these models.
Dimensional reduction of topological terms in NLSMs was also considered in Appendix C of Ref.~\onlinecite{CenkeDual}, but to the best of our 
knowledge the general formula presented in this Appendix has not appeared before in the literature. Finally, we also remark that
the formula presented here can also be used when WZ terms are present, as the form of the WZ term is similar to the form of the
theta term.

To start, recall the theta term
\beq
	 S_{\theta}[\mb{n}]= \frac{\theta}{\mathcal{A}_{D+1}}\int_{\mathbb{R}^{D,1}} \mb{n}^*\omega_{D+1}\ , \label{eq:theta-term-app}
\eeq 
for a NLSM with field $\mb{n}(t,\mb{x})$ in $D+1$ spacetime dimensions, where $t$ represents the time, and 
$\mb{x}= (x^1,\dots,x^D)$ represents the spatial coordinates. Here $\omega_{D+1}$ is the volume form for the sphere
$S^{D+1}$ which was introduced in Eq.~\eqref{eq:volume-form}. The integral is over $(D+1)$-dimensional Minkowski 
spacetime $\mathbb{R}^{D,1}$. To describe the defect configurations considered here, we first decompose the total 
spacetime as
\beq
	\mathbb{R}^{D,1}= \mathbb{R}^{q+1}\times\mathbb{R}^{D-(q+1),1}\ ,
\eeq
and we further decompose the first factor into $(q+1)$-dimensional spherical coordinates as
\beq
	 \mathbb{R}^{q+1}= [0,\infty)\times S^q\ .
\eeq
Here $q$ is a positive integer which is going to be related to the codimension of the defect in the NLSM field. 

We introduce coordinates $r\in[0,\infty)$ and $\mb{s}= (s^1,\dots,s^q)$ to parametrize 
$\mathbb{R}^{q+1}= [0,\infty)\times S^q$. The precise nature of the coordinates $\mb{s}$ for $S^q$ will not be important
to us here. We also use $t$ and $\mb{y}= (y^1,\dots,y^{D-(q+1)})$ to denote the remaining coordinates on 
$\mathbb{R}^{D-(q+1),1}$. The defect configuration we consider takes the form
\beq
	\mb{n}(t,\mb{x})= \{ \sin(f(r))\mb{N}(t,\mb{y}),\cos(f(r))\mb{m}(\mb{s})\}\ , \label{eq:defect-config-q}
\eeq
where $\mb{N}$ is a $(D-q+1)$-component unit vector which depends only on the coordinates $(t,\mb{y})$ for 
$\mathbb{R}^{D-(q+1),1}$, $\mb{m}$ is a $(q+1)$-component unit vector which depends only on the coordinates $\mb{s}$ for 
$S^q$, and where $f(r)$ is a function obeying the boundary conditions
\beqa
	f(0) &=& \frac{\pi}{2} \\
	\lim_{r\to\infty} f(r) &=& 0\ .
\eeqa
Physically, this form of $\mb{n}$ describes a defect of spatial codimension $q+1$ in which the field $\mb{m}$ takes on a non-trivial
configuration on the sphere $S^q$. The field $\mb{N}$ then describes a lower-dimensional NLSM which lives in the core of this
defect, and the core size is controlled by the profile of the function $f(r)$. The non-triviality of the configuration of 
$\mb{m}$ is captured by the winding number $n_q$ of $\mb{m}$ on $S^q$,
\beq
	n_q= \frac{1}{\mathcal{A}_q}\int_{S^q}\mb{m}^*\omega_q\ . \label{eq:defect-winding}
\eeq

After some algebra one can show that the pullback $\mb{n}^*\omega_{D+1}$ of the volume form for the original NLSM field
$\mb{n}$ will reduce on this configuration as
\begin{widetext}
\beq
	\mb{n}^*\omega_{D+1} \to (-1)^{(D-q)(q+1)+1}[\sin(f(r))]^{D-q}[\cos(f(r))]^q f'(r) dr\wedge\mb{m}^*\omega_q \wedge \mb{N}^*\omega_{D-q}\ .
\eeq
\end{widetext}
This formula can be derived from the formula for $\mb{n}^*\omega_{D+1}$ by using the fact that wedge products of the differential
of any coordinate with itself will vanish. This fact strongly constrains the terms which survive in $\mb{n}^*\omega_{D+1}$
once one assumes that $\mb{n}$ is in the defect configuration of Eq.~\eqref{eq:defect-config-q}. Now we just need to do the
integrals over the radial direction (parameterized by $r$) and the sphere $S^q$ to find the reduced theta term form 
$\mb{N}$. For the radial integral we have
\beqa
	I_{r} &\equiv&  -\int_0^{\infty} dr\  [\sin(f(r))]^{D-q}[\cos(f(r))]^{q} f'(r) \nnb \\
	&=& \int_0^{\tfrac{\pi}{2}} df\  [\sin(f)]^{D-q}[\cos(f)]^{q} \nnb \\
	&=& \frac{\Gamma(\frac{D-q+1}{2})\Gamma(\frac{q+1}{2})}{2\Gamma(\frac{D}{2}+1)}\ .
\eeqa
Combining this with Eq.~\eqref{eq:defect-winding} for the winding of the defect in the $\mb{m}$ field, we find that the
theta term of Eq.~\eqref{eq:theta-term-app} for $\mb{n}$ reduces as
\begin{align}
	 S_{\theta}[\mb{n}]&\to \nnb \\
	 &(-1)^{(D-q)(q+1)}\frac{\theta I_r}{\mathcal{A}_{D+1}} \int_{S^q}\mb{m}^*\omega_q \int_{\mathbb{R}^{D-(q+1),1}}\mb{N}^*\omega_{D-q} \nnb \\
	&=\frac{\theta_{eff}}{\mathcal{A}_{D-q}}\int_{\mathbb{R}^{D-(q+1),1}}\mb{N}^*\omega_{D-q}\ , \label{eq:dim-red-theta}
\end{align}
where the effective theta angle for the lower-dimensional NLSM is
\beq
	\theta_{eff}= (-1)^{(D-q)(q+1)} n_q\ \theta\ , \label{eq:theta-eff}
\eeq
and where we used the formula
\beq
	 \frac{\Gamma(\frac{D-q+1}{2})\Gamma(\frac{q+1}{2})}{2\Gamma(\frac{D}{2}+1)}\frac{\mathcal{A}_q}{\mathcal{A}_{D+1}} = \frac{1}{\mathcal{A}_{D-q}}\ .
\eeq

So we see that on this defect configuration the original theta term for $\mb{n}$ has reduced to a theta term for the field
$\mb{N}$ which lives in the core of the defect. In addition, from Eq.~\eqref{eq:theta-eff} we see that the theta angle $\theta_{eff}$ 
for this lower-dimensional NLSM is simply related to the original theta angle by a sign factor $(-1)^{(D-q)(q+1)}$ and by multiplication 
by the winding number $n_q$ of the defect in $\mb{m}$.

\section{Electromagnetic Response of $O(2)$ NLSM in one spacetime dimension}
\label{app:O2}

In this Appendix we derive the electromagnetic response of the $O(2)$ NLSM with theta term, which represents an analog of the 
BIQH state in $1$ spacetime dimension. In the last subsection of Sec.~\ref{sec:BIQH} we presented an alternative derivation of
the electromagnetic response of the $O(2m)$ NLSM with theta term at $\theta=2\pi k$ in $2m-1$ dimensions, 
in which we were able to relate the level $N_{2m-1}$ of the CS term in the response for the $O(2m)$ NLSM to the level $N_1$ for the 
response of the $O(2)$ NLSM at $\theta=2\pi k$. Specifically, we found that the two levels were related as
\beq
	N_{2m-1}= (m!) N_1\ .
\eeq
We now derive the formula
\beq
	N_1= -k\ , \label{eq:O2-response}
\eeq
for the $O(2)$ NLSM with $\theta=2\pi k$, which we then use to complete our alternative derivation at the end of 
Sec.~\ref{sec:BIQH} of the formula $N_{2m-1}= -(m!)k$ for the CS response of the $O(2m)$ NLSM with $\theta=2\pi k$.

We begin the derivation by parameterizing the $O(2)$ field as $\mb{n}= \{\cos(\vphi),\sin(\vphi)\}$ or as
$b_1 = e^{i\vphi}$ in terms of the boson $b_1= n_1 + i n_2$. In terms of the angular
variable $\vphi$ the action for the $O(2)$ NLSM with theta term takes the form
\beq
	S[\mb{n}]= \int_0^T dt\ \left\{ \frac{1}{2g}(\pd_t\vphi)^2 + \frac{\theta}{2\pi}\pd_t\vphi \right\}\ .
\eeq
Here we have made the calculation as concrete as possible by considering a finite time interval $[0,T)$, and we assume
periodic boundary conditions for the boson $b_1$ in the time direction. This leaves open the possibility that $\vphi$ can
wind around the time direction, i.e., we can have configurations in which $\vphi(t+T)=\vphi(t) + 2\pi n$ for an integer $n$. 
As in the higher-dimensional cases, we will be interested in the limit $g \to \infty$. In this one-dimensional case this limit
just projects onto the ground state (or states) of this quantum mechanical system (in higher dimensions $g\to\infty$ corresponds
to the disordered phase of the model).

The $U(1)$ symmetry which acts on $b_1$ as $b_1 \to e^{i\xi}b_1$ then acts on $\vphi$ as
\beq
	U(1):\ \vphi \to \vphi+\xi\ .
\eeq
We would now like to couple $\vphi$ to a $U(1)$ gauge field $A= A_t dt$. For the boundary conditions we are considering we 
can write $A_t$ in the general form
\beq
	A_t= \overline{A}_t + \delta A_t\ ,
\eeq
where $\overline{A}_t= \frac{1}{T}\int_0^T dt\ A_t$ and $\int_0^T dt\ \delta A_t= 0$. This is equivalent to 
the statement that the closed form $A$ can be written as an exact part plus a piece which has a non-vanishing integral around the 
non-trivial one-cycle in the time direction, 
since we assumed periodic boundary conditions in time. We can always remove the exact part $\delta A_t$ from
$A_t$ by a small $U(1)$ gauge transformation $\vphi \to \vphi + \xi$, $A \to A+ d\xi$ with $\int d\xi = 0$. Therefore we will just
work with the constant part $\overline{A}_t$ in what follows. 

The gauged $O(2)$ NLSM action is obtained by the standard minimal coupling procedure,
\begin{align}
	S_{gauged}[\mb{n},A] &= \nnb \\
\int_0^T dt\ &\left\{ \frac{1}{2g}(\pd_t\vphi-\overline{A}_t)^2 + \frac{\theta}{2\pi}(\pd_t\vphi-\overline{A}_t) \right\}\ , \label{eq:O2-gauged}
\end{align}
however, there is one subtle point here. 
This action is invariant under small and large $U(1)$ gauge transformations, where by a large $U(1)$
gauge transformation we mean a transformation in which $\int d\xi \neq 0$. Now if we only cared about invariance under small
$U(1)$ gauge transformations, we could just as well have used the action
\beq
	S'_{gauged}[\mb{n},A]= \int_0^T dt\ \left\{ \frac{1}{2g}(\pd_t\vphi-\overline{A}_t)^2 + \frac{\theta}{2\pi}\pd_t\vphi \right\}\ ,
\eeq
which \emph{does not} involve minimal coupling inside the theta term. This form is more relevant in cases in which one is interested
in enforcing certain discrete symmetries \emph{at the expense} of large $U(1)$ gauge invariance, as could be the case in the 
investigation of global anomalies in discrete symmetries of this theory at $\theta=\pi$ (compare with
the discussion for fermionic systems in one dimension in Ref.~\onlinecite{elitzur1986origins}). 
This could be relevant for studies of the boundary states of SPT phases in two spacetime dimensions. 
In our case, however, we are interested in the $O(2m)$ NLSM in $2m-1$ dimensions as a low-energy description of a bosonic lattice
model which can be coupled to a \emph{compact} $U(1)$ gauge field, 
and so we gauge the theory in such a way as to preserve this large $U(1)$ gauge invariance. With these remarks in mind,
we now proceed with the computation. 

From Eq.~\eqref{eq:O2-gauged} the momentum conjugate to $\vphi$ is 
$p = \frac{1}{g}(\pd_t\vphi -\overline{A}_t) +\frac{\theta}{2\pi},$ and the Hamiltonian is
\beq
	H= \frac{g}{2}\left(p - \frac{\theta}{2\pi}\right) + p\overline{A}_t\ .
\eeq
To quantize, we impose the commutation relations $[\vphi,p]= i$ (we set $\hbar=1$ here), and we use the Schrödinger
representation $p = -i\pd_{\vphi}$. The eigenfunctions of $p$ and $H$ are then the Fourier modes 
$\psi_n(\vphi)= \frac{1}{\sqrt{2\pi}}e^{i n\vphi}$,
$n\in\mathbb{Z}$. We now restrict ourselves to the case of $\theta=2\pi k$ and $g\to \infty,$ which is the case for which we are
trying to calculate the electromagnetic response. Then the ground state is 
$\psi_k(\vphi)= \frac{1}{\sqrt{2\pi}}e^{i k\vphi}\equiv \lan \vphi | G.S.\ran,$
and the partition function (vacuum-to-vacuum transition function) in this case is 
\beq
	Z[A]= \lan G.S. |e^{-iHT} |G.S.\ran = e^{-i k T \overline{A}_t}\ ,
\eeq
or in terms of the original field $A= A_t dt$\ ,
\beq
	Z[A]=  e^{-i k \int_0^T dt\ A_t}= e^{-ik \int A}\ .
\eeq
The effective action is then
\beq
	S_{eff}[A]= -i\ln(Z[A])= -k\int A\ , 
\eeq
from which Eq.~\eqref{eq:O2-response} immediately follows.

%\bibliography{SPT-response-refs}


\end{document}